%% file: WP_07_YoungSNRs.tex
\documentclass[11pt,a4paper]{article}

\input{ASTROHWhitePaperStyle.tex}
\input{WP_AuthorList.tex}

\begin{document}

\newcommand{\WhitePaperTitle}{Young Supernova Remnants}
\newcommand{\WhitePaperAuthors}{
J.~P.~Hughes~(Rutgers~University),
S.~Safi-Harb~(University~of~Manitoba),
A.~Bamba~(Aoyama~Gakuin~University),
S.~Katsuda~(JAXA),
M.~Leutenegger~(NASA/GSFC),
K.~S.~Long~(STScI),
Y.~Maeda~(JAXA),
K.~Mori~(Miyazaki~University),
H.~Nakajima~(Osaka~University),
M.~Sawada~(Aoyama~Gakuin~University),
T.~Tanaka~(Kyoto~University),
H.~Uchida~(Kyoto~University),
H.~Yamaguchi~(NASA/GSFC~\&~UMD),
F.~Aharonian~(DIAS~\&~MPI-K),
S.~Funk~(Stanford~University),
J.~Hiraga~(University~of~Tokyo),
M.~Ishida~(JAXA),
K.~Koyama~(Kyoto~University),
H.~Matsumoto~(Nagoya~University),
M.~Nobukawa~(Kyoto~University),
M.~Ozaki~(JAXA),
T.~Tamagawa~(RIKEN),
H.~Tsunemi~(Osaka~University),
H.~Tomida~(JAXA),
Y.~Uchiyama~(Rikkyo~University), and
S.~Uno~(Nihon~Fukushi~University)
}
\MakeWhitePaperTitle

\def\ah{{\it ASTRO-H}}
\def\suzaku{{\it Suzaku}}
\def\chandra{{\it Chandra}}
\def\xmm{{\it XMM-Newton}}

\newcommand{\av}{\ensuremath{A_{V}}}
\newcommand{\ji}{\textit{J}}
\newcommand{\hi}{\textit{H}}
\newcommand{\ki}{\textit{K}}
\newcommand{\ksi}{\textit{K\ensuremath{_S}}}
\newcommand{\hii}{H\,$_{\rm{II}}$}
\newcommand{\kt}{\ensuremath{k_{\rm{B}}T}}
\newcommand{\lx}{\ensuremath{L_{\rm{X}}}}
\newcommand{\lbol}{\ensuremath{L_{\rm{bol}}}}
\newcommand{\fx}{\ensuremath{F_{\rm{X}}}}
\newcommand{\nh}{\ensuremath{N_{\rm H}}}
\newcommand{\nel}{\ensuremath{n_{\rm e}}}
\newcommand{\msun}{\ensuremath{M_{\odot}}}
\newcommand{\casa}{Cassiopeia~A}
\newcommand{\aap}{Astronomy \& Astrophyics}
\newcommand{\nat}{Nature}
\newcommand{\apj}{ApJ}
\newcommand{\apjl}{ApJL}
\newcommand{\apjs}{ApJS}
\newcommand{\aj}{AJ}
\newcommand{\pasj}{PASJ}
\newcommand{\mnras}{MNRAS}
\newcommand{\etal}{et al.}

\begin{abstract}
Thanks to the unprecedented spectral resolution and
  sensitivity of the Soft X-ray Spectrometer (SXS) to soft thermal
  X-ray emission, \ah\ will open a new discovery window
  for understanding young, ejecta-dominated, supernova remnants
  (SNRs).  In particular we study how \ah\ observations will
  address, comprehensively, three key topics in SNR research: (1)
  using abundance measurements to unveil SNR progenitors, (2) using
  spatial and velocity distribution of the ejecta to understand
  supernova explosion mechanisms, (3) revealing the link between the
  thermal plasma state of SNRs and the efficiency of their particle
  acceleration.
\end{abstract}

\maketitle
\clearpage

\tableofcontents
\clearpage

\section{Using \ah\ to Type Supernovae and Unveil their Progenitors}

\subsection{Background and Previous Studies}

Unlike unresolved X-ray sources, supernova remnants (SNRs), because of
their significant extent, have not benefited from the high-resolution
spectroscopic capabilities provided by the gratings on \xmm\ and
\chandra.  Indeed virtually all studies of SNRs have been carried out
with CCD-type spectral resolution (i.e., $E/\Delta E \sim 10$).  This
is sufficient to resolve He-like and H-like lines of the major
elemental species (O, Ne, Si), but it has not allowed us to perform
true spectral analysis based on individual resolved emission lines.
This is especially frustrating for SNR aficionados, since SNR spectra
are arguably among the most complex in astrophysics with dependencies
on elemental composition, electron temperature, ion temperature,
ionization and recombination timescales, turbulent and bulk velocity
flows, and the significant presence of non-thermal particle
distributions.  Thus high spectral resolution observations with \ah's Soft X-ray Spectrometer (SXS)
will provide a huge ``game-changing'' leap forward for X-ray studies of
SNRs.

As the principal production and dispersal sites of stellar and
explosive nucleosynthesis, SNRs hold important keys to our
understanding of the numerous theoretical processes involved in the
production of the chemical elements (see \citealt{wallerstein97} for
an update of \citealt{burbidge57}). In general, however, this theory has been
tested largely against ensemble-averaged measurements, such as the
relative abundance distribution of the elements in various
environments (e.g., solar), the atmospheres of stars, and so on.
Nucleosynthesis is a rich, complex field that involves many disparate
processes operating in different environments and at different phases
of stellar evolution. Observational tests of specific model
components, especially of the most energetic processes, are woefully
lacking.  In particular, the processes that produce Fe and Fe-group
elements and eject them into the interstellar medium during the
explosions of both core-collapse and thermonuclear supernovae (SNe)
are among the most poorly tested parts of the entire nucleosynthesis
picture.  X-ray studies of young SNRs (see Vink 2012
and references therein for a review) can provide
critical tests of the nucleosynthesis picture especially as regards
the production of Fe and Fe-group elements in specific individual
examples of core-collapse and thermonuclear SNe.  In addition since
the production of Fe and Fe-group elements is at the heart of these
explosions, they offer critical insights into the explosion processes
in SNe.

Core-collapse (CC) SNe make up roughly three-quarters of all observed
SNe.  They come from stars more massive than $\sim$$8 M_\odot$ and from a
nucleosynthetic point of view are the dominant producers of O, Ne, and
Mg, although they do produce a broad spectrum of elemental species
including the Fe-group. They leave compact remnants in the form of
neutron stars (or black holes for the most massive progenitors), 
while their gaseous remnants tend to be
highly structured with dense optically-emitting knots and typically
more diffuse X-ray features.  It is known that the precipitating event
for a CC SNe is the collapse of a stellar core, but the process
whereby the core rebounds and ejects the rest of the star is still
poorly understood with at least two competing ideas currently in
vogue: neutrino-driven convection (e.g., \citealt{herant94, burrows95,
  kifonidis00}), including in its latest development the instability
of the stalled shock (e.g., \citealt{blondin06, foglizzo06, burrows07,
  scheck08}); and jet-induction (e.g., \citealt{khokhlov99}).  It is
also the case that at present nucleosynthesis predictions still rely
on spherically symmetric models with an assumed neutron star/black
hole ``mass cut'' (e.g., \citealt{woosley07}).

The other main class of SNe are the thermonuclear or Type Ia SNe (SN
Ia).  These make up roughly one-quarter of all SNe and are widely
believed to result from the total incineration of a carbon-oxygen
white dwarf that grows close to the Chandrasekhar mass.  How the star
increases its mass is unknown; single degenerate scenarios where the
white dwarf accretes matter from a normal-star companion in a binary
and double degenerate scenarios where two white dwarfs coalesce are
the two favored possibilities.  During the explosion about half of the
star's mass is converted to $^{56}$Ni which decays to stable
$^{56}$Fe.  Even with hundreds of well-observed type Ia SNe and the
intense focus of the theoretical community over the past 15 years, the
SN Ia explosion process, i.e., how nuclear ignition occurs and the
subsequent burning proceeds, remains an unsolved problem (e.g.,
\citealt{ ropke08, jordan08}).  Models that most successfully
reproduce optical spectra of SNe Ia essentially parameterize the speed
of the burning front through the star (e.g., \citealt{iwamoto99}).
These models also usually include a parameterized transition from a
subsonic burning front (deflagration) to a supersonic burning front
(detonation).

Observational results on the X-ray properties of young SNRs obtained
to date, as important and relevant as they are, suffer, fundamentally,
from a lack of precision.  CCD-type spectral resolution is unable to
provide accurate measurements of even the most basic thermodynamic
quantities that characterize the plasma, i.e., electron temperature
and the charge states of the relevant ions. This introduces large
errors in relative abundance measurements.  Studies of dynamics are
limited to only the most extreme speeds (1000's of km s$^{-1}$).
Significant advances require improved spectral resolution in order to
resolve individual lines and derive plasma diagnostics from line ratios.

Although the X-ray emission from SN ejecta provides powerful
diagnostics of the progenitor's nature, we should keep in mind that
what we see is only the part of the ejecta that has been shock-heated
to X-ray emitting temperatures. Our discussions, therefore, must be
influenced by the dynamical state of the particular SNR under
study. In most young Type Ia SNRs (such as Tycho, SN~1006, and
0509$-$67.5), the reverse shock is still propagating through the
Fe-rich central region (where C-O has been burnt to nuclear
statistical equilibrium), while the other elements (Si, S, Ar, and Ca)
at the outer incomplete Si-burning layer have already been fully
heated (e.g., \citealt{hwang97, warrenhughes04}).  Therefore, the
observed abundance ratios between Fe and the intermediate mass species
cannot be used to, for example, infer the transition density from
deflagration to detonation to which it is most sensitive. Similar
issues of ``hidden'' or unshocked ejecta exist for core-collapse SNRs, and in
these cases the assumption of stratified ejecta will be even more
untenable. Nevertheless, \ah\ SXS spectra of young SNRs will allow
us to determine the temperature and ionization age of \textit{each}
species visible and thereby constrain the spatial variation of
thermodynamic conditions throughout the remnant. This will offer more
``grist'' to constrain hydrodynamical simulations, such as those of
\citet{badenes03, badenes06}.

\subsection{Prospects \& Strategy}

Here we present three important scientific research topics that
\ah\ can address regarding young supernova remnants. Most of these
require the high spectral resolution of the SXS.

\subsubsection{Typing Ia supernovae from abundance measurements of their 
remnants: addressing the prompt vs.\ delayed mechanisms}

Original work on typing SNe from their remnants \citep{hughes95}
identified three SNRs in the Large Magellanic Cloud (LMC) as
originating from type Ia explosions using the gross difference in the
expected X-ray spectra of CC (more O-rich) vs.\ Ia (more Fe-rich) SNe.
One of the candidates (SNR 0509$-$67.5) was subsequently confirmed as
a {\it spectroscopic} Ia by measurement of the light echoes from the
original SN \citep{rest08a}.  Another of the candidate SN~Ia remnants
(N103B) is unusual in that it lies near a young, rich star cluster in
the LMC and appears to be interacting with dense circumstellar
gas. There is also some disagreement in interpretation between the
\chandra\ ACIS spectro-imaging data \citep{lewis03} and the
\xmm\ integrated RGS spectrum \citep{vanderheyden02}.  Both studies
find significant emission from oxygen, silicon, and iron, as well as
other species.  \citet{vanderheyden02} chose to interpret their global
spectrum as coming entirely from SN ejecta and concluded that the
presence of large amounts of oxygen argued for a core-collapse origin.
On the other hand \citet{lewis03} found strong evidence for a
shell-like remnant structure in the emission from Si, S, and Fe, while
the O emission was distributed in an entirely different manner and was
more closely correlated with the continuum.  These authors argued that
the oxygen emission was therefore from a different component, namely
the swept-up interstellar medium, and that the SN~Ia hypothesis was
still viable and indeed preferred. Indeed the properties of N103B bear
some resemblance to Kepler's SNR, another putative SN~Ia remnant with
interstellar interaction \citep{reynolds07}.

The Tycho SNR also shows light echoes \citep{rest08b}.  Optical
spectra of these have revealed the originating SN to belong to the
majority class of normal type Ia SN \citep{krause08b}, a result that
confirms earlier predictions by \citet{badenes06} based on their
interpretation of the X-ray spectrum of Tycho.  For the remnant
mentioned above, 0509$-$67.5, on the other hand, light echo spectra
showed that it belongs to the bright, highly energetic subclass of
Type Ia explosions, similar to SN 1991T.  This subclass is widely
suspected of producing a larger mass of $^{56}$Ni during the
explosion, which, again, is consistent with the X-ray properties of
the remnant \citep{badenes08a}.

Studies of large samples of Type Ia SNe have revealed the need for at
least two evolutionary paths to these explosions: a prompt component
(whose rate is proportional to the star formation rate) and a delayed
one (proportional to the total stellar mass) (e.g.,
\citealt{scannapieco05}).  It has been suggested (e.g.,
\citealt{aubourg08}) that the prompt type could have progenitors that
are relatively young massive stars with high metallicity.
Therefore, the single degenerate scenario is often linked to the
prompt population, while the double degenerate to the other. Moreover,
recent systematic observations suggest that star-forming (and/or
high-z) galaxies tend to host bright Type Ia SNe (e.g.,
\citealt{howell09, sullivan10}), implying that single-degenerate
progenitors can more easily induce bright SNe. X-ray observations of
SNRs offer a powerful test of this hypothesis. Some Type Ia SNRs
(e.g., Kepler, RCW~86, N103B) are suspected of being associated with
star-forming regions (e.g., \citealt{rosado96, chu88}). Do they belong
to this recently recognized subclass of Ia SNe (i.e., prompt)? There
are already some pieces of evidence that Kepler and RCW~86 have single
degenerate origins \citep{williams11, williams12}. 
These important questions and details are best addressed with \ah.

For completeness we note that \chandra\ has revealed other, older
Magellanic Cloud SNRs to be type Ia based on the composition of the
ejecta contained within their interiors: DEM L71 \citep{hughes03}; DEM
L238 and DEM L249 \citep{borkowski06}; and 0104$-$72.3 \citep{lee11}.
Others have been shown to be of core-collapse origin: e.g., 0103$-$72.6
\citep{park03}.  Some of these objects may be interesting targets in
conjunction with studies of Magellanic Cloud and Galactic middle-aged ($\sim$10~kyr-old) remnants, where
evidence for metal-rich SN ejecta is being sought or found (e.g. 
0454--67.2 in the LMC, Seward et al. 2006; see also Section 1.3.7).

With \ah\ SXS, we will make precise abundance measurements of 
Type Ia SNRs that include ``normal'' type Ia remnants, such as 
Tycho SNR, SN~1006, and SNR~0519$-$69.0,
and possible ``prompt'' ones like Kepler and N103B.  With
sensitive line strength measurements, combined with accurate
ionization and temperature diagnostics, we will make accurate
abundance determinations. Comparison to theoretical models of
nucleosynthetic yields (e.g., \citealt{iwamoto99,maeda10b}) will
provide insight into the possible explosion models for SNe Ia and the
differences between the prompt and delayed evolutionary channels.

\subsubsection{Mn-to-Cr mass ratios, odd-Z and radioactive species}

The excellent resolving power of \ah's SXS will allow a detection 
and study the chemical elements beyond the most abundant even-Z
species O, Ne, Mg, Si, S, Ar, Ca and Fe with which X-ray astronomers
have been working almost exclusively, to date. Recent exceptions to
this general rule are the detection of the trace abundance species Cr
($Z=24$) and Mn ($Z=25$) in several SNRs using CCD spectra. Cr and Mn
are the species with the highest abundances (DEX abundances of 5.67
and 5.39, respectively, \citealt{ag89}) between Ca and Fe.

W49B was the first remnant in which Cr and Mn emission lines were
detected \citep{hwang00b}; see the Old SNRs+PWNe White Paper (WP \#8) for a discussion
of this SNR.  Currently, thanks to \suzaku\ observations
(see Table 2 in \citealt{yang13}), there are six SNRs with Mn line
detections (Kepler, W49B, N103B, 3C~397, Tycho, and G344.7$-$0.1) and
an additional two SNRs for which only Cr has been detected (Cas~A and
SNR 0519$-$69.0).  In the following Table we list these 8 SNRs in
order of their Cr line strength.  The three columns show the total SXS
count rate in each of the three lines (using
{\texttt{sxt-s\underbar{~}120210\underbar{~}ts02um\underbar{~}of\underbar{~}intallpxl.arf}}
and assuming all of the photons arrive at the mean energy as shown).

\begin{center}
\begin{tabular}{|lccc|}  \hline\hline
SNR name     & Cr rate (5.6 keV)    & Mn rate (6.1 keV)    & Fe rate (6.6 keV) \\ \hline
W49B         &   $1.5\times 10^{-2}$ &   $4.8\times 10^{-3}$ &   $4.0\times 10^{-1}$ \\
Cas A        &   $1.1\times 10^{-2}$ &   $\ldots$           &   $7.8\times 10^{-1}$ \\
Tycho        &   $8.0\times 10^{-3}$ &   $7.7\times 10^{-4}$ &   $2.2\times 10^{-1}$ \\
3C~397        &   $4.8\times 10^{-3}$ &   $2.8\times 10^{-3}$ &   $5.9\times 10^{-2}$ \\
Kepler       &   $1.6\times 10^{-3}$ &   $8.7\times 10^{-4}$ &   $1.3\times 10^{-1}$ \\
G344.7$-$0.1 &   $1.2\times 10^{-3}$ &   $3.5\times 10^{-3}$ &   $8.8\times 10^{-3}$ \\
N103B        &   $5.2\times 10^{-4}$ &   $3.2\times 10^{-4}$ &   $1.0\times 10^{-2}$ \\
0519$-$69.0  &   $2.6\times 10^{-4}$ &   $\ldots$           &   $4.1\times 10^{-3}$ \\ \hline\hline
\end{tabular}
\end{center}

Metallicity is one possible difference in the properties of the
progenitor between the prompt and delayed channels for Type Ia SN
explosions.  A recent powerful advance in this area came from the
recognition that the Mn-to-Cr mass ratio in Type Ia explosion models
tightly correlates with the initial metallicity of the original white
dwarf's progenitor \citep{badenes08b}.  The basic idea is that the
mass ratio of Mn to Cr produced though nuclear burning during the
explosion depends sensitively on the electron-to-nucleon fraction
($Y_e$) in the white dwarf.  \citet{timmes03} have already shown that
$Y_e$ is linearly proportional to the metallicity of the white dwarf
progenitor.  Applying this technique to the \suzaku\ spectrum of the
Tycho SNR where Mn and Cr were both detected \citep{tamagawa09},
\citet{badenes08b} were able to show that the metallicity of Tycho's
progenitor star was supersolar.  Although the errors were large,
values of metallicities much below solar can be rejected.  Since then
deeper \suzaku\ observations of Tycho, Kepler, and several other Ia
SNRs have been taken and refined estimates of the progenitor
metallicities are being obtained (see, for Kepler,
\citealt{parkkepler13}).

It is also claimed that the delayed (and hence dim and probably
double-degenerate) SNe Ia yield lower abundance ratios of Cr/Fe,
Mn/Fe, and Ni/Fe than the prompt (brighter, single-degenerate) SNe Ia
(Tsujimoto \& Shigeyama 2012). Therefore, not only the Mn/Cr abundance
ratio but also the (Cr+Mn+Ni)/Fe ratio should be compared among
different types of Type Ia SNRs.

If the Mn and Cr lines are unbroadened, then the \ah\ SXS will have
greater sensitivity (than CCDs) to the line detection and will have
better ability to measure the charge states of these species for
comparison to the charge state of Fe.  In fact a joint study of the
SXS and Soft X-ray Imager (SXI) spectra of these objects may be most effective in
clarifying the ionization conditions and thereby derive accurate
relative masses for these species.

The detectability of Cr and Mn K-shell lines in CCD spectra is
possible, not only because the cosmic abundance of these species is
sufficiently high, but also because their K-shell line energies
($\sim$5.6 keV and $\sim$6.1 keV) fall in a generally line-free region
of the X-ray band.  Detecting and studying odd-Z species with lower
atomic number is made much more difficult by confusion with stronger
lines at nearby photon energies from the more abundant even-Z species.
The high spectral resolution of the SXS will help overcome
these effects of confusion and make the first robust detections of
these species.

In order of decreasing abundance (assuming solar values from
\citealt{ag89}), the odd-Z species we intend to study are Al
(abundance in DEX = 6.47; He-like K-shell energy = 1.6 keV) and Na
(6.33, 1.1 keV), then P (5.45, 2.1 keV), all of which are more
abundant than Mn (5.39).  Just slightly below this threshold are Cl
(5.27, 2.8 keV) and K (5.12, 3.5 keV). Most of these species K-shell
lines (with the possible exception of Na) are far from the Fe-L blend
and hence it will be possible to establish unique line identifications
for them.  Given that yields of all the neutron-rich elements depend
on the progenitor's metallicity, these odd-Z elements could also be a
useful metallicity probe. Notably, Al emission can be observed around
1.6 keV, where the effective area of the SXS is much larger than that
in the Mn-K$\alpha$ band. Moreover, lighter elements should have lower
ion temperatures, and hence expected thermal broadening will be
smaller (although Doppler broadening from turbulence and bulk motion
will still pertain). A previous {\it Suzaku} observation of a possible Type
Ia SNR, G344.7$-$0.1, already detected strong Al emission from this
remnant (\citealt{yamaguchi12}, although this emission may originate
from swept-up ISM). The SXS will enable us to detect this line from
most of the other Ia SNRs.

If a SNR is young enough, emission from radioactive species can become
detectable. The prime candidate species for this is $^{44}$Ti, which
decays through electron capture to $^{44}$Sc. Electron capture leaves
a K-shell vacancy in the Sc atom that relaxes through the emission of
a fluorescence line at 4.1 keV (which is in the SXS band).  This is fully independent of the
reverse-shock heating and provides an estimate of the amount of
$^{44}$Ti synthesized by the SN. The decay of $^{44}$Ti also produces
nuclear $\gamma$-rays at energies of 67.9, 78.4 and 1157 keV; the
former two lines can, in principle, be detected by the HXI.  Finally
$^{44}$Ti is synthesized at extremely high temperatures in the deepest
layers of the exploding star and, in the case of core-collapse SNe, near the mass
cut.  Recent simulations of off-center delayed-detonation Type Ia
explosions showed a synthesized $^{44}$Ti mass of $\sim$$2 \times
10^{-5}$ \msun\ \citep{maeda10b}. On the other hand, core-collapse SN
models predict a larger mass $> 6 \times 10^{-5}$ \msun\ \citep{timmes96},
or even more if the explosion is asymmetric \citep{nagataki98}.

\subsubsection{Core-collapse remnants: Is the observed diversity of their
 neutron stars linked to their supernova progenitors?}

Here we discuss how to use \ah\ to infer the progenitor masses of SNRs
associated with neutron stars of several different types.

Core-collapse SNRs provide unique laboratories to study, not only the
heavy elements created in the Universe and the physics of shocks and
particle acceleration, but also the origin of the most exotic,
compact, and highly-magnetized stars in the Universe: neutron stars
and their associated nebulae.
When a star with a ZAMS mass in the range $\sim 8-20\, M_\odot$ ends
its life, in addition to expelling several solar masses of gaseous
ejecta, it leaves behind a neutron star \citep{woosley02}. The most
famous example of a CC SNR is the Crab nebula believed to be
associated with SN~1054AD.

While Ia SNRs are believed to expand in relatively uniform media,
core-collapse explosions expand in media strongly modified by
mass loss from the SN progenitor, naturally affecting their
dynamics, morphology and spectral signatures (e.g. Dwarkadas 2005, Lopez et al. 2011). 
The presence of a compact object in the SNR interior, surrounded by a synchrotron nebula
for the rotation-powered neutron stars, further complicates
the analysis of these objects, but also offers another tool for probing
their SN progenitors and explosion properties (e.g. Chevalier 2005).

The past decade has witnessed the discovery of a growing diversity of
neutron stars associated with SNRs (e.g. Mereghetti 2013). These include the ``classical"
rotation-powered pulsars typically discovered at radio wavelengths and
with dipole surface magnetic fields $B$$\sim$10$^{12}$~G (like the
Crab)\footnote{the question on the absence of SNR shells in Crab-like
  nebulae is by itself a puzzling question that is being addressed in
  the ``Old SNRs and PWNe" White Paper \#8}, the X-ray and gamma-ray
discovered `magnetars', commonly believed to be neutron stars with
super-strong $B$$\sim$10$^{14}$--10$^{15}$~G and with two flavors (the
anomalous X-ray pulsars [AXPs] and the soft gamma-ray repeaters), high-B
pulsars (HBPs) with magnetic fields intermediate between the Crab-like
pulsars and magnetars and close to the quantum electrodynamic value of
4.3$\times$10$^{13}$~G (like PSR J1846--0258 in SNR Kes~75), and the
Central Compact Objects (CCOs) discovered in X-rays near the centers
of their hosting SNRs (like the CCO discovered with \chandra\ in the
O-rich and young SNR Cas~A) and dubbed as `anti-magnetars' with
$B$$\sim$10$^{10}$--10$^{11}$~G.  Understanding the origin for this
diversity remains one of the hottest topics in the field today. While
studying any connection between these apparently different classes is
being conducted with dedicated studies of the compact objects
themselves\footnote{see the ``High-Mass X-ray Binaries and Magnetars"
  White Paper \#4 for a detailed discussion on this topic}, another promising
approach to this puzzle is through studying their hosting SNRs.  SNRs
should provide clues on their environment, energetics and the
supernova explosions that created them.  In particular, it is not
clear whether the diversity of neutron stars is related to a diversity
in supernova progenitors of core-collapse supernovae.

Of particular interest, there has been recently growing evidence for
the highly magnetized neutron stars (i.e. HBPs and magnetars) to
originate from \textit{very} massive stars (of mass exceeding $\sim$20
\msun).  Although this remains controversial, X-ray and
multi-wavelength studies seem to be pointing to this conclusion (see,
e.g., \citealt{gaensler05, kumar2012, SSHKumar2013}).  X-ray
spectroscopy of SNRs provides a powerful tool to infer the mass of the
progenitor star. This is achieved through modelling the X-ray spectra
of ejecta-dominated SNRs; and subsequently comparing the
inferred abundances of the metals detected in the 0.3--10 keV band (O
through Fe) to nucleosynthesis models (e.g.
\citealt{WW95, Nomoto06}).

Below we show how to take advantage of the Soft X-ray Spectrometer
(SXS)'s unprecedented spectral resolution and sensitivity to the
emission lines expected from young SNRs to type the progenitors and
determine or pin down their masses in a number of promising targets.
The proposed science will further illustrate the leap that will be
achieved in high-resolution X-ray spectroscopy of \textit{extended sources}
such as SNRs, a task that is challenging with gratings on-board
\textit{XMM-Newton} and \chandra.

\subsection{Targets \& Feasibility}

\begin{figure}[ht]
  \begin{center}
   \includegraphics[height=1.95in]{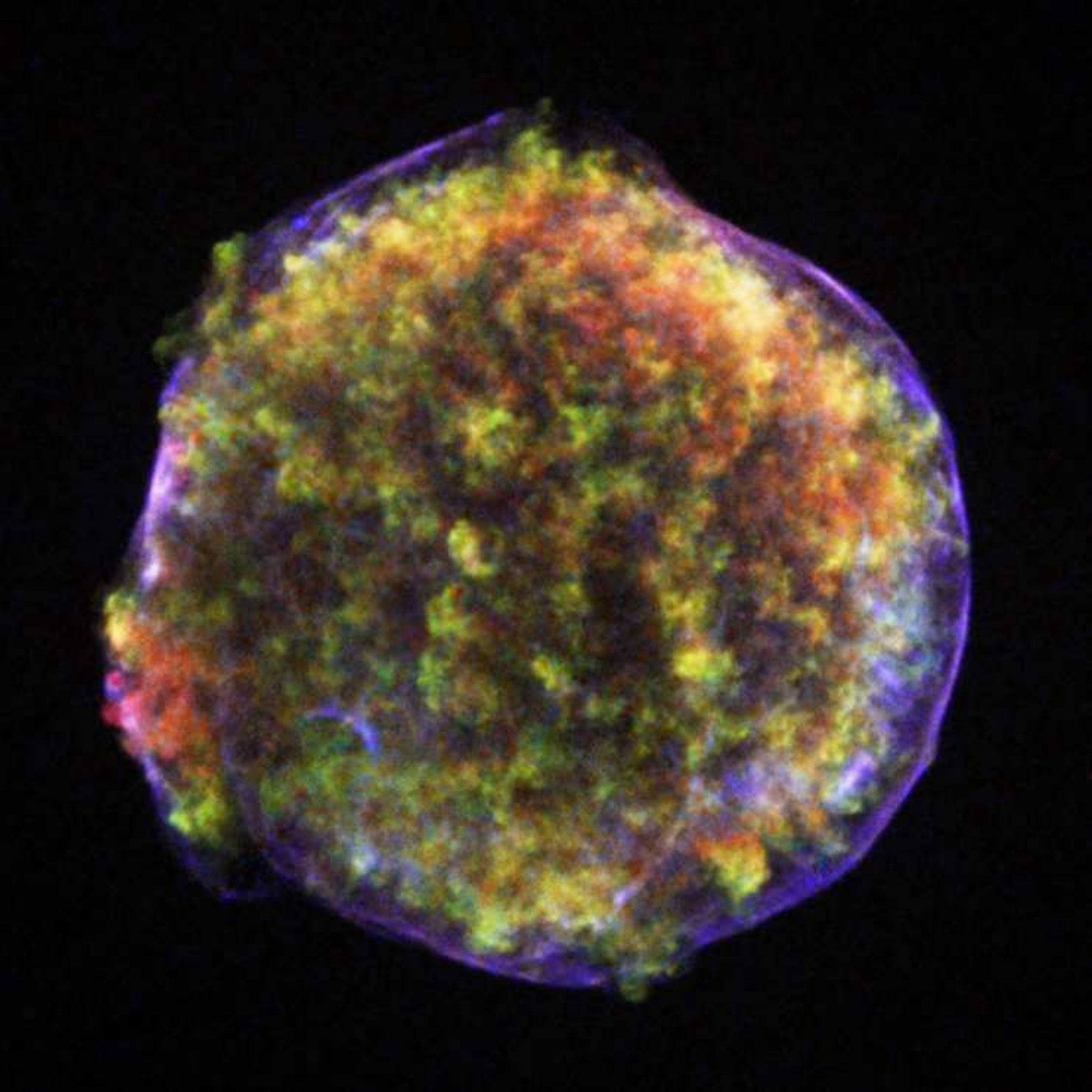}
     \includegraphics[height=1.95in]{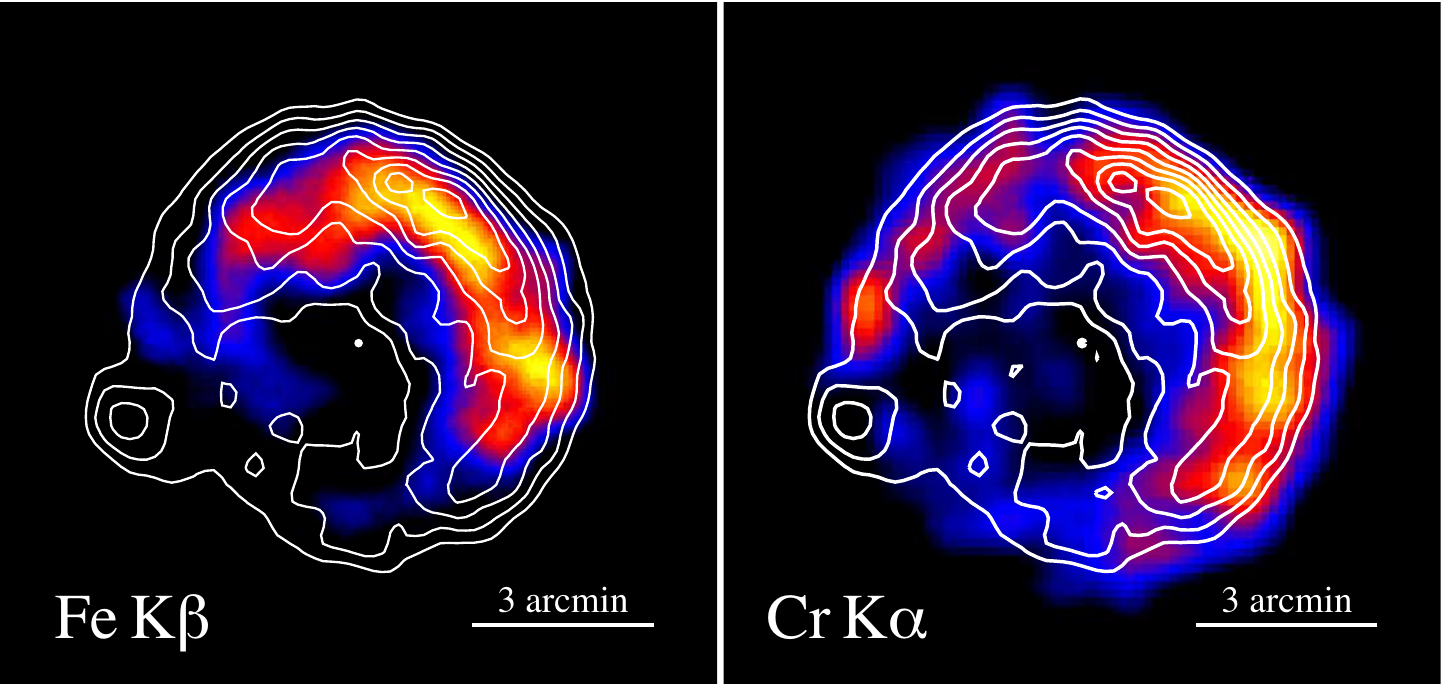}
  \caption{X-ray images of Tycho's SNR from \chandra\ {\it (left
      panel)}, the \suzaku\ XIS in the Fe-K$\beta$ band
    (7.0--7.2\,keV) {\it (middle panel)} and Cr-K$\alpha$ band
    (5.4--5.6\,keV) {\it (right panel)}. The \chandra\ image shows the
    Fe L band in red, Si K in green and 0.4-6.0 keV continuum band in
    blue. The non-thermal continuum is subtracted from the
    \suzaku\ images. The Fe-K$\alpha$ image (6.43--6.53\,keV) is
    over-plotted in contours.  The Fe-K$\beta$ emission peaks at a
    smaller radius than the Fe-K$\alpha$ emission in the bright NW
    rim, while the Cr-K$\alpha$ emission peaks at a larger radius (Yamaguchi et al. 2014).  
    The profile of the
    Fe-K$\beta$ emission reflects the spatial distribution of the
    low-ionized Fe ions (see the NSF White Paper for details), so is also
    important as an indicator of the reverse shock position.  }
  \label{tycho:suzaku}
  \end{center}
\end{figure}

\subsubsection{Tycho's SNR: A unique laboratory for high-resolution spectroscopy} \label{subsec:tycho}

\begin{figure}
  \begin{center}
     \includegraphics[height=2.5in,clip,trim=148 250 149 250]{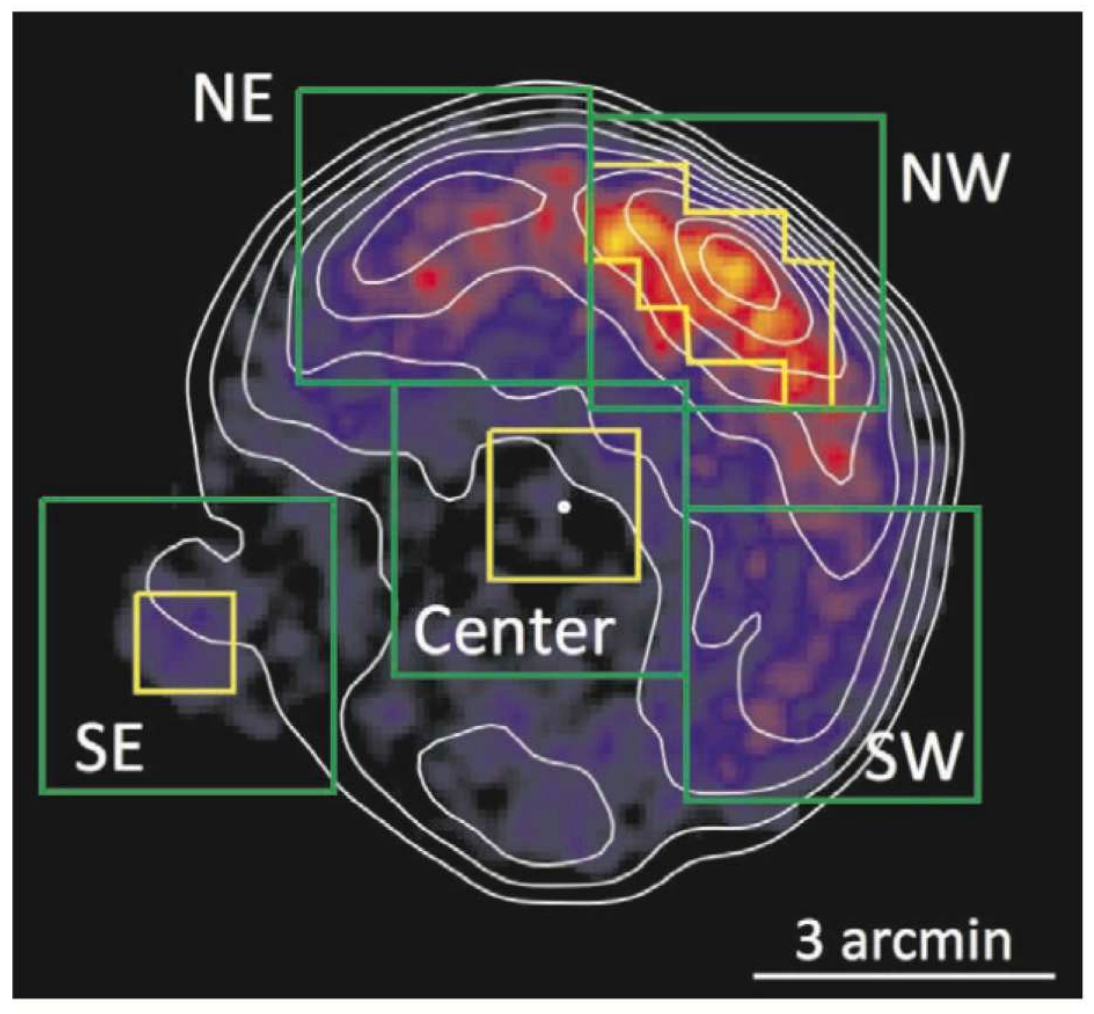}
     \includegraphics[height=2.5in,clip,trim=190 290 190 290]{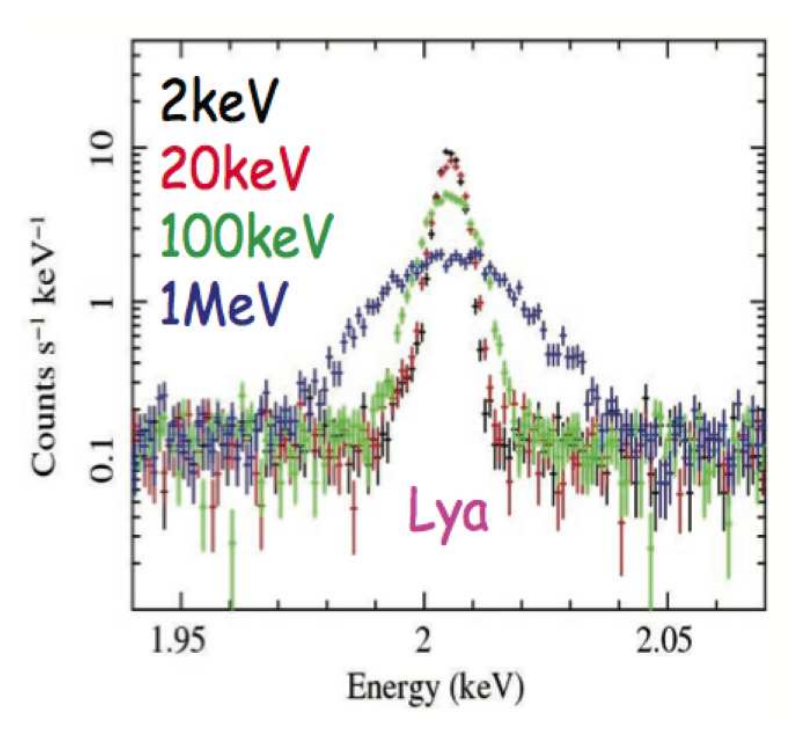}
  \caption{{\it (Left)} {\it Suzaku} XIS image of Tycho's SNR in
    Fe-K$\alpha$ (6.43--6.53\,keV) (in color) with the Si K$\alpha$
    image (1.75--2.05\,keV) over-plotted as contours.  Five possible
    \ah\ SXS pointings are shown with the green boxes. {\it (Right)}
    100-ks SXS simulation of the Si Ly-$\alpha$ line showing the effect of 
   changing the line broadening depending on the Si ion temperature.}
  \label{tycho:ptgs}
  \end{center}
\end{figure}

Tycho is the brightest and most prototypical (in fact ``normal'') Type
Ia SNR in the Milky Way.  The thermal spectrum is fully dominated by
the ejecta with little evidence of emission from shocked ISM
\citep{cassamchenai07}.  Tycho also shows evidence of local
inhomogeneity in the ejecta (e.g., \citealt{vancura95}), specifically
in the form of two compact knots, one Fe-rich, the other Si-rich,
lying at the edge of the SNR rim in the SE quadrant (see
Figure~\ref{tycho:suzaku}--left panel).  The narrowband images of
Tycho's SNR obtained by {\it ASCA} show that the Fe-K$\alpha$ emission is
located interior to the emission lines of intermediate-mass elements
\citep{hwang97}.  This was confirmed by higher-resolution observations
by \chandra\ \citep{warren05} and \xmm\ \citep{decourchelle01}.  With
{\it Suzaku}, {\citet{tamagawa09} detected emission from Cr and Mn,
  which were later shown to be important elemental species with which
  to probe the progenitor's metallicity \citep{badenes08b}.  The 
  {\it Suzaku} data also allowed measurement of the ejecta expansion
  velocity from line broadening due to superposition of red- and
  blue-shifted components.  \citet{hayato10} found the expansion
  velocities of Si, S, and Ar ejecta ($4700 \pm 100$\,km\,s$^{-1}$) to
  be distinctly higher than that obtained from the Fe-K$\alpha$
  emission ($4000 \pm 300$\,km\,s$^{-1}$), which is consistent with
  segregation of the Fe in the inner ejecta.

In 2008 a deeper {\it Suzaku} observation was performed with the
original purpose of accurately measuring the Cr and Mn abundances in
the ejecta.  This new observation also enabled the
first-ever detection of Fe-K$\beta$ emission, which has turned out to
be useful as a diagnostic of the ionization population.
\citet{yamaguchi13} found that the observed intensity ratio of
K$\beta$/K$\alpha$ is larger than that expected from the dominant
charge state of the Fe ejecta (Fe$^{16+}$) determined from the
K$\alpha$ centroid energy. This indicates that a significant fraction
of the Fe ejecta is in an even lower charge state.  It was also
revealed that Cr-$\alpha$ emission clearly peaks at a larger radius
compared to the Fe emission (Figure~\ref{tycho:suzaku} right).  This is
against the recent claim by \citet{yang13}  that Cr and Fe
are always spatially correlated. With a high-resolution observation by
SXS, we can accurately determine the ionization age of each element:
if Cr indeed is distributed in the outer region, it should be more
highly ionized than Fe. The SXS observation will also allow us to
measure accurate abundances of these elements, which are important to
constrain the progenitor's nature.

Tycho is approximately 8$^\prime$ in diameter with a number of
interesting regions (see Figure~\ref{tycho:ptgs}--left panel) that are
worthy of study: (1) the bright NW rim where the X-ray intensity is
highest and the shock front is most circularly shaped, (2) the SE
region where there are two knots of ejecta (one Si-rich, one Fe-rich)
that seem to have expanded further than the rest of the shell (3) the
center, (4) the SW region where the synchrotron emission is most
intense and extends to the highest energies, and (5) the NE region
where there is some evidence for interaction with molecular clouds (Xu, Wang, \& Miller 2011).

With a 100-ks exposure at the NW rim, we can cleanly measure the Si
ion temperature with an estimated error of 10\% (see
Figure~\ref{tycho:ptgs}--right panel). Still the NW rim merits a deeper
observation (400 ks) with the SXS in order to determine the extent of
collisionless heating at the reverse shock, which is propagating into
the Fe-rich ejecta.  We will measure the charge state of Fe from both
the K$\alpha$ and K$\beta$ lines and measure the ion temperature from
the line widths.  We will be able to distinguish Fe temperatures
varying from 10 keV up to 1 MeV and detect the low ionization
population, although this will be quite challenging in the high
temperature case (see Figure~\ref{tycho:nwsims}).  Note that the long
exposure time is driven by the need to measure the faint Fe-K$\beta$
line complex which comes predominantly from the lower Fe ionization
states.

\begin{figure}
  \begin{center}
    \includegraphics[width=5.0in,clip,trim=180 220 60 260]{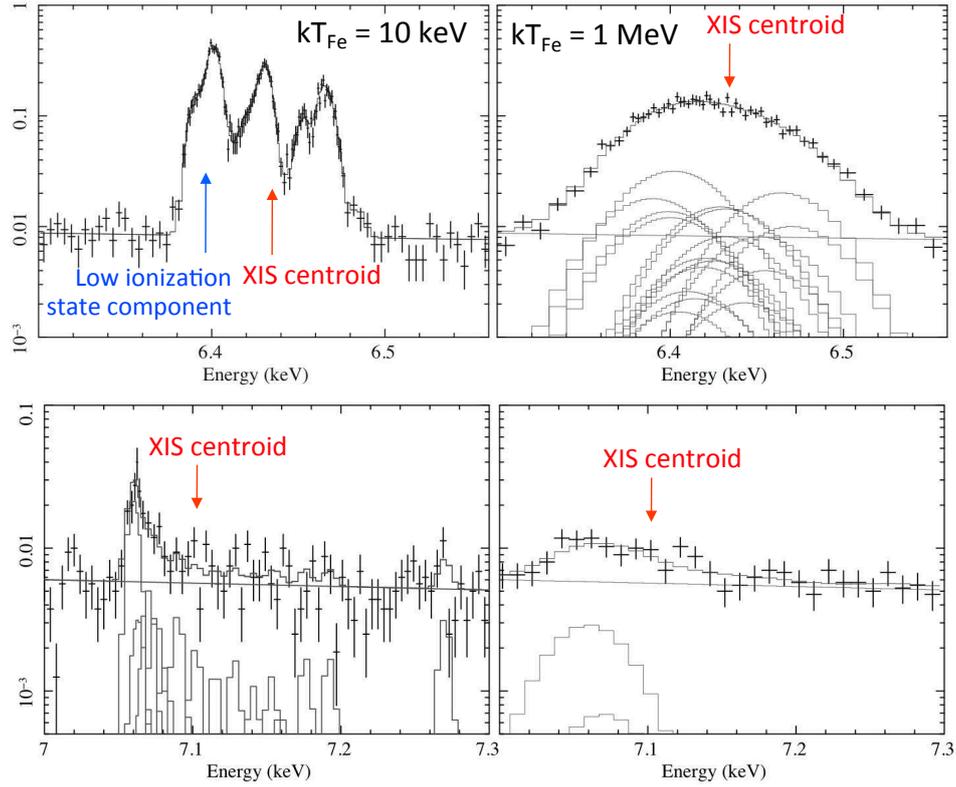}
  \caption{SXS simulation (400~ks) of the NW rim of Tycho for Fe-K$\alpha$
    {\it (top panels}) and Fe-K$\beta$ {\it (bottom panels)} 
    assuming two different ion temperatures: 10 keV  {\it (left panels)} 
    and  1 MeV  {\it (right panels)}. }
  \label{tycho:nwsims}
  \end{center}
\end{figure}

For the SE region, where the ejecta knots are, we simulated a 100 ks
exposure and showed that the Fe ion temperature can be determined to
10-15\% accuracy for Fe ion temperatures in the range 200 keV to 3 MeV
(see Figure~\ref{tycho:seknot}).  Since these knots are at the rim
(moving perpendicular to our line-of-sight) and fairly isolated, bulk
motion broadening should not be a significant concern. The origin of
these ejecta knots is still mysterious and the aim of this pointing is
to measure the ionic charge states and electron and ion temperatures
for comparison with the more ``normal'' ejecta on the NW rim. 

\begin{figure}
  \begin{center}
    \vspace{-0.25in}
    \includegraphics[width=5.5in,clip,trim=55 205 55 215]{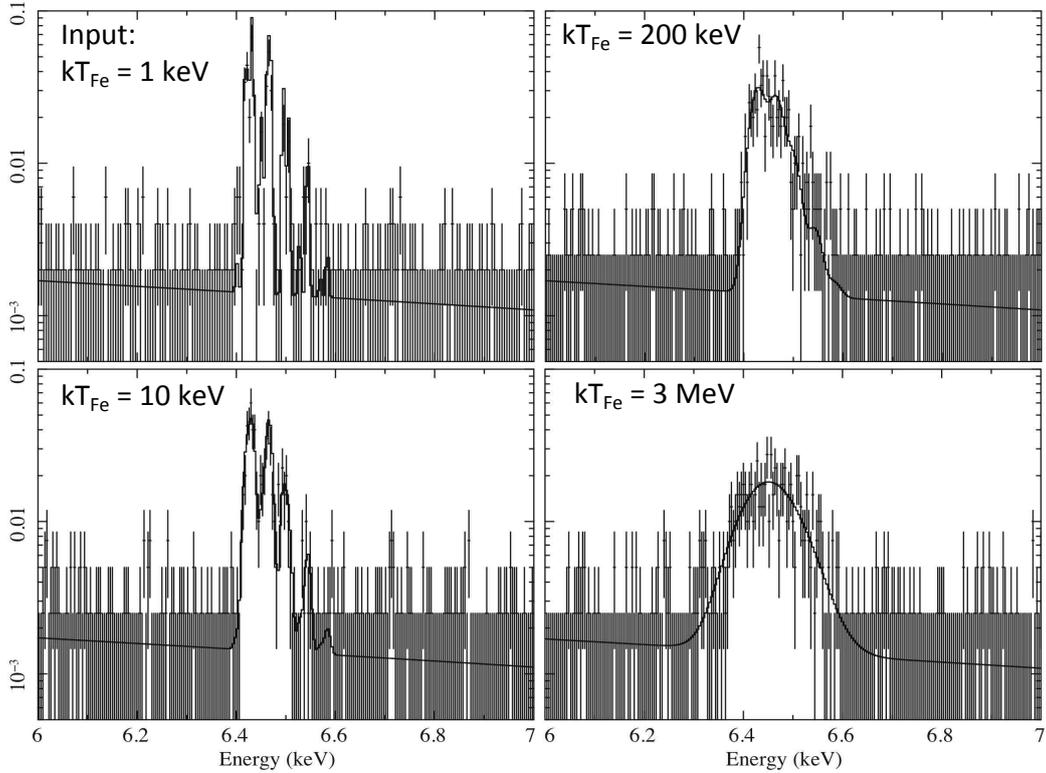}
  \caption{ SXS simulations (each 100-ks) of the Fe-rich knot on the
    SE rim of Tycho assuming four different ion temperatures from 1
    keV to 3 MeV as labeled. Only the energy band around the Fe-K$\alpha$ 
    is plotted.}
  \label{tycho:seknot}
  \end{center}
\end{figure}

The center pointing is important to determine the physical expansion
velocity for comparison to the known angular expansion rate.  This
will yield an accurate and precise measurement of the SNR's
distance. This pointing will also allow for better determination of
the reverse shock velocity, which will be important for interpreting
the thermal line broadening we measure. The statistical uncertainty on
the centroids of the strong Si and S lines will quickly reach the
systematic error level in short exposures (10's of ks), but to
adequately sample the Fe line and other fainter features (i.e., trace
odd-Z elements) will require exposures of the order of 100~ks.

The three regions mentioned above are of the highest scientific priority
for advancing our understanding of the thermal emission of Tycho.
Additionally, the SW region will be of interest also to White Paper \#18, the
Shock Acceleration task force team; the main goal will be to see if
the thermal properties of the hot shocked plasma are different here
compared to the rest of the remnant due to the more significant and
efficient acceleration processes occurring in the SW.  
Finally, a complete mapping of Tycho, which can be achieved with
some 9 SXS pointings, will further allow us to 1) do a precise determination of
the ionization state and abundance measurements, search for odd-Z and low-abundance
elements, map Cr and Mn across the SNR, 
and 2) map the hard (above 7 keV) X-ray emission with SXS and HXI.
This, together with the measurement of the velocity structure through Doppler shifted lines 
from the central regions, will address all main topics covered in this White Paper.

\subsubsection{SN~1006: odd-Z trace elements} \label{subsec:sn1006}

We aim for several ``pencil'' beam SXS measurements of this remnant.
Its large size (approx.~30$^\prime$ in diameter) means that the modest
angular resolution of \ah\ will not be a large negative factor.  One
pointing should be at or near the center to determine the expansion
velocities of the approaching and receding hemispheres (this is
presented in Section \ref{subsec:sn1006}). A second pointing should be
near the SE where \suzaku\ observations suggest a high Si abundance
and the possible detection of Cr and Mn.  The NW rim is an excellent
candidate for studying the interaction of the forward shock with
interstellar material to compare the thermal broadening of the O and
Ne lines (likely interstellar) and Si and S (likely ejecta).  An
additional pointing near one of the bright synchrotron rims to study
the efficiency of cosmic ray acceleration is alway worth pursuing.

SN~1006 also figures prominently in section 2 of this white paper.
Here we restrict ourselves to presenting some simulations relevant to
the detection of low abundance elemental species. \citet{uchida13}
used the \suzaku\ XIS data to study SN~1006's spectral and spatial
variations.  One of their findings is highlighted in
Figure~\ref{sn1006:cmncr} (left panel) which indicates the possible
presence of Cr and Mn emission lines from the SE region of the
remnant.  As shown in the right panel of the same figure, a 200 ks
exposure with the SXS would permit detection of these trace species
assuming the best-fit model to the XIS data.

\begin{figure}
  \begin{center}
     \includegraphics[width=0.9\hsize]{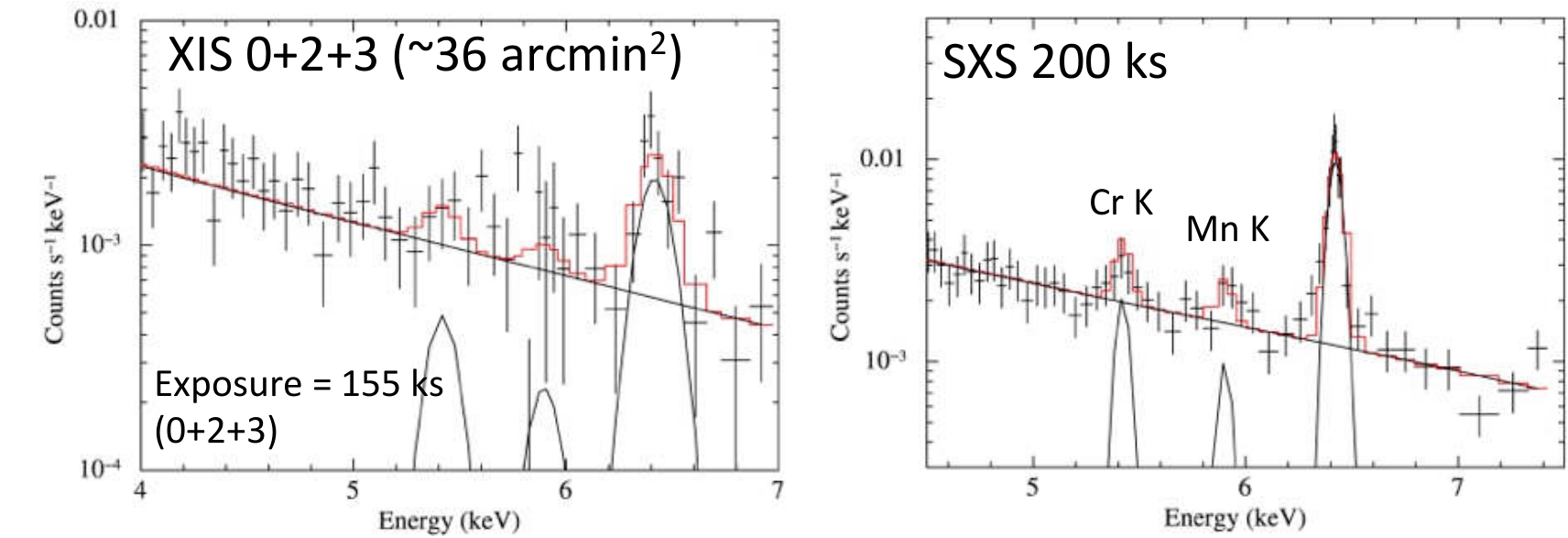}
  \caption{{\it (Left)} \suzaku\ XIS spectra extracted from a
    $3^\prime \times 3^\prime$ region at the SE rim of SN~1006 showing
    the clear detection of the Fe-K$\alpha$ line and evidence for Cr
    and Mn emission. {\it (Right)} SXS simulation of the best fit XIS
    model assuming a 200 ks exposure with \ah. All the lines were
    assumed to be broadened by 34 eV. }
  \label{sn1006:cmncr}
  \end{center}
  \begin{center}
    \vspace{-0.01in}
     \includegraphics[width=0.9\hsize]{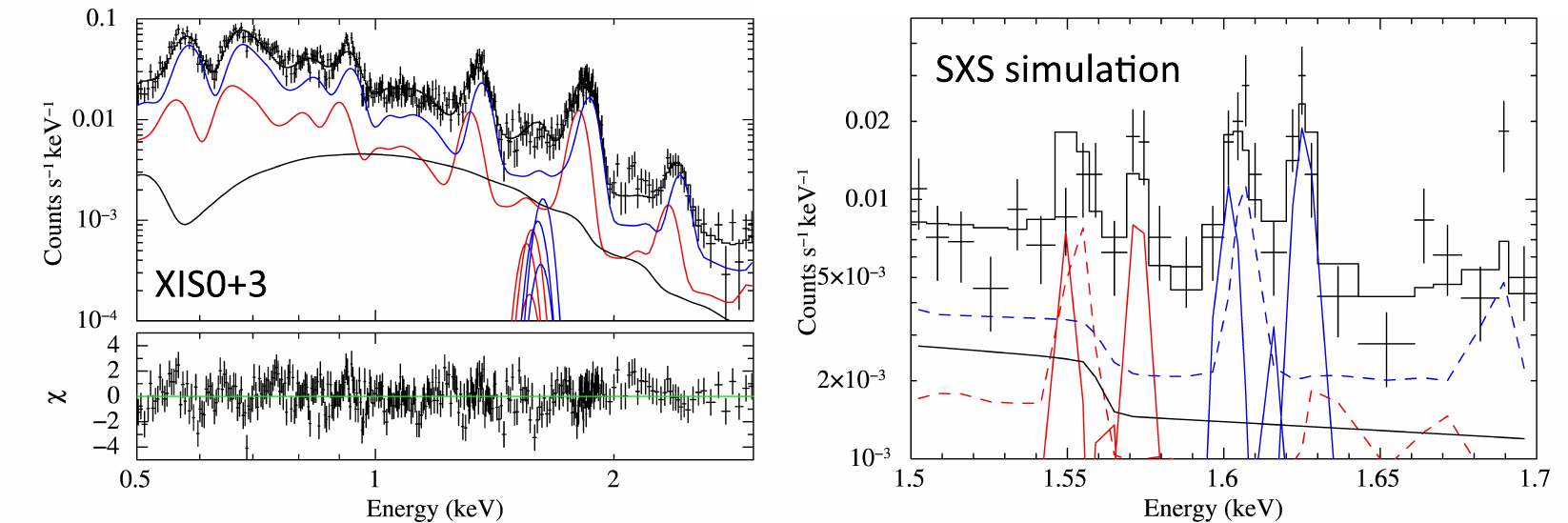}
  \caption{{\it (Left)} \suzaku\ XIS spectra extracted from a
    $3^\prime \times 3^\prime$ region near the center of SN~1006
    showing the two-component spectral model fit needed to account for
    the approaching (blue) and receding (red) hemispheres of the SNR.
    {\it (Right)} SXS simulation of the best fit XIS model
    highlighting the region about the Al lines (the He$\alpha$
    resonance and forbidden lines are shown).  This simulation assumed
    a 200 ks exposure with \ah. }
  \label{sn1006:al}
  \end{center}
\end{figure}

Along similar lines, an investigation into the possibility of
detecting lower atomic number odd-Z species was made.  This is shown
in Figure~\ref{sn1006:al}.  The left panel shows the \suzaku\ XIS data
fitted to a two component model to account for the approaching and
receding parts of the SNR. The parameters of the 2 vpshock models were
the same except for the intensity scaling which needed to be different
for the red- and blue-shifted components. Evidently there is an
asymmetry in SN~1006 since the amount of X-ray emitting shocked
material in the front and back halves is noticeably different. Al
emission was included as three Gaussian lines (the forbidden,
inter-combination, and resonance lines) for each velocity
component. The intensities were determined using the APEC code at a
temperature of 1 keV.  Al was assumed to be enhanced by a factor of 3
above solar values.  The right panel of Figure~\ref{sn1006:al} shows
that a significant detection of Al is possible in a 200 ks SXS
exposure.

\subsubsection{Cas A: the famous oxygen-rich remnant with a central 
compact object}\label{subsec:casa}

 \begin{figure}
  \begin{center}
  \includegraphics[height=2.75in]{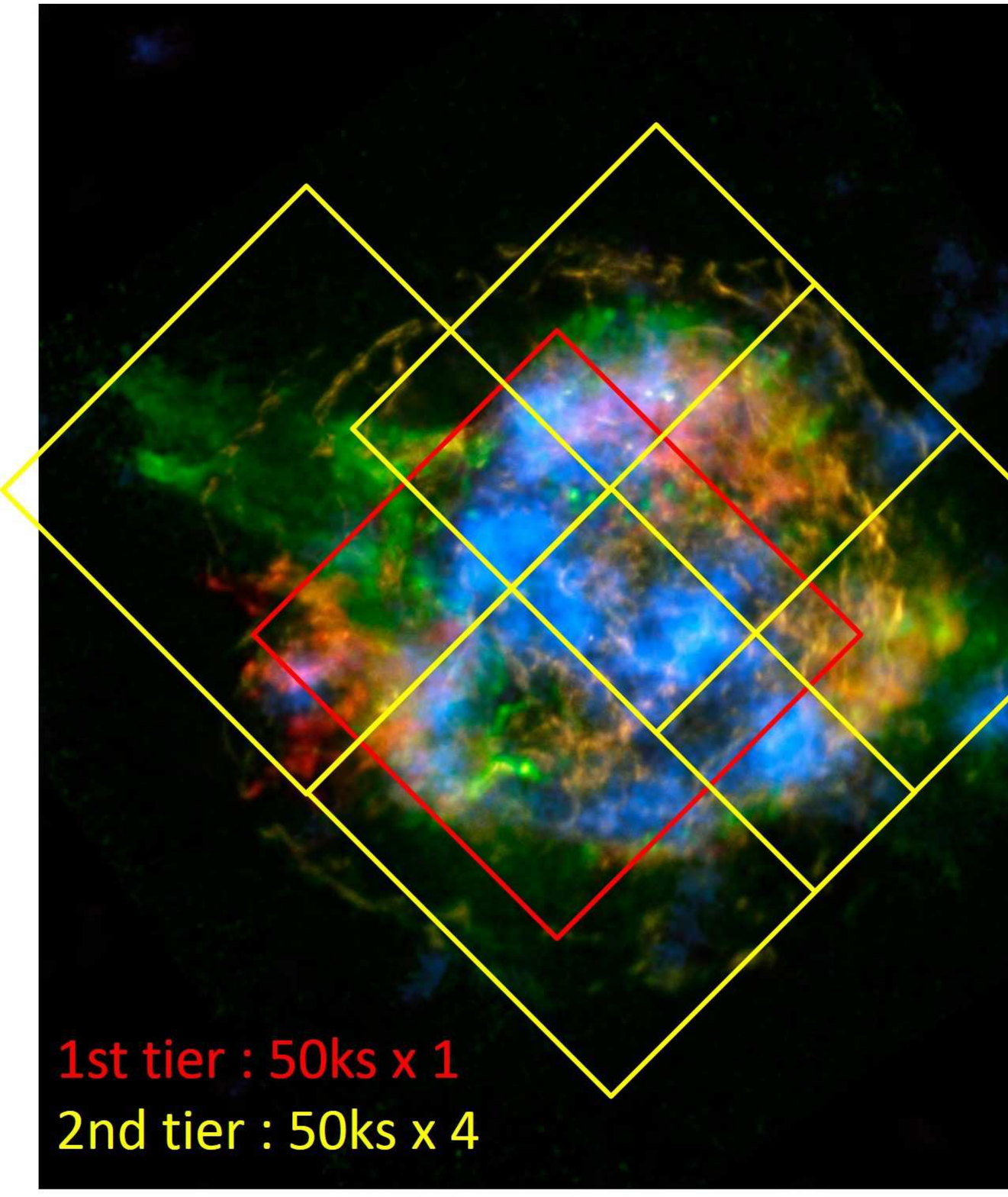}
     \includegraphics[height=2.75in,clip,trim=189 255 180 175]{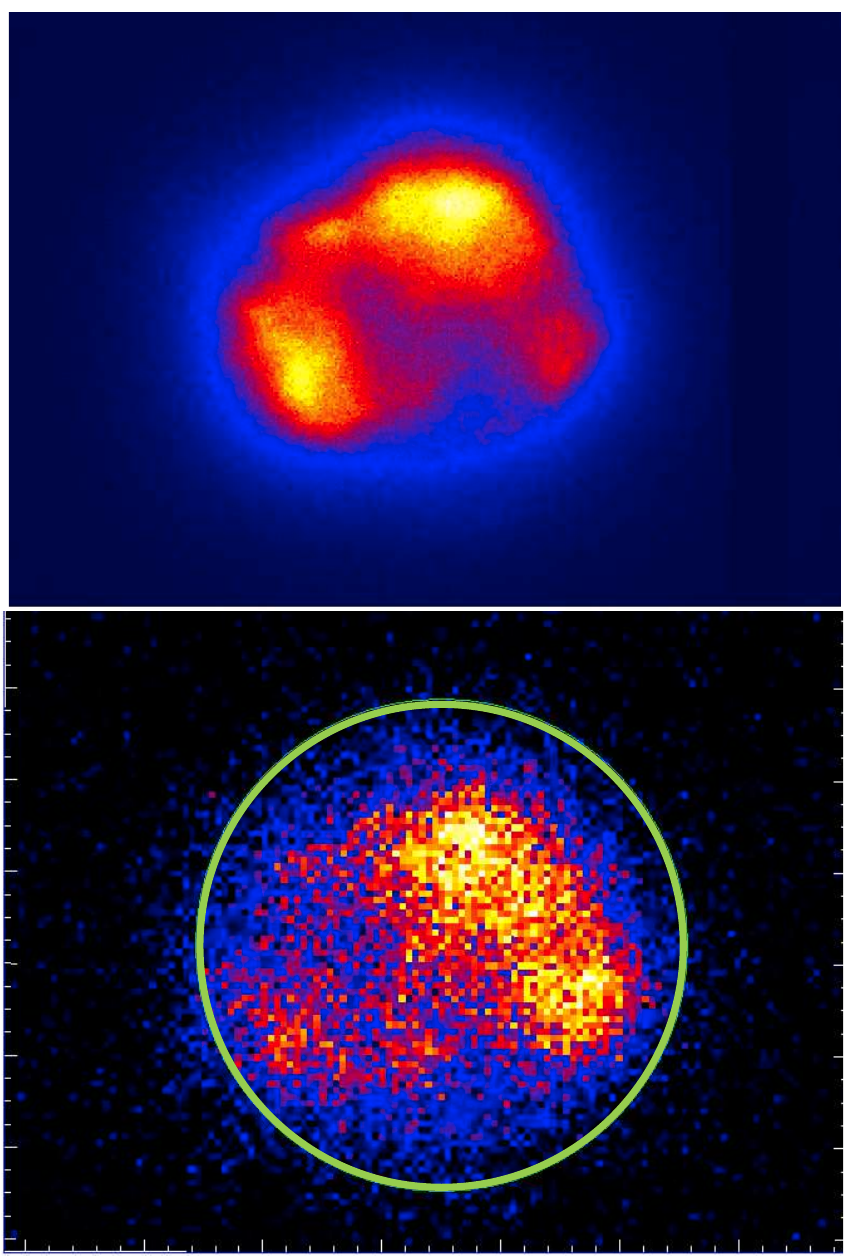}
  \caption{{\it (Left)} \chandra\ ACIS (red: Fe; green: Si/Mg) and {\it NuSTAR} (blue: $^{44}$Ti; Grefensette et al. 2014) 
  image of \casa\  with 5 SXS pointings overlaid covering the full SNR.
  The red box shows a top priority pointing for odd-z measurement.
 {\it (Right)} \textit{sim-x}   simulations of \casa\ for the SXI (top) and HXI (bottom).  Both
    simulations use the \chandra\ ACIS image: the 0.7--7 keV band for
    the SXI simulated image and the 4--7 keV band for the HXI image. }
  \label{casa:ptgs}
  \end{center}
\end{figure}

The young ($\sim$330 yr-old) supernova remnant \casa\ is a bright
core-collapse remnant whose X-ray emission clearly reveals large-scale
spatial composition differences. \chandra\ observations show distinct
regions of Si- and Fe-rich ejecta where Fe lies beyond Si at the
eastern limb of the remnant \citep{hughes00b}. The Fe was formed in
the innermost ejecta but was evidently overturned on large scales
relative to the Si layer that was originally produced at higher
locations in the exploding star.  Doppler velocity measurements using
\xmm\ show that the Fe and Si emission in the northern part of the
remnant are segregated in a same way as at the eastern rim
\citep{willingale02}.  Optical spectra of light echoes indicate that
the Cas~A SN was a Type IIb that originated from the collapse of the
helium core of a red supergiant that had lost most of its hydrogen
envelope before the explosion \citep{krause08a}.  A central compact
object was revealed by the first \chandra\ image of Cas~A
\citep{tananbaum99}.  Evidence from subsequent \chandra\ observations
spanning many years indicates that this CCO is a neutron star with a
carbon atmosphere whose surface temperature is declining with time
\citep{heinke10}.  It has been proposed that this decline is due to
neutrino emission from Cooper pair formation as the neutrons in the
core become superfluid \citep{shternin11}. However, recent {\it NuSTAR} observations
of \casa\ \citep{casa_nustar} led to the interesting suggestion that Cas~A's 
neutron star may have experienced a transition to a quark star
by undergoing a second explosion a few days after the supernova itself
\citep{laming14, ouyed14}.

The complete explosive burning of Si to $^{56}$Ni also forms trace
quantities of the radioactive isotope $^{44}$Ti. Cas~A is the only SNR
for which decay lines from both steps in the decay of $^{44}$Ti to
$^{44}$Sc to $^{44}$Ca have been unambiguously detected
\citep{Iyudin94,Vink01}. The recent 1.2~Ms {\it NuSTAR} study 
 has {\it imaged} \casa\ in the light of the $^{44}$Ti
radioactive decay line at 67.9 keV \citep{casa_nustar}, revealing a clumpy,
centrally-located distribution that does not correlate well with the
X-ray emitting Fe-K shell emission \citep{hughes00b,hwang04}.  This is
a surprising result given the presence of highly pure Fe
X-ray emitting clumps \citep{hwang03} that were likely formed by
complete Si burning and would be expected to also display radioactive
$^{44}$Ti lines.  Recently, Cr K-shell line emission has been detected
from \casa\ using \chandra\ \citep{Yang09} and
\suzaku\ \citep{Maeda09}. However Mn K-shell lines have not been
detected yet.

\begin{figure}
  \begin{center}
 \includegraphics[width=6.5in]{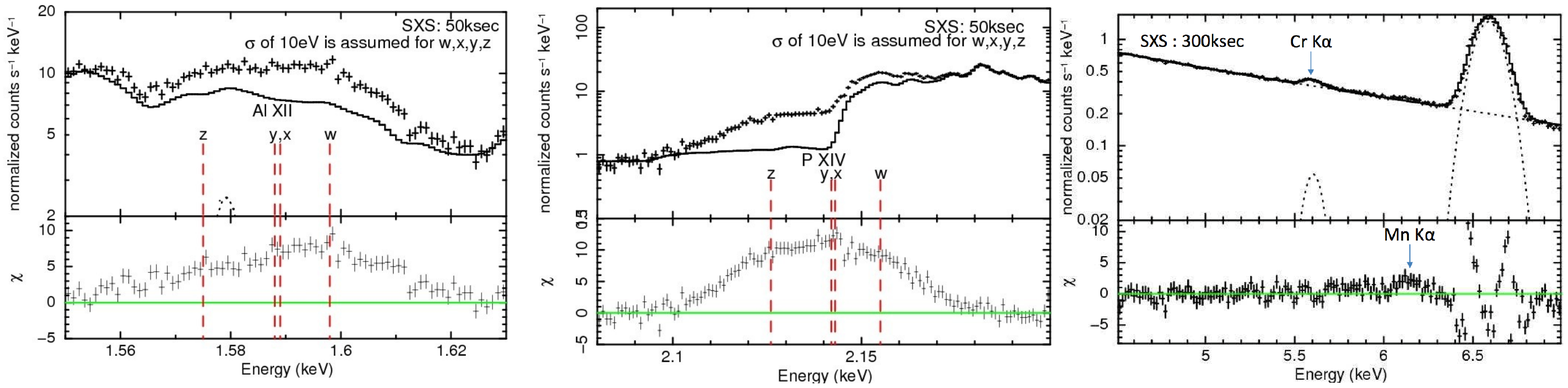}
  \caption{{\it (Left)}
Simulated SXS spectrum of \casa\ from a 
50~ks observation showing a close-up of the band containing the He-like w, x, y, z lines of Al (an odd-Z element).
 {\it (Middle)} Same as the left panel but for P (an odd-Z element formed in O-rich layer).
  {\it (Right)}
    Simulated 300 ks SXS spectrum of \casa\ showing the band
    encompassing the Cr, Mn and Fe-K shell lines.}
  \label{casa:spectra1}
  \end{center}
  \begin{center}
 \includegraphics[width=5.5in,clip,trim=60 405 70 200]{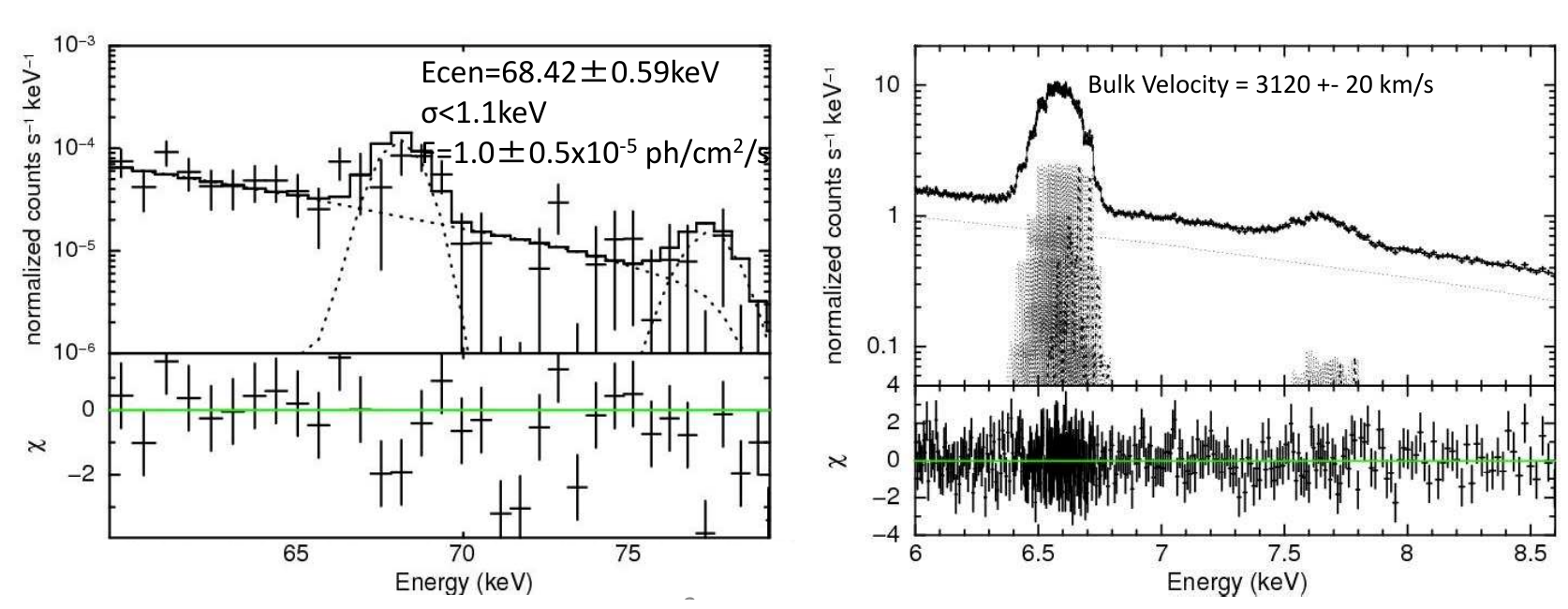}
  \caption{{\it (Left)} Simulated HXI spectrum of \casa\ from a 300 ks
    observation.  {\it (Right)} Simulated 100 ks SXS spectrum of the
    nothern region pointing of \casa\ showing the band encompassing the
    Fe-K$\alpha$ and Fe-K$\beta$ lines.}
      \label{casa:spectra2}
  \end{center}
  
\end{figure}

Figure~\ref{casa:ptgs} (left panel) shows five pointings that will allow
a complete coverage of \casa\ with the SXS. The central pointing's
primary objective is to address the odd-Z elements measurement,
providing key information on the nucleosynthesis in this core-collapse remnant.
In particular, phosphorus (P), predominantly formed in the O-rich layer,
is a very useful diagnostic as it doesn't suffer absorption nor contamination
by the Fe-L shell lines.
 These pointings have
been also targeted on the Fe-rich regions at the SE and N rims and the
spectrally hard (synchrotron-dominated) region to the SW.  In the
right two panels of Figure~\ref{casa:ptgs}, we show simulated SXS and HXI images.
For all \casa\ simulations we will require 50~ks for each pointing for a
total of 300 ks on-source exposure.

For the SXS spectral simulations, our model closely follows that of
\citet{casa_hetg}, which is derived from \chandra\ HETG data.  We
include bulk-motion line broadening determined from the \suzaku\ XIS.
Figure~\ref{casa:spectra1} (left panel) shows the energy band containing
the strong He-like transitions of Al around 1.6 keV for the combined
total 300 ks exposure. The 10 eV broadening assumed in the simulation
corresponds to a velocity broadening of $\sim$2000 km s$^{-1}$.  For
illustrative purposes we show the spectrum including the Al component
as the data points and the spectrum without Al as the solid histogram.
The Al lines account for up to 30\% of the total flux in this band and
will therefore be detected with high significance.  As well, in the middle
panel of Figure~\ref{casa:spectra1} , we highlight the simulated emission from P
around 2.15~keV.
Being the brightest thermal X-ray supernova remnant, \casa\ is the best target to search
for and study the low abundance odd-Z elements, especially from a core-collapse explosion.

The right panel of Figure~\ref{casa:spectra1} shows the energy band
containing Cr, Mn, and Fe emission and, as above, assumes the total
combined exposure of 300 ks.  The Fe and Cr lines are well detected,
but Mn will be detected only if the Mn to Cr intensity ratio is
$>$0.22, which is about in the midpoint of measured values to date
(see Table above).  These simulations assumed a conservative velocity
broadening of 3000 km s$^{-1}$. We may have higher sensitivity to Cr
and Mn emission by restricting the analysis to regions near the rim
where the shocked material will be moving transverse to our line of
sight, reducing the bulk-motion velocity broadening.

For the HXI spectral simulations, we assume the {\it NuSTAR} measured line
centers and fluxes \citep{casa_nustar} and a velocity broadening of
3000 km s$^{-1}$. Figure~\ref{casa:spectra2} (left panel) shows the
resulting HXI spectrum for a 300 ks exposure.  We will be able to
detect the $^{44}$Ti radioactive decay lines, but to improve on the
{\it NuSTAR} results a much deeper exposure will be required.  It may be
possible to build up a deeper exposure over the course of the
\ah\ mission.

\subsubsection{SN1987A: the youngest known supernova remnant} \label{subsec:sn87a}

\chandra\ and \xmm\ continue to monitor SN1987A as it transitions into
a supernova remnant. The coverage with \chandra\ is roughly twice a
year for typically 50 ks of exposure and, since 2009, have been
carried out with the HETG inserted as a pile-up mitigation strategy.
Recent \chandra\ observations \citep{helder13} show the object to be
still barely resolved (diameter of $\sim$1.5$^{\prime\prime}$) with
X-ray fluxes of $F_X = 7.6\times 10^{-12}$ erg s$^{-1}$ cm$^{-2}$
(0.5--2 keV band) and $F_X = 8.9\times 10^{-13}$ erg s$^{-1}$
cm$^{-2}$ (3--8 keV band).  These fluxes continue to increase at a
rate of about 13\% per year.  Deep observations with the ACIS/HETG
\citep{dewey12} in 2007 and 2011 (exposure durations of $\sim$350 ks
and $\sim$200 ks) were fitted with a complex multi-parameter
three-shock model that included some line broadening (width of $9300
\pm 2000$ km s$^{-1}$).  The HETG data have good signal-to-noise over
the 0.6-2.2 keV band and clearly show emission lines of O VIII, Ne IX
and X, Mg XI and XII, and Si XIII and XIV, in addition to the forest
of Fe-L lines.

\begin{figure}[h]
 \centering \includegraphics[width=0.49\textwidth]{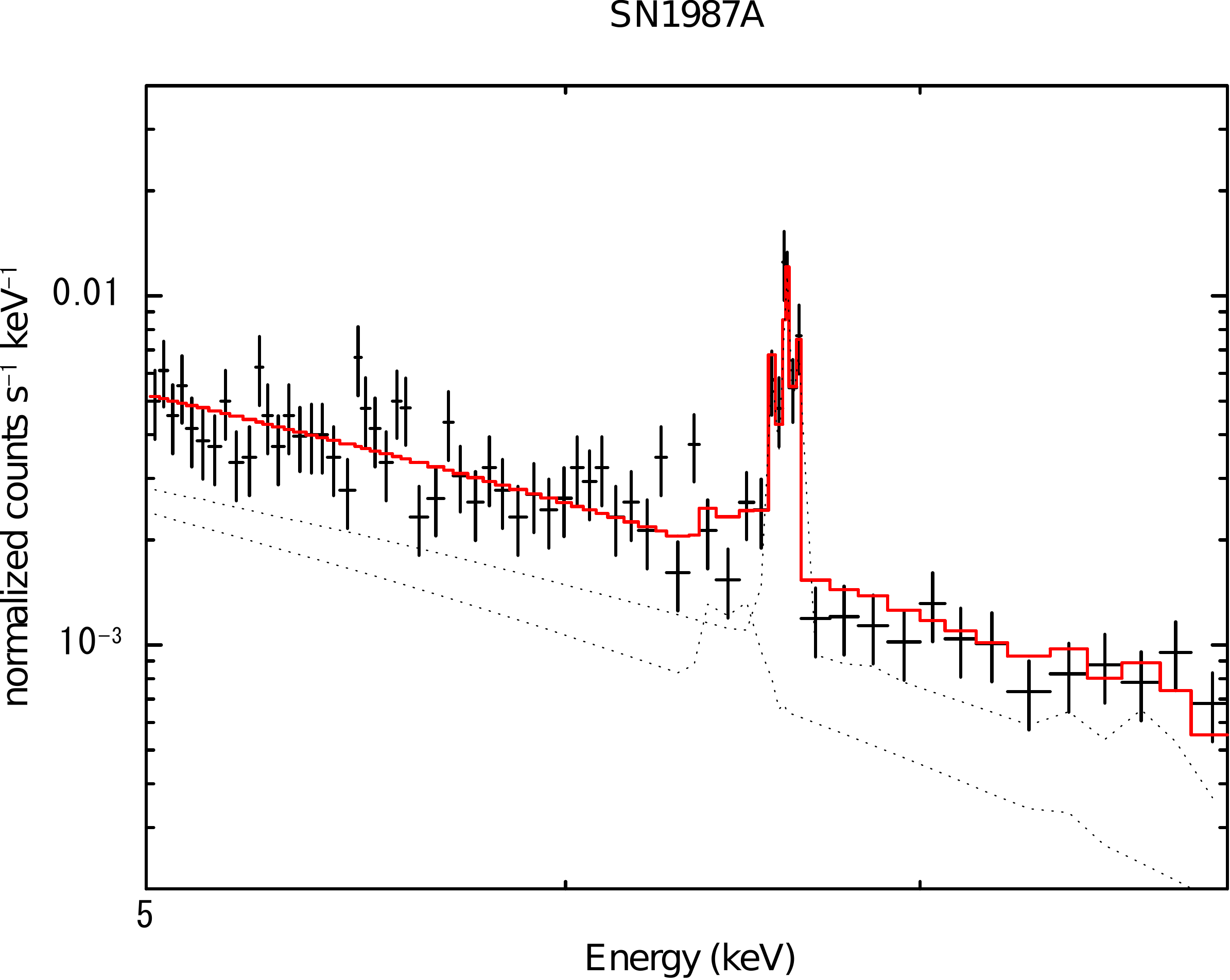}
 \includegraphics[width=0.49\textwidth]{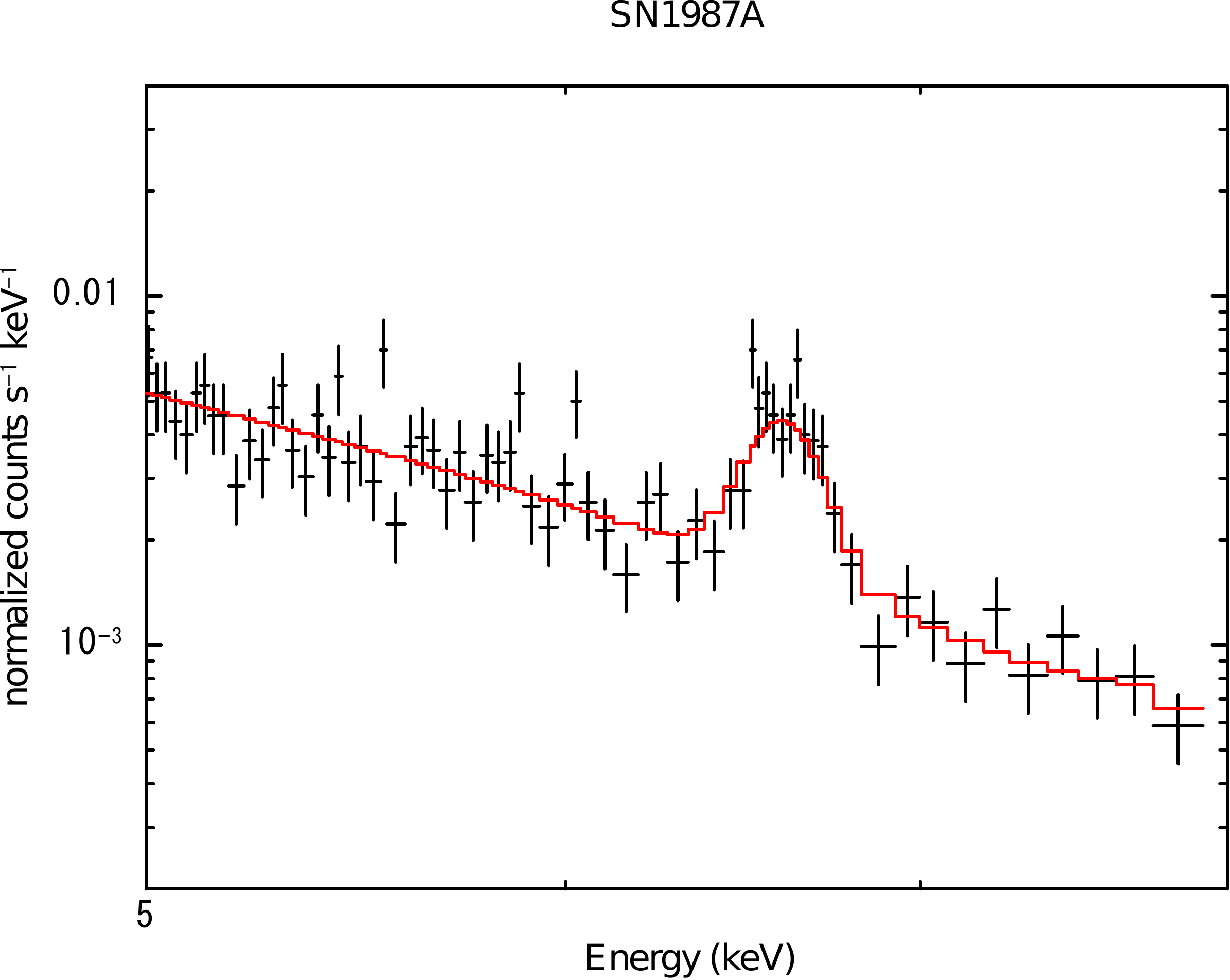} 
   \caption[]{SXS simulations of the Fe-K band of SN1987A for two
     different cases: a mixture of plasmas with different ionization
     timescales {\it (left panel)} and a very broad line component
     with a FWHM of order 10$^{4}$~km~s$^{-1}$ {\it (right panel)}.
     This simulation assumes an equivalent width of 300 eV for the Fe
     line.}
 \label{fig:87a}
\end{figure}

While the \chandra/HETG and \xmm/RGS have already revealed emission
line structures in SN1987A over the soft X-ray band, the \ah\ SXS will
allow us to perform Fe-K line diagnostics in the higher energy
band. Fe-K lines are sensitive to the hottest thermal component and
thus to the fastest shocks propagating now at this early stage of the
remnant's evolution. A series of \xmm/EPIC spectra from 2007 to 2011
detected a Fe-K line blend \citep{maggi12}. The line width was
apparently broad, which has been interpreted as being due to a mixture
of plasmas with different ionization timescales, possibly including
fluorescence and/or low ionization Fe \citep{maggi12}. On the other
hand, \chandra/HETG observations show a very broad line component with
a FWHM of order 10$^{4}$~km~s$^{-1}$ and indicate that most of the
3--10~keV emission would originate from it \citep{dewey12}. If this is
the case, Fe-K lines should be intrinsically broad.
Figure~\ref{fig:87a} shows the SXS simulations of the two cases. With
the \ah\ SXS, we can clearly distinguish between these two scenarios.

\subsubsection{Radioactive species in the youngest 
Galactic remnant: SNR G1.9+0.1 } \label{subsec:g1p9}

\begin{figure}[h]
\begin{center}
\includegraphics[height=1.6in]{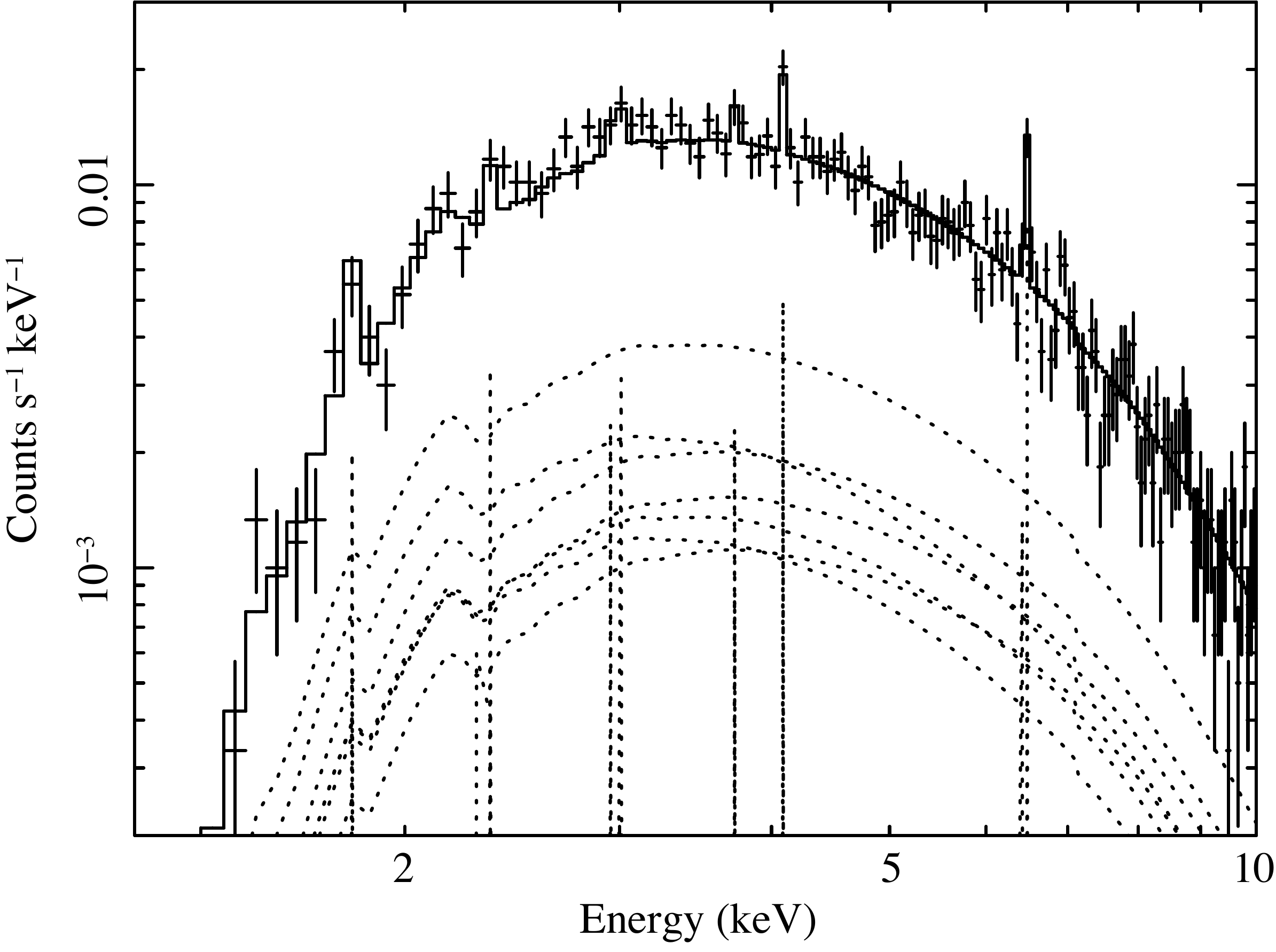}
\includegraphics[height=1.6in]{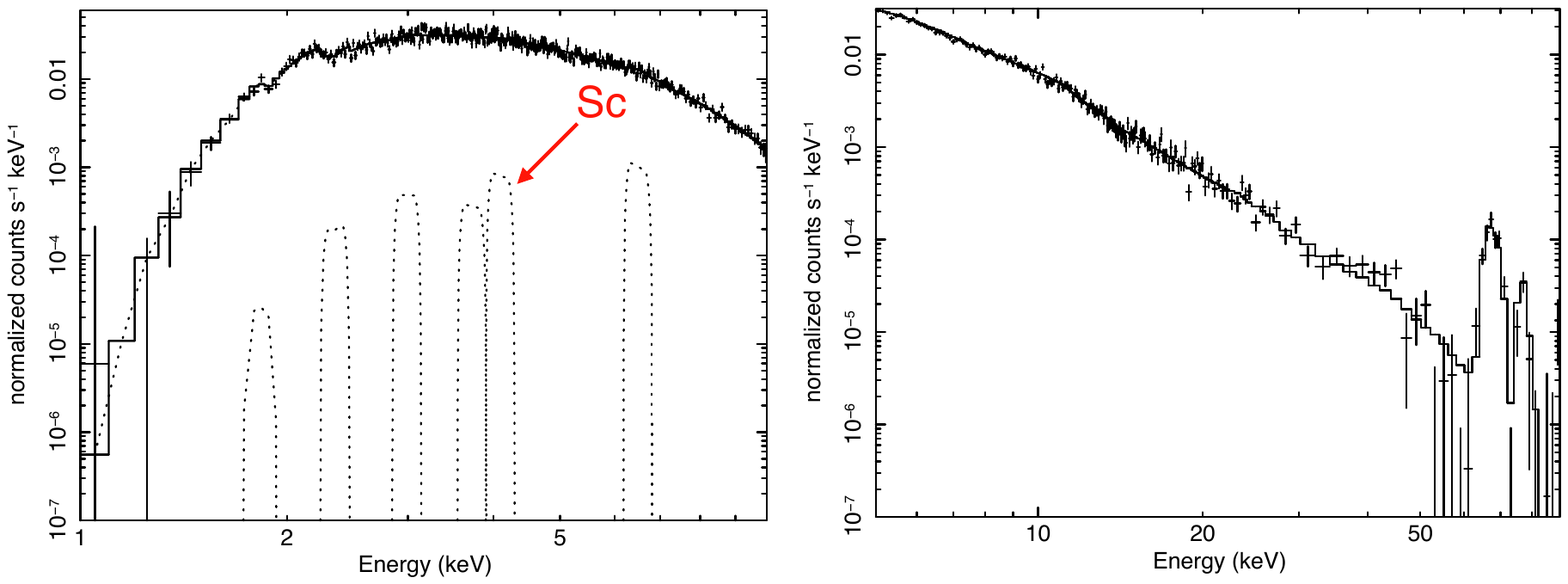}
\caption{{\it (Left)} A simulated 100-ks \ah\ SXS spectrum of G1.9+0.1
  {\it assuming no bulk motion velocity broadening} shows a clear
  detection of the 4.1 keV Sc line from the radioactive decay of
  $^{44}$Ti.  The Fe-K line is clearly detected too. {\it (Middle)} A
  simulated 300-ks \ah\ SXS spectrum of G1.9+0.1 now including the
  line broadening expected from the remnant's rapid expansion (bulk
  motion of the ejecta at a speed of $\sim$14,000 km s$^{-1}$).  No
  lines are detected above the remnant's strong non-thermal synchrotron
  continuum emission.  {\it (Right)} The HXI spectrum shows a
  significant detection of the $^{44}$Ti radioactive hard X-ray lines.
}
\label{fig:g1p9sxs}
\end{center}
\end{figure}

G1.9+0.3 is the youngest known Galactic SNR and is located near the
center of the Milky Way. The age was originally estimated to be of the
order of 100 years \citep{reynolds+08}; additional strong support for
the remnant's youth was provided by \chandra\ measurements of a rapid
expansion rate \citep{carlton+11}.  The angular expansion rate of
0.642\% $\pm$ 0.049\% corresponds to a shock velocity of about 13,000
km s$^{-1}$ (assuming a 8.5 kpc distance, as indicated by the SNR's
high absorbing column density) consist with the spectroscopically
deduced velocities of order 14,000 km s$^{-1}$ \citep{borkowski+10}.
The SNR shows a shell-like morphology with strong synchrotron emission
\citep{reynolds+09}.  A deep observation by \chandra\ successfully
separated a thermal-dominated region from the strong non-thermal
shells \citep{borkowski+10}. The spectrum exhibits emission lines from
low-ionization states of Si, S, Ar, Ca, and Fe, suggesting that it is
evolving in a low-density environment. In addition, the $^{44}$Sc
emission at 4.1\,keV, produced by electron capture from $^{44}$Ti, was
clearly detected. For the remnant age of 100\,yr, the measured line
strength indicates synthesis of $(1-7) \times 10^{-5}$ \msun\ of
$^{44}$Ti. Although this is in the range predicted for both Type Ia
and core-collapse SNe, it is significantly smaller than that reported
for Cas~A ($2 \times 10^{-4}$ \msun).

Although the \chandra\ detection of the 4.1\,keV line was significant,
the statistical uncertainty on the flux was large: the 95\% confidence
range is $(0.35,2.4) \times 10^{-6}$\,ph\,cm$^{-2}$\,s$^{-1}$. One
reason is that the Sc-K$\alpha$ emission was not clearly resolved from
the nearby Ca-K$\alpha$ line at 3.9\,keV. If we ignore any line
broadening then the \ah\ SXS can detect this line even though the
non-thermal emission cannot be resolved
spatially. Figure~\ref{fig:g1p9sxs} (left panel) shows a simulated SXS
spectrum of G1.9+0.3 with an exposure of 100\,ksec. The 4.1\,keV line
counts at a rate of $(2.8 \pm 1.2) \times 10^{-4}$ s$^{-1}$. Refining
the $^{44}$Ti mass estimate will help determine the SNR's origin
(i.e., Type Ia or core-collapse), which remains uncertain.

However, considering the remnant's extremely rapid expansion rate (Borkowski et al. 2013),
line broadening from bulk motion will likely render any emission lines
undetectable (see~Figure~\ref{fig:g1p9sxs} middle panel) even in a
much deeper (300 ks) observation.  \chandra\ was able to detect these
lines by isolating the thermal emission spatially from the strong
non-thermal rims, which \ah\ cannot do.

The HXI can make a significant detection of the radioactive decay lines
of $^{44}$Ti under the plausible assumptions that the synthesized mass
of $^{44}$Ti is $3.3\times 10^{-5}\, M_\odot$ and the remnant's age is
100 yr.  Whether this remains an interesting \ah\ project depends on
the results from the {\it NuSTAR} observations carried out for this remnant.

\vspace{-0.3cm}
\subsubsection{Kes~73: a core-collapse remnant associated with an anomalous
 X-ray pulsar (magnetar)} \label{subsec:kes73}

\begin{figure}
\begin{center}
\includegraphics[width=1.0\textwidth]{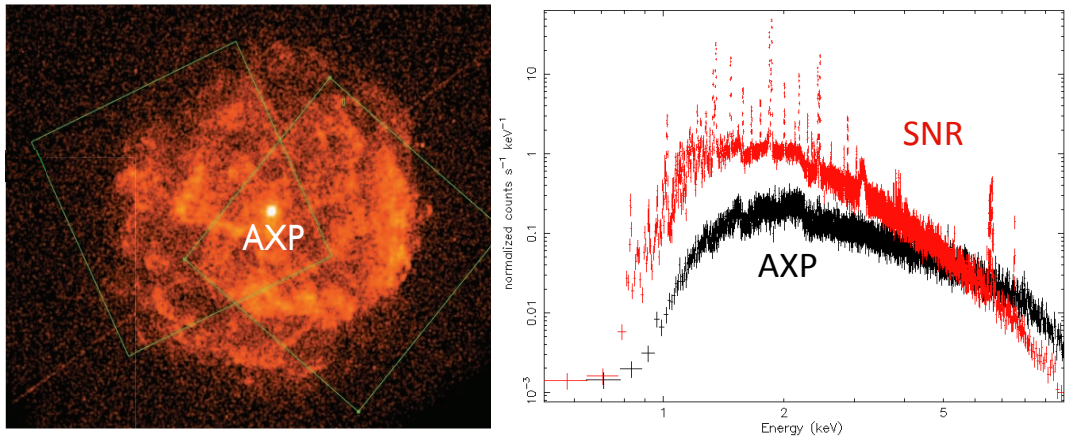}
\caption{{\it (Left)} The \textit{Chandra} image of the SNR Kes~73 with the
  SXS field of view overlaid. The two fields shown cover the eastern
  and western parts of the SNR with the AXP towards the edge of SXS to
  minimize contamination of the SNR signal.  {\it (Right)} 100~ks SXS
  simulation with \textit{sim-x} of the western field showing the relative
  contributions of the AXP and SNR emission. The line emission from
  the SNR clearly dominates over the emission from the AXP. The AXP's emission
  will dominate in the hard band (HXI+SGD).}
\label{fig:kes73fov}
\end{center}
\end{figure}

\begin{figure}
 \begin{center}
 
\includegraphics[width=1.0\textwidth, height=0.35\textheight]{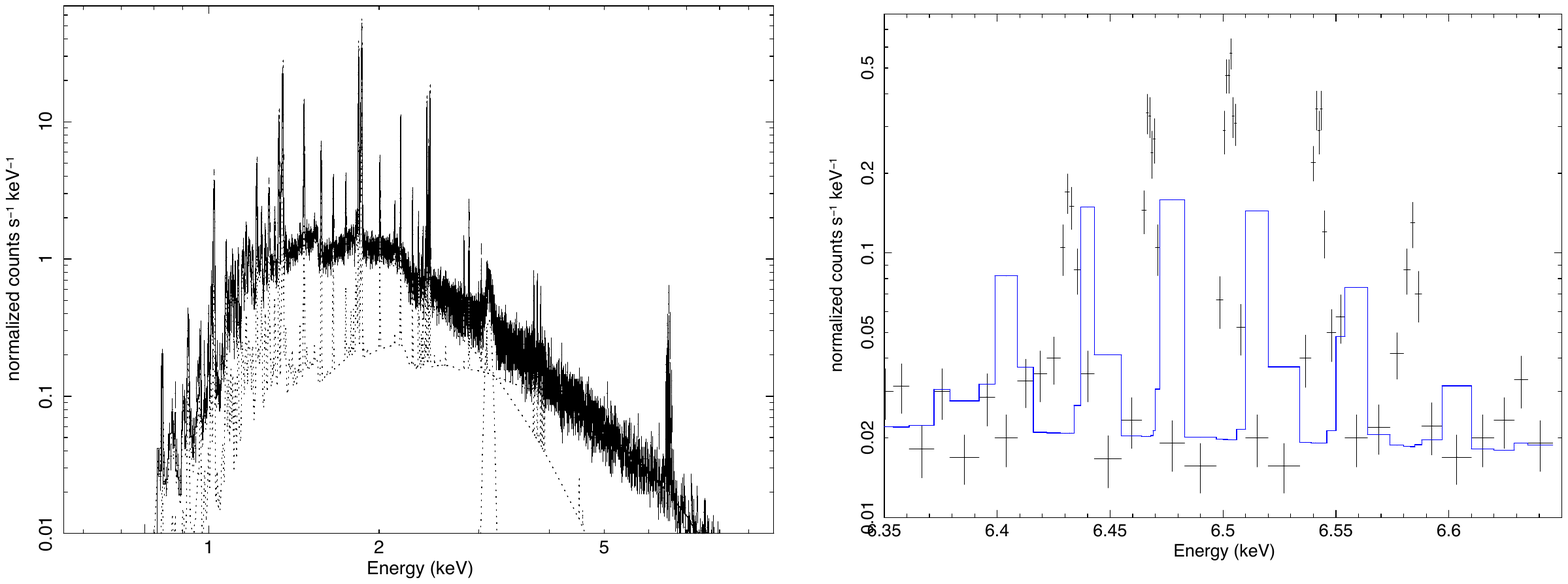}
\caption{(Left): A 100~ksec SXS simulated spectrum of the SNR emission
  (excluding the pulsar's emission) using \textit{sim-x} and based on the
  \textit{XMM-Newton} and \textit{Chandra} spectral fits.  The model
  components are shown as dashed lines (an absorbed two-component
  thermal model plus a Gaussian line to account for Argon) and
  correspond to the best fit for the CCD-type spectra.  In addition to
  the power of SXS to constrain the abundances, particularly of Mg,
  Si, and S in the soft X-ray band, a strong Fe-K line associated with
  the hot component will be detected.  (Right): Zoomed-in SXS
  simulated spectra of the Fe-K line region, with no Doppler shift,
  and with the model (blue line)
  corresponding to a thermal non-equilibrium ionization $pshock$ model
  with a Doppler shift corresponding to an inferred shock shock velocity of
  $\sim$1200~km~s$^{-1}$.}
\label{fig:kes73sxs}
\end{center}
\end{figure}

SNR Kes~73 is a relatively bright and young SNR that hosts the young
anomalous X-ray pulsar AXP 1E 1841$-$045.  The remnant is
ejecta-dominated, showing strong line emission from Mg, Si and S
whose relative abundances make a plausible
case for the SNR arising from a very massive progenitor.  Kes~73
has an angular size ($\sim$4$^\prime$ diameter) that is well matched
to the \ah\ field of view. 

A recent \chandra\ and \xmm\ study of Kes73 \citep{kumar2014} shows
that the 0.5--10 keV emission is characterized by two components: a
soft component with a temperature $kT_s$ of $\sim$0.5 keV and a
high ionization timescale, a hard component with a temperature $kT_h$
of $\geq$1.6 keV and a low ionization timescale, and a total
luminosity of $\sim$ 3~$\times$~10$^{37}$~ergs~s$^{-1}$ (0.5--10 keV,
at an assumed distance of 8.5 kpc).  The soft and hard components
are attributed to shock-heated ejecta and ISM/CSM, respectively. 
The age estimated from the
properties of the SNR is $\leq$2.6~kyr (depending on the phase of
evolution and the environment in which the SNR is expanding), a factor
of $\geq$2 lower than the pulsar's characteristic age (of 4.7~kyr)
which represents an upper limit, indicating a young SNR.  The SNR's
spectrum clearly shows line emission from Mg, Si, and S with enhanced
abundances associated with the soft component, indicating that the SNR
is dominated by shock-heated ejecta.  The abundances inferred from
fitting the global (and spatially resolved spectra) suggest a very
massive ($\geq$20\,\msun) progenitor.  The abundances were however
poorly constrained, and no line emission was detected from O and Fe-K,
both needed to better constrain the progenitor's mass.

Figs.~\ref{fig:kes73fov} and \ref{fig:kes73sxs} illustrate the
sensitivity of the SXS to the line emission from several elements
including Mg, Si, S, Ar, Ca and Fe which will provide much better
constraints on their abundances and thus the progenitor mass.
In particular, SXS will allow the detection of a strong Fe-K line.

Ideally, with two 100~ksec pointings (Figure~\ref{fig:kes73fov}, left), 
SXS will cover the eastern and western sides of the SNR, including
the ejecta-dominated regions and the bright western limb, with the AXP
close the edge of the SXS's field of view.  
Alternatively, a pointing towards the western part will cover the bulk of X-ray
emission and the brighter western limb, with the AXP closer to
the field's center.
Since SXS does not have the
spatial resolution needed to resolve the PSR from the SNR emission,
the pulsar's spectrum (known in quiescence) will be subtracted from
the total source spectrum (addressed in the HMXB+Magnetars White Paper~\#4).  
The right panel of Figure~\ref{fig:kes73fov} shows the simulated SXS 100~ks
spectra (of the western field versus the AXP) illustrating the strong
line emission from the SNR which dominates the emission from the AXP.
We note that the SNR will be fully covered by SXI due to its larger 
field of view.

Fig~\ref{fig:kes73sxs} (left panel) shows a \textit{sim-x} simulated 100~ks SXS
spectrum of the western field with the two-component model, and with
the AXP's spectrum excluded.   In such an observation
the ejecta abundances can be measured to $\sim$10\%
(depending on the species), and compared to nucleosynthesis models 
to confirm or refute the very massive progenitor origin.

Another primary goal of the observation is to confirm the blast wave
velocity which has been estimated from CCD spectra to be $\sim$1200
km~s$^{-1}$ \citep{kumar2014}, with the estimated velocity depending on the SNR's
evolutionary phase and the ambient medium in which it's expanding
(e.g. expansion into the late red supergiant wind phase of its massive
progenitor will yield a higher velocity than expansion into a homogenous ISM). 
This can be addressed by examining the Fe-K$\alpha$ line (associated with the hot component)
which is easily visible in the SXS data, but barely detected in
existing CCD data. Fig~\ref{fig:kes73sxs} (right panel) shows that a
1200 km/s velocity will correspond to a Doppler shifted line
detectable in a 100 ksec exposure with the SXS.  The velocity estimate
will also help constrain the SNR's age and thus address the factor of
$\geq$2 discrepancy between the SNR and pulsar ages. 
This will in turn address the question of magnetic field decay in magnetars.
The AXP's science and simulated broadband spectrum (with HXI and SGD) are discussed in 
detail in White Paper \#4.  In summary, the study of this young magnetar-SNR system will take full advantage of \ah's broadband capability
and SXS's spectral resolution, to address fundamental questions related to the formation of magnetars and 
their link to the other classes of compact objects in core-collapse SNRs.

\subsubsection{Other \ah\ targets}
Other targets (young to middle-aged SNRs) worth exploring with \ah\ to address the science questions raised in this section
include the following SNRs, the study of which is relevant to other core-collapse SNRs investigated in detail
in this White Paper and in the Pulsar Wind Nebulae section of WP\#8:\\
(a)  Kes 75 is associated with HBP J1846?0258, long thought to be a rotation-powered pulsar, 
but, having been caught once revealing itself as a magnetar, now blurs the distinction between the rotation-powered pulsars and magnetars
\citep{KumarSSH2008, Gavriil2008, Ng2008}. Kes~75 has been proposed to
be also associated with a very massive, Wolf-Rayet type, progenitor
\citep{Morton2007}. \ah\ will allow
a precise determination of abundances for the progenitor mass estimate.\\
 (b) G292.0+1.8, an O-rich remnant like Cas~A, 
but older and harbouring a rotation-powered (more `normal') pulsar powering a pulsar wind nebula.
This is a unique remnant, likely originating from an asymmetric explosion, 
and showing evidence for interaction of ejecta fragments and
the blast wave with a CSM produced by a very massive progenitor's stellar winds
 \citep{hughes01, gonzalez03, park07}. Recent {\it Suzaku} observations allowed the first detection of hot ejecta
 through Fe-K shell line emission at 6.6~keV \citep{kamitsukasa14}.
  \ah\ will allow precise abundance measurements for confirming the progenitor mass estimate.
  Section 2 further elaborates on the science that can be further addressed with velocity structure measurements (particularly for Mg, Si, S, and Fe-K).\\
  (c) MSH~15--52, a middle-aged core-collapse SNR powered by the high-magnetic field, rotation-powered, pulsar
PSR~B1509-58. Surrounded by a bright pulsar wind nebula prominent in the hard X-ray band (recently observed with {\it NuSTAR},
\citet{An14}, and referred to as the ``Hand of God"),
the pulsar/PWN have been suggested to be  interacting with the surrounding medium or SNR shell (e.g. through precessing jets) forming the RCW~89 nebula;
see e.g. \citet{yatsu05}.
The emission from this nebula peaks in the soft X-ray band and is an excellent target for \ah\ SXS for a precise measurement of the velocity distribution, and thus
to confirm or refute the a precessing jet or blast wave origin.\\ 
 (d) 3C~397, a bright SNR with an unusual morphology, over-abundance in Fe-peaked elements, 
 and whose nature as a type Ia or core-collapse SNR is still being debated \citep{chen99, ssh00, ssh05, yamaguchi14}.
 \ah\ will allow precise determination of abundances, leading to a certain determination of the
progenitor type and mass, as well as the search for Cr and Mn lines.
In relation to Section 2 below, the study of the velocity
 structure of the ejecta (particularly Fe) will help probe the explosion mechanism.\\
(e) SN~1987A: As one of the youngest known and best probed SNRs outside our Galaxy (see Section 1.3.4 for details and Fe-K line diagnostics),
SN~1987A's progenitor mass estimate can be used to calibrate progenitor mass estimates of other SNRs.
Studying the progenitor type can be achieved through a precise measurement of the abundances and probing the ionization states of Si, S and Fe.

\subsection{Beyond Feasibility}

{\bf Search for the characteristic X-rays from $^{44}$Sc in Cas A}

Initial SXS simulations for measuring the characteristic line at 4.1
keV from $^{44}$Sc (a daughter product of the decay of $^{44}$Ti) in
  \casa\ suggested that only an upper limit was possible in 100 ks.
  This simulation, however, assumed the radioactive Ti was distributed
  over the entire SNR and the line was broadened by bulk motion at the
  expansion speed of the ejecta.  {\it NuSTAR}'s localization of the
  $^{44}$Ti hard X-ray emission lines to the center of \casa\ moving
  with the expansion speed of the unshocked ejecta, motivates a new
  look at the detection feasibility.  Clearly a central pointing would
  be required.  Successful detection of $^{44}$Sc would valid the decay
    chain and allow for constraints or measurements of the expansion
    velocity of the radioactive Ti, which cannot be obtained in any
    other way.  This would provide a rare insight into the inner
    workings of the core-collapse explosion mechanism.

\vspace{0.25cm}

\noindent {\bf On the possible non-detection of Mn }

The single-degenerate scenario has a requirement that the progenitor's
metallicity must be higher than $\sim 0.1$\,solar. Otherwise a white
dwarf cannot eject over-accreted matter (from its companion) via a
radiation-driven stellar wind, and hence the progenitor system
undergoes a common-envelope phase (so stable burning of the white
dwarf is no longer expected) before the white dwarf obtains the
Chandrasekhar mass (Hachisu et al.\ 1996). Thus, if we find a Type Ia
SNR with a metallicity lower than this threshold, this will be direct
evidence of its double-degenerate origin. At the moment, there is no
plausible candidate of such a type Ia SNR, because even the Small
Magellanic Cloud has an average metallicity of $\sim 0.15$\,solar
(Russell \& Dopita 1992). However, local inhomogeneities in the ISM
abundances may allow some SNR progenitors' metallicity to be lower
than the mean value of its host galaxy. We should therefore keep in
mind that non-detection of the Mn line is important to constraining
the progenitor type.


\section{Ejecta Distribution in Space and Velocity }

\subsection{Background and Previous Studies}

The questions to be addressed here are twofold: (1) How are the ejecta
distributed in core-collapse and SN Ia remnants?  (2) What does this
tell us about the explosion mechanisms and the physics of shocks in
these objects?

Measurement of the distribution of the ejecta in space and velocity
provides unique information about both the progenitor star and the
details of the explosion.  The ejecta from core-collapse SNe and
deflagration/detonation (Type Ia) SNe have distinct abundance
patterns. Specifically, the ejecta from core-collapse SNe have high
ratios compared with cosmic abundances of oxygen to iron; those from
Type Ia's have the opposite. X-rays are the ideal band for seeking
this signature, as broad band X-ray spectrometers cover the K lines
from oxygen and iron, as well as the iron L band. The complications of
abundance determination make SN typing challenging. Compounding the
difficulty is our inability to know what fraction of the ejecta is
visible in the X-ray band, and the fact that the oxygen abundance is
difficult to determine in an unambiguous way as the oxygen line
strength correlates with the column density in spectral fitting in
moderate spectral resolution detectors.  This is particularly
problematic for Galactic SNRs with heavy interstellar absorption.

Despite these difficulties, there were some notable early successes in
measuring abundances, and deriving physical information from the
measurement. The most successful early attempt was the use of 
{\it Einstein} FPCS observations of Puppis A to show an overabundance of
oxygen with respect to iron, requiring the progenitor to have a mass
of more than 25 \msun\ \citep{canizares81}.  Other early observations
of note are comprehensive analysis of Tycho data requiring ejecta
\citep{hamilton86} and the analysis of an {\it EXOSAT} observation of
W49B requiring the presence of a substantial amount of shocked ejecta
\citep{smith85}. However, it is with the advent of spatially resolved
spectroscopy through which shock structures can be isolated that most
of the advances have taken place in our knowledge of the abundances of
reverse shocked ejecta and forward shocked ISM. {\it ASCA} was the
pathfinder mission in this regard (e.g., \citealt{holt94, hughes95}), but
\xmm\ and \chandra\ excel in these studies. The recent studies are
providing real insight into the ejecta masses and their degree of
mixing, as well as the explosion mechanism (Vink 2012).

{\bf Type Ia Remnants:} The two most thoroughly studied Type Ia
remnants are Tycho and SN~1006.  The \xmm\ image of Tycho shows
that the Fe-K emission peaks at a smaller radius than the Fe L
emission, verifying that the temperature with the ejecta increases
toward the reverse shock \citep{decourchelle01}. The narrow band Si
image corresponds well with the radio image, and is thought to mark
the contact discontinuity, distorted by Rayleigh-Taylor
instabilities. This latter conclusion is reinforced in a dramatic
fashion by the \chandra\ image of the Si emission, which shows plume
like structures throughout the interior \citep{hwang02, warren05},
whose structure is reproduced by 3-D hydrodynamic modeling of the
ejecta \citep{warren13}. Some of the plumes viewed tangentially reach
the outer shock.

The line emission in SN~1006 has been a secondary consideration to the
non-thermal emission arising from the bright limbs. Line emission is
clearly observed throughout the remnant, except in the bright
non-thermal limbs. Along the northwestern rim, \chandra\ imaging
spectroscopy shows a clear separation between the forward-shocked
material and the ejecta \citep{long03}. The forward shock shows
material at ordinary solar abundances, shock-heated to electron
temperatures of $\sim$0.6--0.7 keV. Interior to both the
non-thermal northeast shock and the thermal northwest shocks are
plume-like structures similar to those observed in Tycho. Their
presence invites the speculation that such structure is common in Type
Ia remnants. Spectral analysis of these structures reveals enhanced O,
Mg, Si, and Fe abundances.  No quantitative X-ray based analysis of
the ejecta mass has been performed for SN~1006. Abundance measurements
in SN~1006 using X-rays are compromised by the known presence of a
substantial amount of high-velocity, unshocked ejecta (Fe, Si, S, and
O) interior to the reverse shock \citep{wu83}.

In addition to these prominent examples, there are a number of other
young Type Ia remnants accessible for study using \ah.  These
include the Galactic remnants Kepler, G1.9+0.1, and RCW~86, and the
LMC remnants 0509, 0519, and N103B.

{\bf Core-Collapse SNRs:} Core-collapse SNRs generally exhibit highly
inhomogeneous ejecta in X-rays (e.g., \citealt{hughes00b, hwang04,
  park07}).  SXS measurements of the radial velocities of different
elements as a function of position provides a three-dimensional
picture of the ejecta dynamics needed to distinguish between various
explosion scenarios. For example, jet-like models can be distinguished
from models involving hydrodynamic instabilities based on their
velocity maps. By measuring the abundances of the ejecta produced near
the mass cut, SXS also provides insight into the core-collapse that
initiates the explosion.

The first \chandra\ images revealed clear and unexpected differences
in the distribution of metals (e.g., \citealt{hughes00b,
    hwang00a}). The Si, S, Ar, and Ca maps are similar to each other,
  and to the distribution of fast optical knots. Equivalent width maps
  reveal the distribution of the prominent ejecta constituents. The
  structures in these maps contrast sharply with the 0.5--10.0 keV
  broadband map and the 4--6 keV continuum map. The northeastern jet,
  known from optical studies to contain Si group ejecta, shows up
  strongly in these maps. The Fe-K emission has a very different
  morphology. In particular, in the southeast of the remnant, the Fe-K
  emission is located at larger radii than the Si. This suggests that
  the inner Fe ejecta layers have been overturned and propelled beyond
  the Si group ejecta in this part of the remnant (Figure \ref{casa:ptgs}). Such
  overturning is consistent with recent models of core-collapse
  explosions (e. g., \citealt{burrows95}). \citet{willingale02} have
  used the \xmm\ imaging data to infer the global metal abundance
  ratios and compare them with supernova models. They show that the
  ratios with respect to Si of a large number of lines is most
  consistent with the theoretical nucleosynthesis yield for a 12 \msun\
  progenitor.

Detailed studies of individual knots show that they have a variety of
compositions.  Features with distinct composition can be found on the
smallest size scales. While most knots show a mix of ejecta, some are
dominated by Si group elements and others by Fe. At least one knot
emits Fe lines exclusively, and apparently is devoid of lower mass
material. The knots also show a variety of ionization conditions,
which have been used in the context of analytic hydrodynamical models
to constrain the ejecta density profile, the location of the knots in
mass coordinates, and the degree of explosion asymmetry
\citep{laming03, hwang03}.  The ejecta show a range of density
profiles, from very shallow ($\propto r^{−n}$, where n$\sim$6) to
very steep, (n$\sim$30--50). The ejecta close to the jet show the
shallowest profile, possibly due to an asymmetric explosion in which
more of the energy is directed along the jet than elsewhere. For a
total ejecta mass of 2 \msun\ expected from a 20 \msun\
progenitor, the Fe-rich clumps are found to arise in a layer 0.7--0.8
\msun\ from the center. The observed composition appears to be
possible only if Si burning products are mixed with O burning
products.

The overall appearance of Cas~A contrasts starkly with the young Type
Ia remnants. Cas~A consists of small knots and thin filaments, not the
emission plumes observed in Tycho. The similarity between the
structures in Cas~A and the prediction of models involving Fe bubbles
has been noted. \citet{laming03} argue that the knots are not
especially over-dense compared with their surroundings, and their high
ionization ages and the proximity of some to the forward shock are the
result of early passage through the reverse shock.

The \chandra\ image of the other O-rich and young ($\sim$1600 yr) SNR G292.0+1.8, reveals
a thin, nearly circular outer shell of hard emission filled with an
array of knots and filaments rivalling Cas~A in complexity and
contrasting starkly with the Type Ia remnants \citep{hughes01, park02, gonzalez03}. The
composition and distribution of the shocked ejecta are different from
Cas~A. The ejecta consist primarily of O, Ne, and Si, with less S and
Ar, and very little Fe (although a recent {\it Suzaku} study led to the detection
of hot Fe ejecta; \citealt{kamitsukasa14}) , and are distributed primarily around the
remnant's periphery. An X-ray bright equatorial band has normal
composition, and is thought to be associated with pre-supernova mass
loss (e.g. \citealt{park07}). Measuring the velocity structure of the ejecta could be addressed \ah\
(see also Section 1.3.7).

Compared with the Type Ia remnants there is a dearth of young or
historical Galactic core-collapse remnants.  As with the Type Ia
remnants, however, the \ah\ sample can be expanded through the
study of bright Magellanic Cloud remnants, including 1E 0102.2$-$7219
and N63A.

\subsection{Prospects \& Strategy}

Specific questions that \ah\ can provide unique information about
include: (i.) what is the large-scale velocity structure and metal
abundance spatial distribution of SN ejecta; (ii.) how much turbulent
velocity structure is there; (iii.) what are the ion temperatures in
the ejecta and how do they compare to the thermal, bulk, and turbulent
velocities; i.e., can we distinguish between the different
constituents of a SNR (e.g., ejecta, ISM) using velocity or line width
information; (iv.) can we detect evidence for radioactive decay of
short-lived isotopes – the lines of $^{44}$Ti, its daughter product
$^{44}$Sc (4.1 keV) or Co-K (6.9 keV); (v.) can we detect evidence for
supra-thermal electrons to address the injection problem in particle
acceleration at shocks?

All of the above questions can be addressed using the high resolution,
spatially-resolved spectroscopy enabled by the SXS and/or the broad
band imaging spectroscopy provided by the HXI.  The SXS allows us to:
measure the velocity structure, abundances and spatial distribution of
SN ejecta; measure thermal, bulk, and turbulent velocities in resolved
ejecta features; search for weak lines from radioactive decay
products; and use Fe-K diagnostics and line shapes for to search for
supra-thermal electrons.  The HXI allows a measurement of the hard
continuum to search for non-thermal Bremsstrahlung from supra-thermal
electrons, and should provide sufficient sensitivity at the high end
of its band to detect the $^{44}$Ti lines at 68 and 78 keV in some
remnants.

\subsubsection{Structural diversity in Ia remnants}

Type Ia SNe are important astronomical phenomena because of their use
for a study of cosmology; the large and relatively homogeneous
luminosity make them to be standardizable candles to measure
cosmological distance, which contributed to revealing the accelerating
universe \citep{riess98, perlmutter99}
The SNe Ia also made an important role in chemical evolution of the universe, as
they are major sources of Fe-peak elements (e.g., 
\citealt{nomoto84}).  Despite effort in the last decades, however,
detailed explosion mechanism of SNe Ia is still unsolved. It is widely
known that type Ia SNe show diversity in their optical spectra and
lightcurve (e.g., \citealt{phillips99, benetti05}).  A primary origin
of the diversity is under the debate. One interpretation is that
spherically asymmetric explosion is primarily responsible (e.g.,
\citealt{kasen09}).  \citet{maeda10a} systematically studied SNe Ia
and found that random viewing angles for such asymmetric (but almost
identical) explosions can quantitatively explain the observed spectral
diversity. Besides them, recent multi-dimensional simulations have
suggested that thermonuclear ignition in Type Ia progenitors is offset
from their center (e.g., \citealt{woosley04, kuhlen06}).  If this
interpretation is the case, a SNR should show nonuniform ejecta
distribution as the result. Some observations of SNRs in X-rays indeed
revealed such asymmetric distribution of ejecta.
Figure~\ref{sn1006:suzaku} shows {\it Suzaku}/XIS narrow-band images of
SN~1006 \citep{uchida13}.  Si and other intermediate-mass elements
concentrate in the southeast region, while the lighter elements (i.e.,
O, Ne, Mg) distribute relatively homogeneously.  Tycho also shows
evidence of local inhomogeneity in the ejecta (e.g.,
\citealt{vancura95}).

\begin{figure}
  \begin{center}
     \includegraphics[scale=0.8]{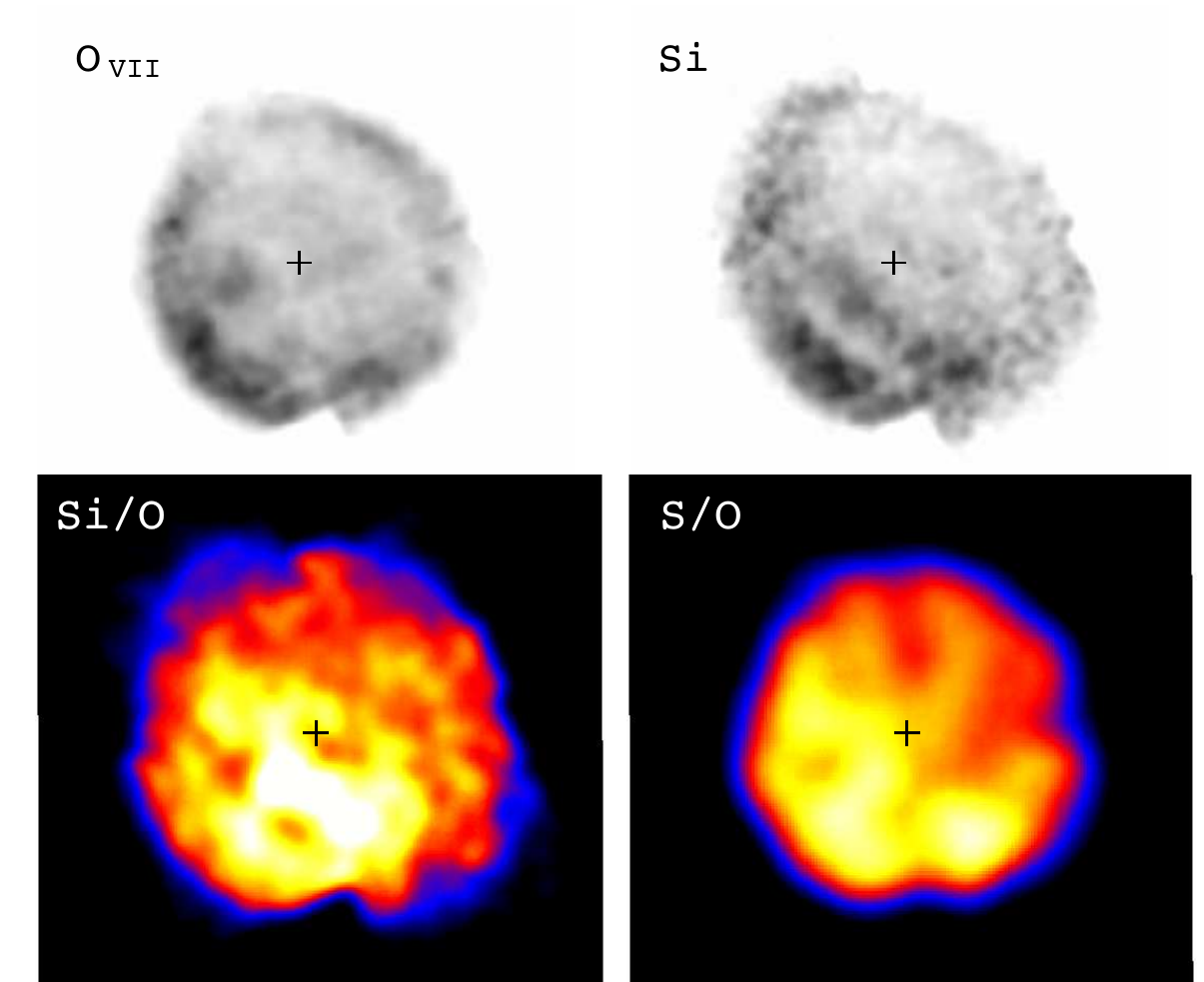}
  \caption{{\it Suzaku} XIS images of SN~1006 in the O (top-left) and Si
    (top-right) bands.  Equivalent-width-ratio images of Si/O and S/O
    are shown in bottom-left and bottom-right, respectively (Uchida et al. 2013). These
    indicate asymmetric distribution of the intermediate-mass
    elements.  }
  \label{sn1006:suzaku}
  \end{center}
\end{figure}

There is, however, another interpretation for the major origin of the
diversity. \citet{mazzali07} showed that the summed masses of
$^{56}$Fe and the intermediate-mass elements are almost uniform among
various SNe Ia, even though the $^{56}$Fe mass is widely spattered in
the range of 0.2\,\msun--0.9\,\msun.  They thus argued that the
diversity mainly originates from the difference in explosion energy
and/or density when a deflagration-to-detonation transition (DDT)
occurs (see Section 1). Theoretically, both the explosion energy or
DDT density affect the expansion velocity of the SN ejecta (e.g.,
\citealt{iwamoto99}), which can be investigated by high-resolution
spectra of SXS.

Investigation into global distribution and velocity structure of
ejecta is also important to understand how the explosive
nucleosynthesis wave propagates in the exploding white dwarf. A
delayed-detonation scenario, well accepted as a standard SN Ia model,
predicts layered composition of nucleosynthesis products with little
effect of mixing (e.g., \citealt{maeda10b}).  On the other hand, a
full deflagration explosion predicts significant mixing among the
elements (e.g., \citealt{ropke07}).  X-ray observations of Ia SNRs
show evidence of layered ejecta distribution with Fe commonly
concentrated toward the SNR's center with respect to the lighter
high-abundance elements, such as Si and S (e.g., \citealt{hwang97}),
supporting qualitatively the delayed-detonation scenario.  However,
the distribution among other elements (e.g., Si vs Cr, Cr vs Fe, Fe vs
Ni) are still unknown.  How does DDT occur? Recent multi-dimensional
simulations showed that a deflagration-dominant SN Ia produces larger
turbulent instabilities than detonation-dominant one (e.g.,
\citealt{kasen09}), suggesting that dimmer SNe Ia should have larger
velocity dispersion in their ejecta. \citet{howell01} showed that
sub-luminous SNe Ia are more spherically asymmetric than luminous
ones. These would be important subjects for high-resolution
observations with SXS.

So far, distribution of ejecta in SNRs has been investigated mostly by
X-ray images.  This is, of course, the most straightforward way. The
high-resolution imaging capability of \chandra\ and \xmm\ has largely
improved our knowledge of spatial structure in SNRs during this
decade. Unfortunately, the angular resolution of \ah\ is not as good as
that of \chandra\ or \xmm. We can, however, still investigate global
ejecta distribution with a different way, making most use of the
capability of SXS.  Given that a SNR's reverse shock heats and ionizes
the ejecta from the outer layer inward, the inner ejecta should have a
lower ionization age than the outer ejecta. In fact, Fe ejecta in
Tycho and SN~1006 are known to have a lower ionization age compared to
the intermediate-mass elements \citep{hwang97, yamaguchi08}.  In
addition, recent {\it Suzaku} observations of Tycho with deep exposure
has revealed that Fe-K$\beta$ emission has a smaller peak radius than
the Fe-K$\alpha$ emission (\citealt{yamaguchi13};
Figure~\ref{tycho:suzaku}). Since the K$\beta$ emission is induced by
$3p$$\rightarrow$$1s$ transition following $1s$-shell ionization, this
is likely to originate from the more-recently-shocked ejecta with
extremely low ionization state (Fe$^{\sim 8+}$; see also the New Spectral Features White Paper \#17),
showing clearly the correlation between the radius and ionization age.
With high-resolution spectra of SXS, we can more accurately measure an
ionization age of {\bf each independent element} from the line
centroid or ratio of their emission (which is also essential to
determine accurate abundance; Section 1).

As mentioned above, it was suggested that dimmer SNe Ia tend to have a
larger velocity dispersion \citep{kasen09}.  We can explore the
correlation between the Fe abundance and turbulence velocity of
several SNRs.  This can also be done with the SXS. The SXI also help
investigate large-scale ejecta asymmetry.

\subsubsection{SN~1006: An exceptional target}

Here we describe how these measurements provide insight into some of
the key questions using SN~1006 as an example.  SN~1006 is an
exceptionally important target for \ah.  It is the remnant of the
brightest supernova ever observed, a Type Ia explosion.  In contrast,
today its remnant is under-luminous at all wavelengths, due to its
location in a very low ambient density high above the Galactic plane.
Its distance is reasonably well determined to be 2.2 kpc, and its
$\sim$30-arcminute extent makes it well suited for spatially resolved
spectroscopy.  While its emission is dominated by bright bands of
non-thermal emission in its NE and SW quadrants, the X-rays from the
interior arise from reverse shocked ejecta, characteristic of a Type
Ia remnant.  The proper motion in the NE and NW filaments has been
directly measured using \chandra\ observations at 5000 and 3000 km
s$^{-1}$, respectively.

\citet{wu83} detected high velocity, UV absorption lines from low
ionization states of Fe and Si using IUE observations of the
Schweitzer-Middleditch (SM) star that lies close to the center of SN
1006 in projection.  This has provided invaluable information on the
properties of the {\it unshocked} ejecta in SN~1006.  With the SXS on
\ah\ we can measure the velocity structure of the {\it shocked} ejecta
along very nearly the same line-of-sight.

SN~1006 has a reported measurement of an ion temperature based on
X-ray data compared to an optically determined velocity.  In the
X-ray, \cite{vink03} used the RGS on \xmm\ to measure the widths of
the O VII and VIII K-shell lines coming from a knot on the
northwestern edge.  The width they measured, $\sigma = 3.4 \pm 0.5$
eV, indicates an oxygen temperature of $kT\sim$500 keV. This
temperature, if assumed to arise from the thermalization of the shock
velocity according to the Rankine-Hugoniot relations, $kT_i =
(3/16)m_i v_s^2$, implies a shock velocity of $\sim$4000 km s$^{-1}$,
broadly consistent with the value from optical studies. The main
weakness of this result is the assumption that turbulent velocities
are negligible (i.e., the observed line broadening is purely thermal).
This concern can be addressed by the \ah\ SXS by studying the line
widths across different elemental species.

SN~1006's integrated X-ray spectrum is dominated by non-thermal
synchrotron emission from relativistic electrons with energies
approaching the TeV range \citep{koyama95}.  But this emission only
comes from two bright ``caps'' located on the northeast and southwest
rims of the remnant.  As shown by \citet{gamil08}, the southeastern
rim shows very faint radio emission and no non-thermal X-rays based on
\chandra\ observations.  \citet{gamil08} make the case for azimuthal
variation of the efficiency of CR acceleration around the rim of
SN~1006: minimum in the SE and maximum in the NE and SW bright lobes.
Although there is no evidence for non-thermal emission from the
southeast (and northwest rims), the emission there appears to be
entirely from shock-heated ejecta (e.g., \citealt{katsuda13}).  Thus
it is clear that there is significant variation in the density of
relativistic (non-Maxwellian) electrons around the rim of SN~1006.
This difference provides a unique opportunity to search for the
effects of a non-Maxwellian electron distribution on the emission line
properties.

SN~1006 observations can therefore be used to address many of the
questions of interest:
\begin{enumerate}

\item Identifying lines present in the X-ray spectrum, and using them
  to determine the abundances and state of the plasma across the
  remnant (temperature, degree of equilibrium), as discussed in detail
  in Section 1 above.

\item Measuring the expansion velocity along the line of sight by
  separation of red and blue shifted components.  Measuring the
  velocity along the rims through line broadening.  In principle it
  will be possible to search for velocity differences among the
  individual elements (e.g., O vs. Si).  Combining the expansion
  measurement with the X-ray proper motions already determined by
  \chandra\ will provide an independent measurement of the distance to
  SN~1006. This will not be subject to uncertainties in the
  interpretation of the widths of optical Balmer lines, as the current
  most precise distance measurements are.

\item Determining the electron-ion temperature non-equilibration at
  various locations along the rim.  It will be possible to search for
  differences in ion temperatures among distinct ionization states and
  elements.

\item Detecting the effects of non-thermal particles on thermal
  emission: This could manifest itself in one of two ways: either the
  lines are narrower than expected given the velocity, due to energy
  leakage into cosmic-rays (e.g., \citealt{hughes00a, helder09}), or
  the lines are broader than expected due to atomic interactions with
  accelerated ions \citep{tatischeff98}.

\end{enumerate}

\subsection{Targets \& Feasibility}

Many of these questions will be addressed by observing
the same targets as mentioned in Section 1 above.
We present next specific feasibility estimates for two illustrative targets:
SN~1006 and the LMC SNR 0519--69.0. 

\subsubsection{SN~1006: determining the expansion velocity in multiple
elemental species} \label{subsec:sn1006}

For SN~1006 \ah\ SXS simulations were performed for the center, NW rim
and NE rim, based on emission models derived from fitting
\chandra\ ACIS and \suzaku\ XIS data.  The inferred count rates for
these regions are 0.11, 0.09, and 0.31 cts/s for the center, NW rim
(outer half of the FOV), and the NE rim, respectively.

The center field spectrum included expansion velocities of +6000 km
s$^{-1}$ (redshift) and -5000 km s$^{-1}$ (blueshift) for O, and
+5000 km s$^{-1}$ (redshift) and -4000 km s$^{-1}$ (blueshift) for Si
(see Figure~\ref{sn1006:center}). An exposure time of 200 ks gives
fairly tight constraints: $\pm$3\% for O He$\alpha$ and $\pm$10\% for
Si He$\alpha$.  In reality, multiple-ionization plasma along the line
of sight may cause some additional uncertainties.

\begin{figure}
\hfill
\subfigure[Oxygen]{\includegraphics[width=7.5cm]{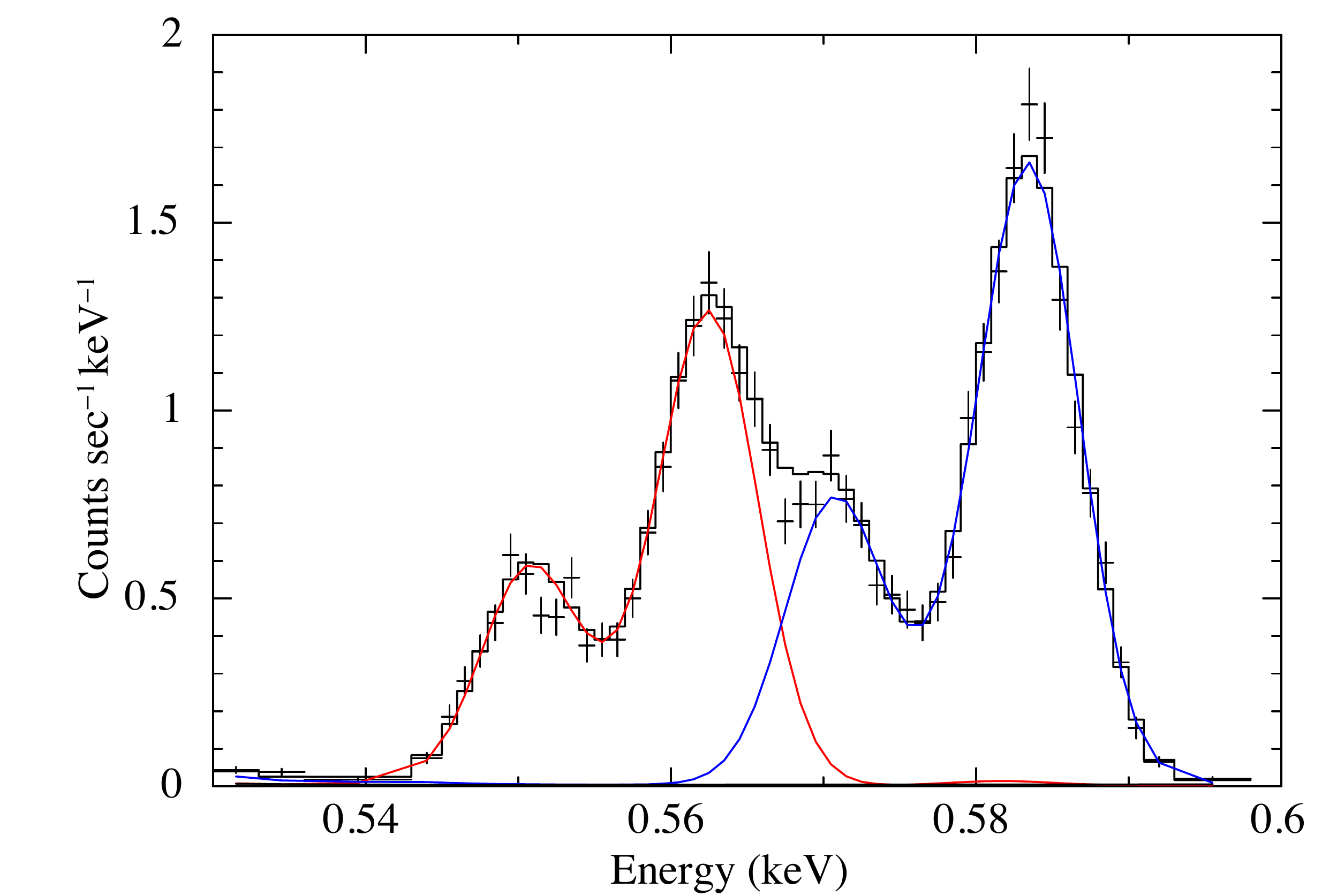}}
\hfill
\subfigure[Silicon]{\includegraphics[width=7.5cm]{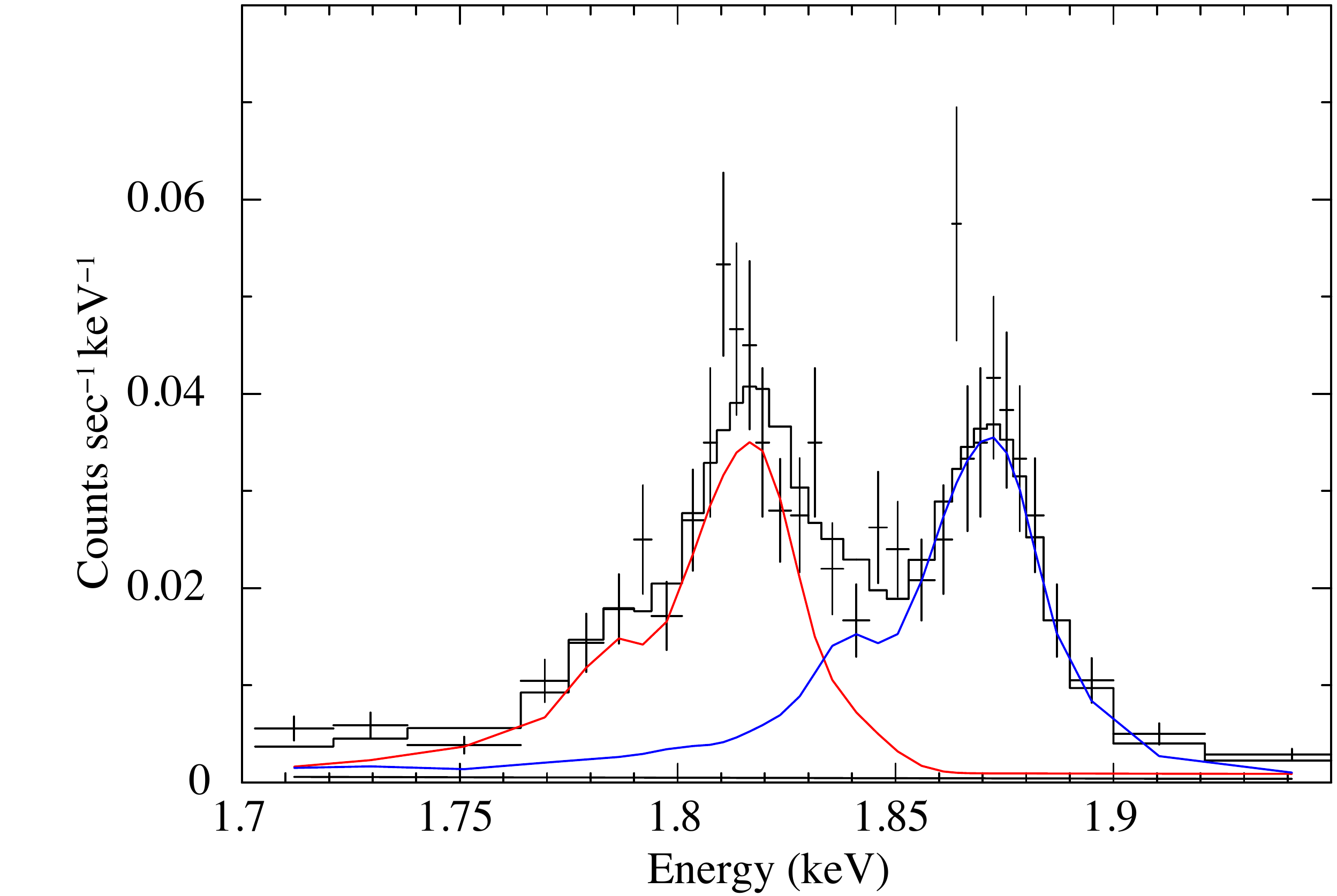}}
\hfill
\caption{Simulated SXS observations of a pointing near the center of
  SN~1006 showing Oxygen ({\it Left}) and Silicon ({\it Right}) that
  demonstrates how cleanly we will be able to separate the emission
  from the approaching (blue curves) and receding (red curves)
  hemispheres of the expanding shell of ejecta.}
 \label{sn1006:center}
\end{figure}

Thermal Doppler broadening was introduced into the NW rim and center
spectra by adding thermal width to the lines of $\sigma$ = 2.4 eV
\citep{vink03}.  This width is clearly resolved in the 200 ks exposure
of the center and a 100 ks exposure of the NW rim.

\subsubsection{SNR 0519--69.0: bulk motion expansion} \label{subsec:0519}

SNR 0519--69.0 is the second brightest (in soft X-rays) LMC remnant of
Type Ia SN origin.  In the optical band this remnant shows a pure
Balmer line spectrum \citep{tuohy+82} with a broad H$\alpha$ component
width of $2800 \pm 300$ km s$^{-1}$. Similarly broad H Ly$\beta$ was
detected in FUSE observations, which were used to constrain the
velocity of the forward shock to be 2600--4500 km s$^{-1}$
\citep{ghavamian+07}.  The RGS spectrum \citep{rasmussen02} shows
broadened X-ray lines, likely due to the expansion velocity of the
shell.  Figure~\ref{0519:sxs} shows a simulated SXS spectrum of SNR
0519--69.0 for a 60\,ksec exposure, assuming a bulk motion velocity
broadening consistent with previous work.  The widths of Si lines can
be determined with an uncertainty of $\sim$10\%.

\begin{figure}
  \begin{center}
     \includegraphics[scale=0.3]{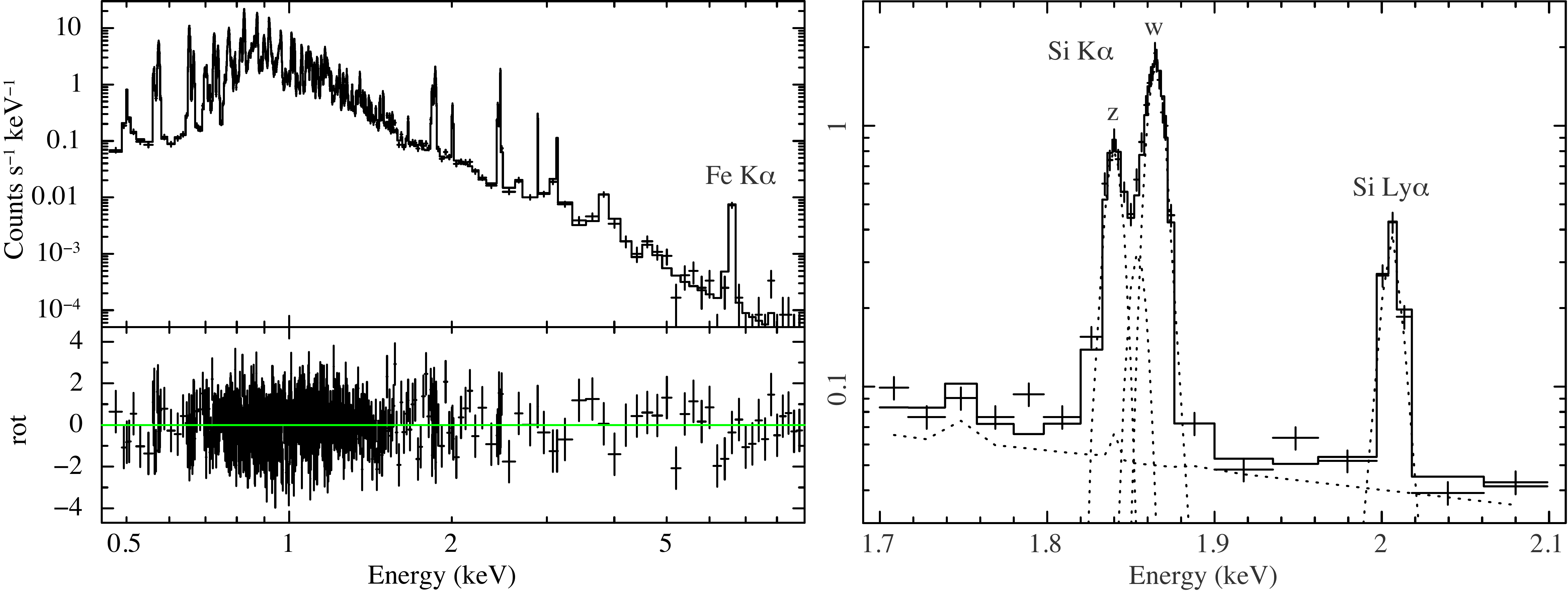}
  \caption{Simulated SXS spectrum of SNR 0519--69.0 with an exposure of 60\,ksec. 
  The widths of Si lines can be determined with an error of $\sim$10\%.}
  \label{0519:sxs}
  \end{center}
\end{figure}

\subsubsection{Other \ah\ targets}

Many of the questions discussed in this section will be addressed by observing 
the following SNRs: Tycho, SN~1006, Cas~A, SN~1987A, 
G1.9+0.1. Additional targets include Kepler, RCW~86,
G292.0+1.8, 0509$-$67.5, 0519$-$69.0, and N63A.

Furthermore, SN~1987A (see Section 1.3.4) is a promising target for
this science topic, particularly for measuring the velocity structure of
Mg, Si, S and Fe lines.  Multiple pointings spanning several years will allow us to
track the time variability of all emission components, and further address the topics
discussed in Sections 1 and 3 as well.

\subsection{Beyond Feasibility}

\noindent {\bf The radial velocity structure of Cas A from the profile
  of the Fe-K line}

For this ``Beyond Feasibility'' section we reach back to the
\casa\ simulation presented in Section 1.3.3.  The general idea is to
use a specific line profile to inform the radial velocity structure of
that particular species.  It can be applied to any emission line from
any remnant, but Cas A's exceptional brightness makes it the best case
on which to attempt such a measurement.

Figure~\ref{casa:spectra2} (right panel) shows the detailed structure of
the line emission from different charge states contributing to the Fe
K line for an NEI model that matches the \suzaku\ \casa\ observation.
The expected bulk motion of 3000 km s$^{-1}$, included in the full
spectrum that is plotted above the individual line contributions,
smears out the charge state structure.  With some assumptions (e.g.,
that the thermodynamic state does not vary strongly with radius) and
sufficient signal, it may be possible to extract information on the
radial velocity distribution from the Fe-K line profile. Arriving at a
robust result from this measurement will be challenging.

\section{How Does the Thermal Plasma State of a Supernova Remnant
 Link with the Efficiency of its Particle Acceleration?  }

\subsection{Background and Previous Studies}

Key questions to address in this topic area are: Where is the thermal
emission in synchrotron-dominated SNRs?; what is the partition of
shock energy into bulk motions, thermal energy, and relativistic
particles?

Supernova remnants have long been considered to be the primary
acceleration sites of cosmic-ray particles below the energy of the
so-called {\it knee} in the cosmic ray spectrum, $\sim$$10^{15}~{\rm
  eV}$.  The first evidence for multi-TeV acceleration was the
discovery of synchrotron X-ray emission from the shell of the
supernova remnant SN~1006 \citep{koyama95}. 
Recent high quality morphological
and spectral studies with the HESS TeV imager combined with the X-ray
imager such as {\it ASCA}, found a good keV-TeV correlation from RX~J1713.7$-$3946 \citep{Aharonian04} 
and RXJ~0852.0$-$4622 \citep{Aharonian07}, indicating that both the X-rays and the TeV
$\gamma$-rays are emitted by the TeV particles in the SNR shell. 
The extremely thin X-ray filaments with a 0.02~pc width in SN~1006
\citep{Bamba03,long03}, as well as the X-ray variability on time
scales of a year from RX~J1713.7$-$3946 \citep{Uchiyama07} strongly
support the efficient acceleration of particles by SNR shocks.  Up to
now, many observational results support that supernova remnants (SNRs)
are major sources of galactic TeV cosmic rays.

Efficient acceleration of cosmic rays (CR) at a supernova shock has
several observational consequences.  First is the increase in the
compression factor above the typical factor of four assuming the
Rankine-Huguniot jump conditions.  This has the effect of modifying
the structure of the shocked region, particularly the gap between the
forward shock and the contact discontinuity \citep{decourchelle00}, an
effect that has now been observed in both Tycho \citep{warren05} and
SN~1006 \citep{gamil08, miceli12}. Directly imaging the gap requires the high
angular resolution of \chandra, but the increased density enhances the
ionization state of the post-shock gas \citep{patnaude09} and may
therefore be observable by the \ah\ SXS.  

Furthermore, recent 3D hydrodynamical simulations coupled
with a non-linear acceleration model to account for efficient particle acceleration
show that the integrated thermal X-ray emission is reduced with particle back-reaction, 
with the effect being more significant for the highest photon energies (e.g. Fe-K vs. O-K to Ne-K energy band) \citep{ferrand12}.
As well, the computed non-thermal emission maps and broadband spectra (radio, X-rays and $\gamma$-rays)
can be used to probe the presence of energetic \textit{ions} at the shock \citep{ferrand14} 
since, in addition to impacting the dynamics of the shock and thermal X-ray emission (as mentioned above),
they impact the evolution of the magnetic field and thus the non-thermal emission from electrons.
High magnetic fields ($>$100$\mu$G) directly impact the synchrotron emission from electrons,
by restricting their emission to thin rims, and indirectly impact the inverse Compton
emission from electrons and the pion decay emission from protons by shifting
their cut-off energies to respectively lower and higher energies.
Such high magnetic fields have been in fact inferred from X-ray observations of
non-thermal rims of bright and young SNRs such as Tycho (e.g., \citealt{slane14})  
and of synchrotron-dominated, efficient particle accelerators, such as RX~J1713.1$-$3946 \citep{Uchiyama07}.
For RX~J0852.0--4622 (Vela Junior), however, the level of magnetic field amplification remains an open question \citep{lee13, Aharonian07, 
berezhko09, bamba05b}.
Such effects can be observed with \ah\ SXS and HXI+SGD observations, combined with 
high-resolution (\chandra\ or \xmm) observations and
multi wavelengths studies from radio to TeV energies.

The level to which we can observe such effects strongly depends on the efficiency of CR acceleration at the supernova shock. 
This efficiency can be as high as 50\% (e.g. \citealt{morlino13, helder09}), which represents a sizeable drain
of energy from the thermal population. In this case the temperature of
the ions and electrons will be reduced with a measurable effect. 
In particular, for SNRs with accurately
known shock velocities (like those in the Magellanic Clouds or with
shock velocities determined from broad Balmer lines) it will be
possible to combine shock velocity measurements with temperature
measurements to search for evidence of a reduced temperature due to
efficient particle acceleration.  This was done early on during the
\chandra\ mission for the young oxygen-rich SNR 1E 0102.2$-$7219
\citep{hughes00a}, where a significantly lower electron temperature
was measured compared to the shock velocity (lower even when
non-equipartition of electron-ion temperatures was included).

Another related topic concerns the lack of thermal X-ray emission from the
synchrotron-dominated shell-type SNRs, e.g., RX~J1713.7$-$3946 and RX~J0852.0$-$4622, 
for which CR acceleration efficiency is believed to be high.  The
\ah\ SXS should take deep exposures near the non-thermal rims of these
SNRs to detect or set the best constraint on the thermal emission.
Detecting a thermal signature will be a major discovery since it will
provide an estimate for the ambient medium density and thus allow us
to confirm or refute the hadronic vs.\ leptonic origin of their TeV
$\gamma$-ray emission.

We also wish to highlight the injection mechanism of cosmic ray
electrons.  Electrons that enter into diffusive shock acceleration
must be pre-accelerated to supra-thermal energies (above $\sim$10~keV)
to cross the shock front. We currently do not understand how these
energies are attained. The observational constraint of the energy
spectrum of supra-thermal electrons will bring critical information on
the injection mechanism.

\subsection{Prospects \& Strategy}

In order to carry out a study of the loss of shock energy to
relativistic particles at the expense of the thermal particles, we
require ``clean'' shocks where the thermal emission is clearly from
the post-shock flow (and not further back or from a different
component like the ejecta when studying the forward shock).  The best
possible strategy for this involves observing a large Galactic remnant
with well studied Balmer shocks.  SN~1006 may not be ideal, because of
its strong ejecta emission and lack of X-ray emission from its
interstellar forward shock. RCW~86 may be the best candidate.

To address the injection problem, we have two approaches to detect
supra-thermal electrons at the early stage of the acceleration process:
plasma diagnostics by high-resolution spectra with SXS and hard X-ray
mapping with HXI.

With the SXS, we can resolve the He-like triplet and Li-like satellite
lines.  The He-like resonance line "w" is sensitive to both thermal
and supra-thermal population, while the dielectronic recombination (DR)
satellite lines such as "j" and "d13" are only sensitive to discrete
energies in thermal domain \citep{1979MNRAS.189..319G}.  For thermal
plasma, the line ratios "d13"/"j" and "w/j" are determined only by the
temperature. If supra-thermal electrons exist, however, the ratio "w/j"
is enhanced. The excess of "w/j" above purely thermal case gives the
quantitative constraint to the energy spectrum of supra-thermal
electrons. This method is originally proposed and succeeded for solar
flares \citep{1979MNRAS.189..319G, 1987ApJ...319..541S}. The similar
idea using H-like resonance and the He-like satellite lines is also
proposed for cluster plasma \citep{2009A&A...503..373K}.

Hard X-ray imaging spectroscopy with HXI can also put constraints on the
supra-thermal distribution of electrons. In solar flares, hard X-ray
bursts have been observed preceding the rise of supra-thermal electrons
\citep{1987ApJ...319..541S}.  Such direct bremsstrahlung would be
detected from SNR shocks well above the cut-off energy of synchrotron
X-rays.  While past hard X-ray observations were not able to bring
significant results \citep{2006ApJ...644..274K, 2008A&A...486..837V},
a deep observation with the superior imaging capability of HXI may detect
the emission for the first time. Mapping the bremsstrahlung emission
from supra-thermal electrons would reveal their spatial distribution
behind the shocks, which gives additional information on the injection
mechanism.

\subsection{Targets \& Feasibility}

\subsubsection{Cas A: non-thermal X-ray emission } \label{subsec:casanontherm}

Cas~A is one of several SNRs from which non-thermal X-rays and TeV
$\gamma$-rays have both been detected
(X-rays: \citealt{Allen97,Uchiyama08}, TeV: \citealt{Aharonian01,Albert07}).
In X-rays \casa\ seems to consist of a number of thermal and
non-thermal X-ray emitting knots/filaments
\citep{hughes00b,hwang04,Bamba05a}. Although some non-thermal emission
is associated with the forward shock, the dominant source of
non-thermal emission may be identified with the reverse shock regions
\citep{Helder08,Maeda09}.  It therefore is a unique object in which
we can study the particle acceleration by the reverse shock, because
for the other SNRs the acceleration seems to originate from the
forward shock region only (e.g., \citealt{Parizot06}).

A long observation of Cas~A will be gradually made over several
years. Such a long-term observation gives us a bonus to make a flux
and spectral shape monitor of the synchrotron emission from the
non-thermal high energy electrons. \citet{Patnaude11} showed that the
soft X-ray continuum of Cas~A has declined between 2000 and 2010. At
the same time the spectrum has softened. This can be understood
assuming a loss-limited synchrotron spectrum, as the exponential
cut-off energy depends solely on the braking of the shock velocity. This
is a reasonable interpretation, but should be clarified with a
follow-up observation since the spectral slope and flux is also
dependent on an assumption of the limit (age or loss) and evolution of
magnetic field, etc. (see \citealt{Vink12}). Above $\sim$8~keV, the
X-ray flux is dominated by the synchrotron emission (c.f.,
\citealt{Maeda09}). The wide band spectroscopy with
the HXI and the SXI covers a band above the cut-off energy at possibly
$\sim$3~keV (c.f., \citealt{Maeda09}) up to $\sim$80 keV\@. 
The band is very sensitive to test if the cut-off origin fully explain 
the spectra slope and flux seen in \chandra.

Cas~A is also a plausible target for searching for supra-thermal
electrons.  The remnant has both the bright radio emission from
energetic electrons and strong K-shell lines from highly ionized Fe
ions.  As the reverse shock is responsible for the accelerated
particles as well as the metal-rich plasma in Cas A
\citep{Helder08}, the interaction between the supra-thermal
electrons and Fe ions is naturally expected.

\begin{figure}[h]
\begin{center}
\includegraphics[width=0.48\textwidth,clip,angle=0]{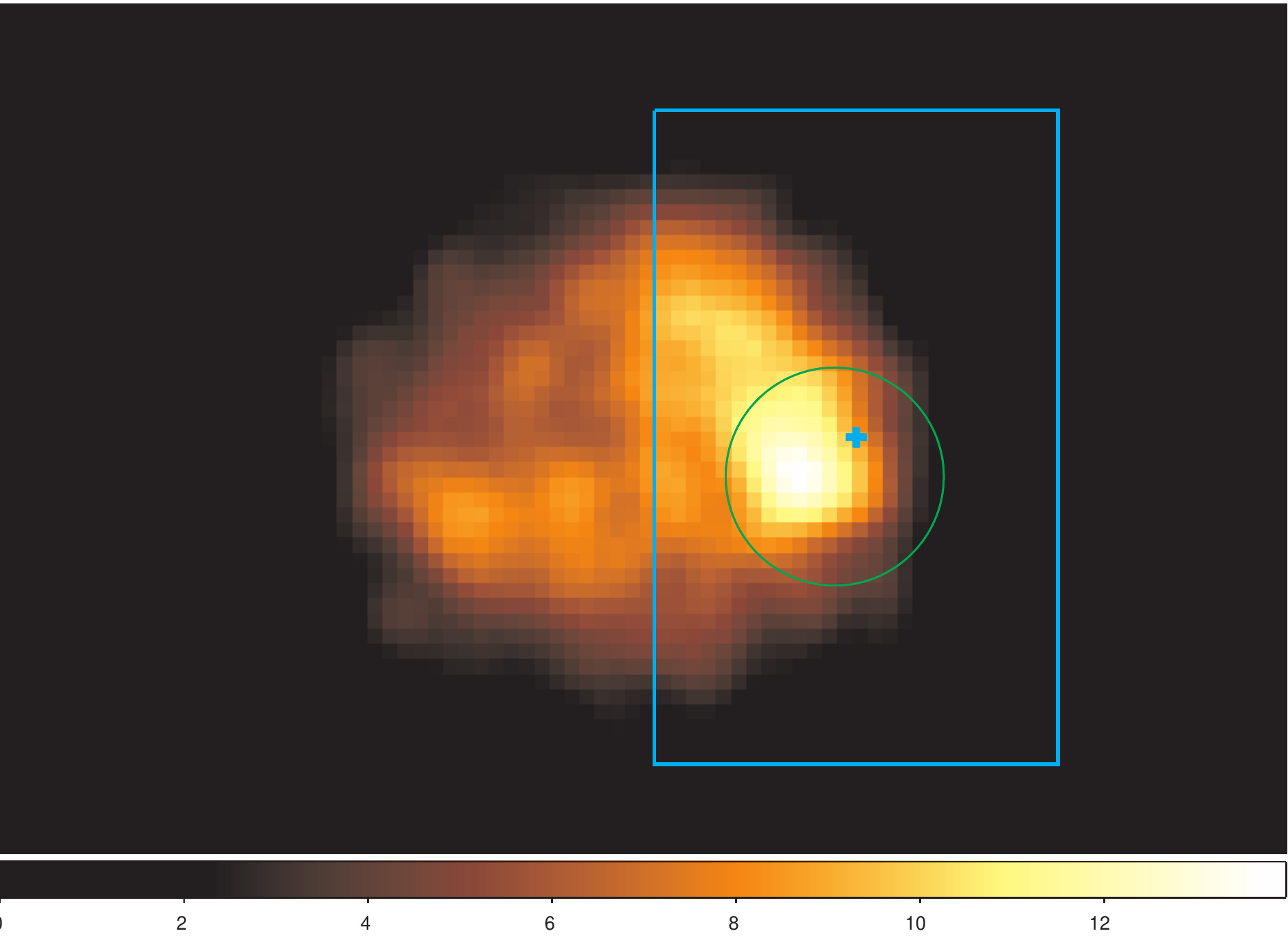}
\includegraphics[width=0.4\textwidth,clip,angle=0]{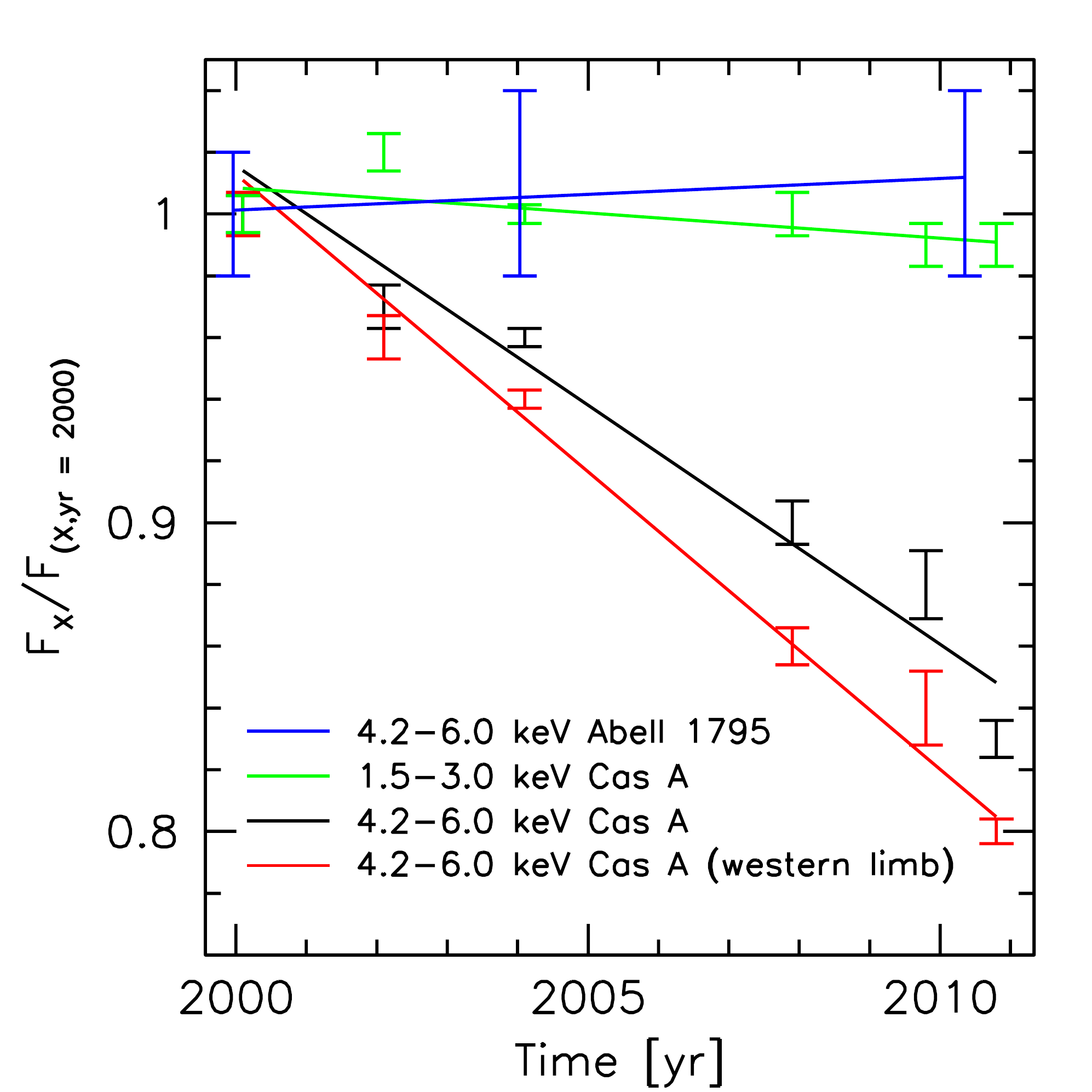}\\
\normalsize
\caption{ Left: {\it Suzaku} XIS 8--11 keV band
  image (\cite{Maeda09}). The green circle is the region where the
  continuum emission dominates.  Right: Comparison of 4.2--6.0 keV
  flux in \casa\ compared to the year 2000 observations by
  \citet{Patnaude11} using \chandra. The black curve and data
  correspond to changes in the whole SNR, while the red curve and data
  correspond to changes in the western portion of \casa\ only (Green
  circle in left figure).  For reference, we show the 1.5--3.0 keV flux from
  \casa\ (fluxed at 1.85 keV) as well as the 4.2--6.0 keV emission
  from the cluster Abell 1795. The observed decline in the 4.2--6.0
  keV emission in \casa\ corresponds to a fractional decline of
  $-$(1.5$\pm$0.17)\% yr$^{-1}$ across the whole SNR, and
  $-$(1.9$\pm$0.10)\% yr$^{-1}$ in the western limb
  \citep{Patnaude11}.  }
\label{fig:long}
\end{center}
\end{figure}

\subsubsection{RX J1713.1--3946: the search for thermal line emission} \label{subsec:rxjnontherm}

The primary goal of an \ah\ observation of this powerful cosmic ray accelerator is to detect the expected thermal line
emission from the hot post-shock plasma, which has not been detected
previously.  In the absence of a thermal component we are left without
many useful diagnostic tools to characterize the shock properties
(i.e., ambient medium density, temperatures, ionization timescales).
The high resolution spectral capabilities of the SXS will
allow the line emission to stand out above the strong non-thermal
continuum.  Two ``big picture'' science questions that we aim to
address are: (1) what is the ambient density -- this gets at the
interpretation of the TeV gamma-ray emission (leptonic vs.\ hadronic)
and (2) what is the thermal pressure behind the shock -- which gets at
the fraction of shock energy going into thermal and relativistic
particles.

\begin{figure}[h]
\begin{center}
    \vspace{-0.25in}
\includegraphics[height=2.15in,clip,trim=60 265 60 360]{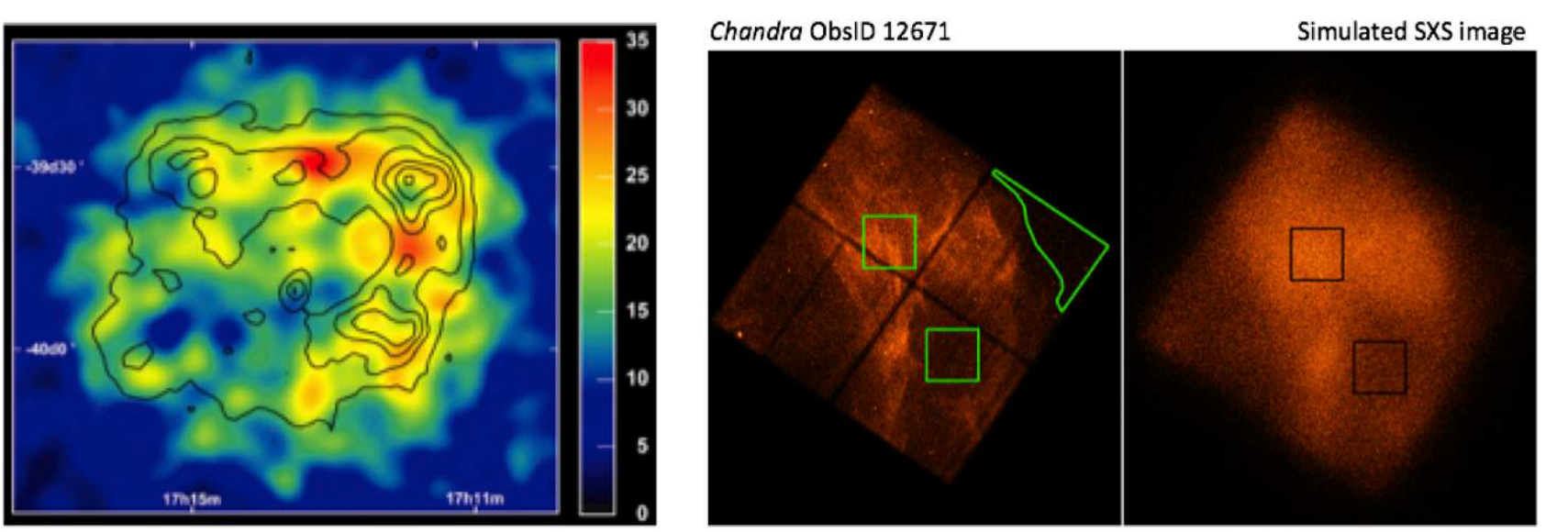}
\normalsize
\caption{{\it (Left)} Images of RX J1713.1$-$3946 from H.E.S.S.  and
  {\it Suzaku} XIS (contours) (Aharonian et al. 2004).  {\it (Middle)} \chandra\ observation
  of the NW rim showing the two regions used to simulate SXS spectra
  (boxes) and the region used for background.  {\it (Right)}
  \chandra\ smoothed to approximate the \ah\ point-spread-function.}
\label{fig:rxjimages}
\end{center}
\end{figure}

Given the large size of this remnant ($\sim$1$^\circ$ in diameter),
the obvious question is where to point the SXS.  The NW rim is an
attractive option because it is one of the X-ray brightest regions
(see \suzaku\ XIS contours on the H.E.S.S.\ image in the left panel of
Figure~\ref{fig:rxjimages}), it is where the twinkling X-ray filaments
are \citep{Uchiyama07}, there are radio filaments in the vicinity, and
it is close to a molecular cloud \citep{slane99}.  

This SNR has been extensively studied with \textit{XMM-Newton} \citep{cc04, hiraga05}
and {\it Suzaku} \citep{takahashi08, tanaka08, sano13}. We have verified, using these archival data,
 that the NW region is the best candidate for the thermal emission search.
 We note that the less absorbed interior would be a good site for the detection of
 thermal X-ray emission from shock-heated ejecta;
  however the SXS simulations show that the analysis will be complicated
 by the contamination from the Galactic ridge X-ray emission and the low-surface brightness of the SNR interior.

We used a \chandra\ observation of the NW rim (ObsId: 12671, 90 ks
nominal exposure) to estimate the level of possible thermal
contribution to the spectrum.  Spectra were extracted from two
regions, each the size of the full SXS array
(3.05$^\prime$$\times$3.05$^\prime$), from one location with bright
filamentary emission (NW1) and another with fainter diffuse emission
(NW2) as shown in Figure~\ref{fig:rxjimages} (middle panel). A modified
version of the \chandra\ data was made by spreading the detected
events around by the point-spread-function (PSF) of the \ah\ soft
X-ray telescope, assumed to be the same for all photon energies.  The
smoothed image is the right panel of Figure~\ref{fig:rxjimages}.  The
bright region yielded fewer extracted events (90\%) while the faint
region produced more events (130\%) when using the data set with the
simulated PSF of \ah. This level of contamination should not be a
concern.

Fits were initially done with a naive model consisting of an absorbed
power-law model ({\tt phabs+pow}). For the bright NW1 region the
$\chi^2$ of the fit is acceptable (421/463 d.o.f.), although a pattern
of low energy residuals was left (see top left panel of
Figure~\ref{fig:rxjspectra}). Note that similar patterns of low energy
residuals are apparent in high signal-to-noise \suzaku\ spectra
\citep{takahashi08}.  Inclusion of a thermal component (single
temperature, solar abundance {\tt mekal} model) resulted in a
significant reduction of $\chi^2$ (397/461 d.o.f.) for a temperature
$kT=0.21\pm0.3$ keV and norm $4.9\times 10^{-3}$ (see top right panel
of Figure~\ref{fig:rxjspectra}).  Next we introduced a more conservative
model to describe the continuum, the so-called {\tt srcut} model which
produces a gently curving spectrum throughout the X-ray band, which is
believed to more accurately reflect the underlying physics of the
emission.  This model by itself does a better job at fitting the
continuum than the pure power-law case (407/463 d.o.f.). Again the fit
is improved if we include a thermal component and we get essentially
the same $\chi^2$ and best-fit temperature as before.  However the
spectral norm of the {\tt mekal} model is reduced by about a factor of
two (norm = $1.8\times 10^{-3}$).

\begin{figure}[h]
\begin{center}
  \vspace{0.25in}
\includegraphics[width=5.5in,clip,trim=80 310 88 320]{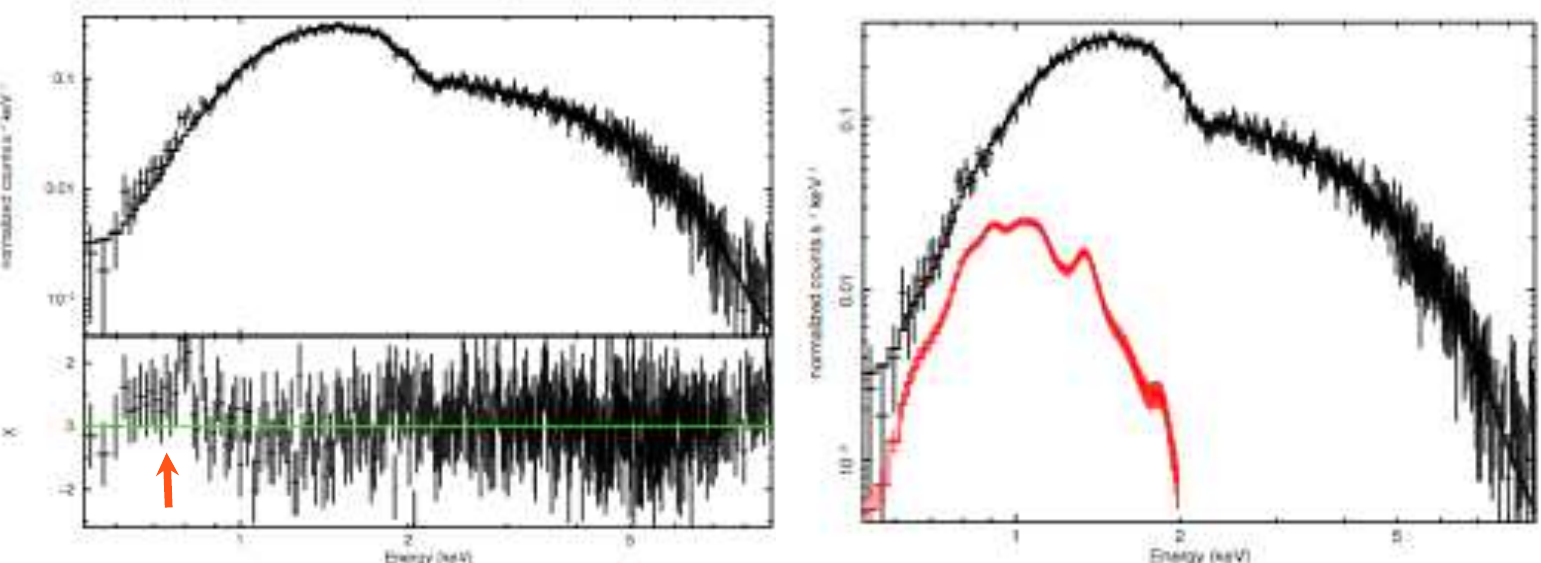}
\includegraphics[width=5.5in,clip,trim=56 290 72 320]{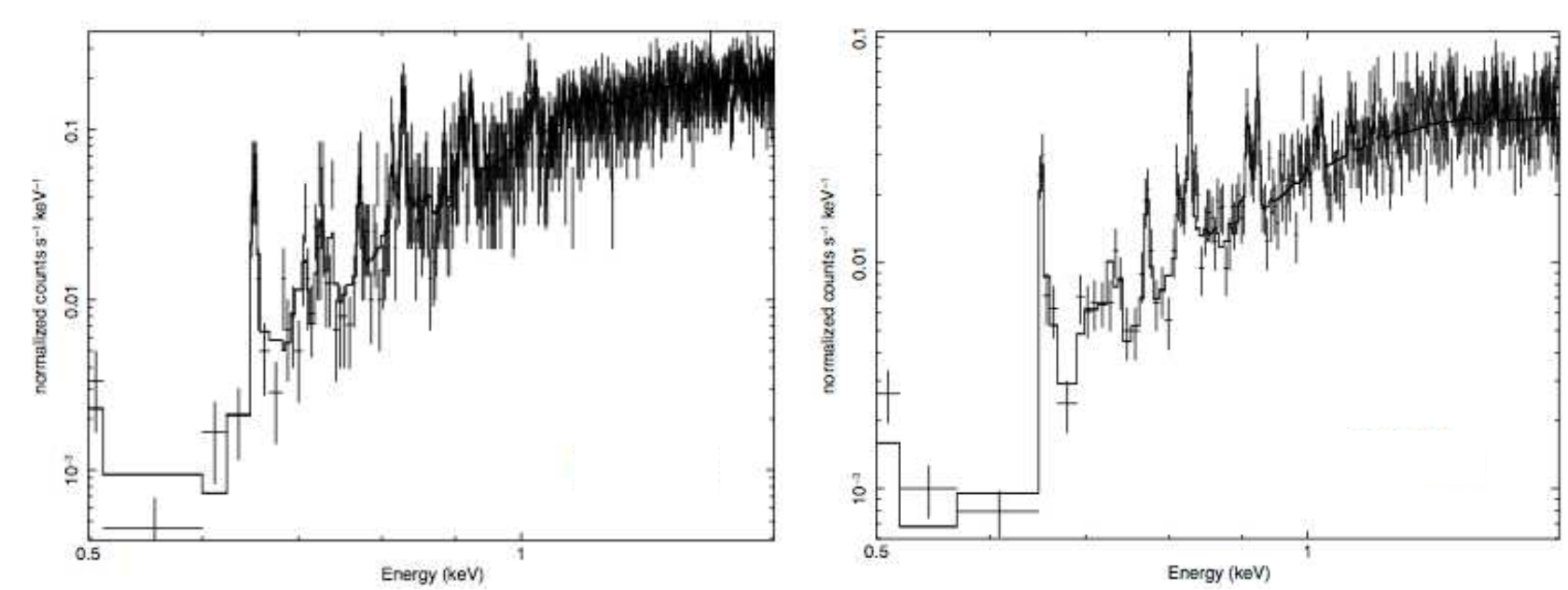}
\normalsize
\caption{{\it (Top left)} \chandra\ spectrum of region NW1 in RX
  J1713.1$-$3946 with best fit absorbed power-law fit.  {\it (Top
    right)} The same spectrum now shown with the best-fit model after
  the inclusion of an additive {\tt mekal}  thermal model.  The best fit
  thermal component is plotted in red.  {\it (Bottom left)} Simulated
  SXS spectrum of the NW1 region assuming a 100 ks exposure using the
  best fit absorbed power-law plus {\tt mekal}  fit from the \chandra\ data.
  {\it (Bottom right)} Simulated SXS spectrum of the fainter NW2
  region assuming a 300 ks exposure using the best fit absorbed
  power-law plus {\tt mekal}  fit from the \chandra\ data.}
\label{fig:rxjspectra}
\end{center}
\end{figure}

Simulated SXS spectra show prominent lines that stand out above the
continuum from O VIII (1s-np series), Ne IX (forbidden,
inter-combination, and resonance), Ne X, and the strong 15 \AA\ line
blend of Fe XVII (see bottom panels in Figure~\ref{fig:rxjspectra}).
According to our simulations such high signal-to-noise spectra are
going to require 100 ks to 300 ks of exposure, assuming that our
estimates of the thermal emission inferred from the \chandra\ data 
accurately describe the true thermal emission from the forward shock in 
RX J1713.1$-$3946.

Note that there is strong interest in this target from White Paper \#18, the Shock
Acceleration task force team.  We have not investigated here the
constraints on the broadband continuum emission from joint fits to the
SXS, SXI, and HXD, since this is addressed in WP\#18.

\subsubsection{Other \ah\ targets}
Other promising \ah\ targets for addressing this science topic include RCW~86, RX~J0852.0$-$4622, and SN~1006.
For  RCW~86, 1--2 \ah\ pointings at the northeast rim will allow a measurement of the degree of equipartition.
Measuring the line widths in that region will determine the effect of non thermal particles.
For RX J0852.0$-$4622 (Vela Junior), \ah\ pointings at the rim and interior will allow us  (like for RX J1713.1$-$3946)
to characterize the properties of the ambient medium (and possibly ejecta) 
by searching for line emission from this synchrotron-dominated SNR.
For SN~1006 which has both thermal and non-thermal X-ray emission, 
the bright non-thermal limbs, strongly correlated with TeV emission observed with HESS (Acero et al. 2010),
 will make a good \ah\ target to address the acceleration efficiency in this Ia SNR,
 in comparison to the synchrotron-dominated SNRs like RX J1713.1$-$3946.
In addition, mapping the synchrotron-dominated limbs in all these SNRs with HXI will help determine 
the shape of the electron energy distribution beyond the synchrotron cutoff
(of relevance to WP\#18).

\subsection{Beyond Feasibility}

{\bf Narrow Fe-K emission lines in Tycho: Evidence for cosmic ray
  acceleration at the reverse shock }

This ``Beyond Feasibility'' section refers back to the Tycho
simulation presented in Section 1.3.1.  The reverse shock speed in
Tycho can be estimated to be $\sim$4000 km s$^{-1}$ using the
expansion velocity of the Fe-K$\alpha$ line measured by
\suzaku\ \citep{hayato10}.  With the Rankine-Hugoniot shock jump
conditions we derive an ion temperature of $kT_{\rm Fe} = 3/16 M_{\rm
  Fe} v_{\rm RS}^2 \sim 2\, \rm MeV$.  This seems to be the most
likely situation for Tycho and, therefore, the simulations presented
above included this amount of line broadening. However, if cosmic ray
acceleration is efficient at the reverse shock in Tycho, then the ion
temperature could be much lower. Therefore, if we detect Fe-K emission
lines much narrower than $\sim$1 MeV, it would suggest that cosmic ray
acceleration is efficient at the reverse shock, which would be a
major breakthrough in this field.

\section{Summary}

Young SNRs represent the best astrophysical probes for studying a wide range of fundamental questions, from the formation of the chemical elements
to the acceleration of cosmic rays to very high energies, to the creation of some of the most exotic objects in the Universe.
We have highlighted in this White Paper some of the important topics for young SNR studies that we expect to carry out with \ah, in particular
with the SXS that will open a new discovery window to tackle fundamental questions in supernova remnant and cosmic ray astrophysics.
We have not stressed here \ah's broadband capability to be provided with combined SXI, HXI and SGD studies, filling an important energy gap between current X-ray
missions and gamma-ray missions; however this important aspect of the \ah\ mission is addressed in the Acceleration White Paper (\#18). 
As well, the Old SNRs+PWNe White Paper (\#8) highlights science topics for older SNRs and Pulsar Wind Nebulae (which would be also
relevant to the ISM and new physics), 
and the Highly Magnetized Neutron Stars White Paper (\#4) focuses on their (associated) compact objects' science.

\clearpage
\begin{multicols}{2}
\end{multicols}

\end{document}

%% file: ASTROHWhitePaperStyle.tex
\usepackage{ascmac}
\usepackage{amsmath}
\usepackage{amssymb}
\usepackage{bm}
\usepackage[footnotesize,bf]{caption}
\usepackage{fullpage}
\usepackage{color}
\usepackage{float}
\usepackage{graphicx}
\usepackage[]{natbib}
\usepackage{subfigure}
\usepackage{txfonts}
\usepackage{threeparttable}
\usepackage{wrapfig}
\usepackage[]{multicol}
\usepackage{wrapfig}
\usepackage{authblk}

\usepackage{floatflt}

\title{\large Complete list of the ASTRO-H Science Working Group}
\date{\vspace{-0.5cm}}
\newcommand{\MakeWhitePaperTitle}{
	\begin{center}
		\begin{figure}
			\vspace{1cm}
			\begin{center}
				\includegraphics[width=0.2\hsize]{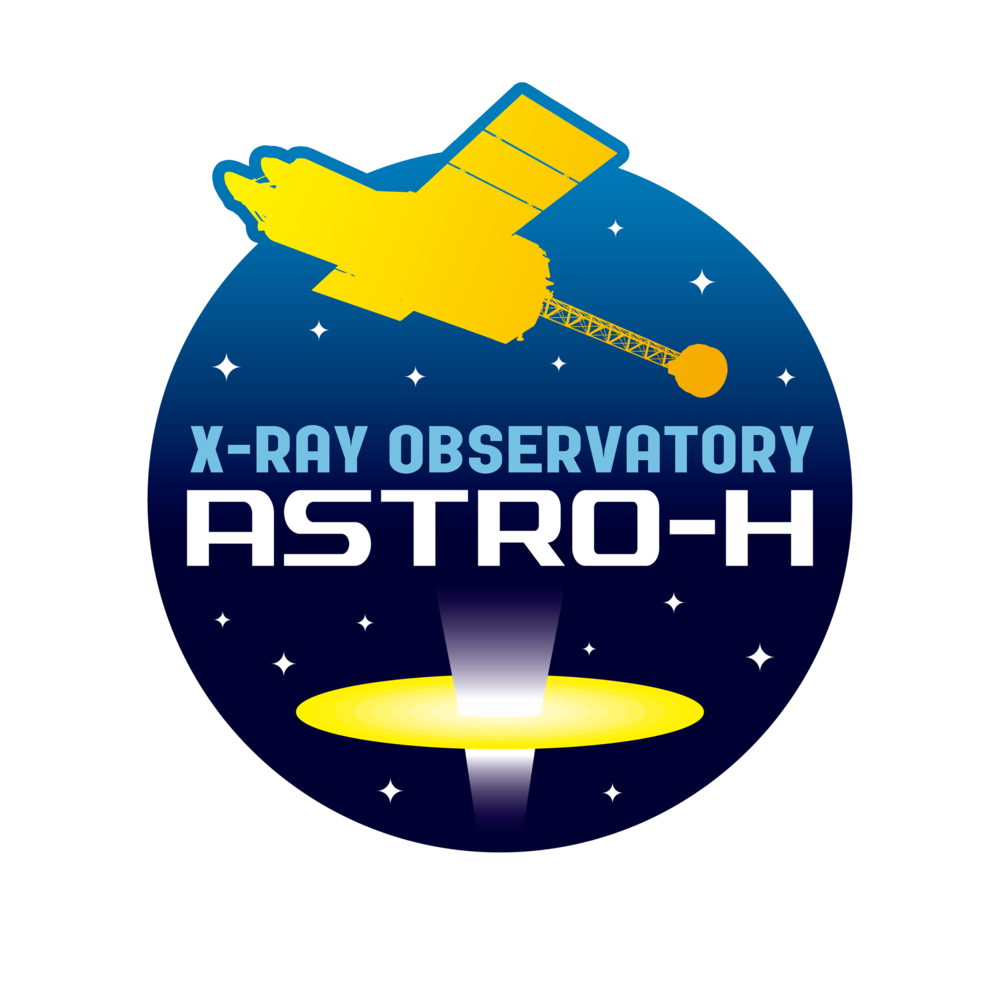}
			\end{center}
		\end{figure}
		\vspace{1cm}
		{\LARGE
		ASTRO-H Space X-ray Observatory\\
		White Paper\\
		}
		\vspace{5mm}
		{\large
		\WhitePaperTitle\\
		}
		\vspace{1cm}
		{
		\WhitePaperAuthors\\
		on behalf of the ASTRO-H Science Working Group
		}
	\end{center}
}

%% file: WP_AuthorList.tex
\usepackage{authblk}
\author[a]{Tadayuki~Takahashi}
\author[a]{Kazuhisa~Mitsuda}
\author[b]{Richard~Kelley}
\author[c]{Felix~Aharonian}
\author[d]{Hiroki~Akamatsu}
\author[e]{Fumie~Akimoto}
\author[f]{Steve~Allen}
\author[g]{Naohisa~Anabuki}
\author[b]{Lorella~Angelini}
\author[h]{Keith~Arnaud}
% ...
\affil[a]{Institute of Space and Astronautical Science (ISAS), Japan Aerospace Exploration Agency (JAXA), Kanagawa 252-5210, Japan}
\affil[b]{NASA/Goddard Space Flight Center, MD 20771, USA}
\affil[c]{Astronomy and Astrophysics Section, Dublin Institute for Advanced Studies, Dublin 2, Ireland}
\affil[d]{SRON Netherlands Institute for Space Research, Utrecht, The Netherlands}
\affil[e]{Department of Physics, Nagoya University, Aichi 338-8570, Japan}
\affil[f]{Kavli Institute for Particle Astrophysics and Cosmology, Stanford University, CA 94305, USA}
\affil[g]{Department of Earth and Space Science, Osaka University, Osaka 560-0043, Japan}
\affil[h]{Department of Astronomy, University of Maryland, MD 20742, USA}
% ...


%% file: WP_07_YoungSNRs.bbl
{\footnotesize

\begin{thebibliography}{99}

\bibitem[Acero et al.(2010)]{acero10}
Acero, F. et al. 2010, A\&A, 516, 62


\bibitem[Aharonian et al.(2001)]{Aharonian01} Aharonian, F., et
  al.\ 2001, \aap, 370, 112

\bibitem[Aharonian et al.(2004)]{Aharonian04} Aharonian, F.~A., et
  al.\ 2004, \nat, 432, 75

\bibitem[Aharonian et al.(2007)]{Aharonian07} Aharonian, F., et
  al.\ 2007, \apj, 661, 236

\bibitem[Albert et al.(2007)]{Albert07} Albert, J., et al.\ 2007,
  \aap, 474, 937

\bibitem[Allen et al.(1997)]{Allen97} Allen, G.~E., et al.\ 1997,
  \apj, 487, L97
  
\bibitem[An et al.(2014)]{An14} An, H. et al. 2014, \apj, 793, 90

\bibitem[Anders \& Grevesse(1989)]{ag89} Anders, E., \& Grevesse,
  N.~1989, Geochimica et Cosmochimica Acta 53, 197

\bibitem[Aubourg et al.(2008)]{aubourg08} Aubourg, {\'E}., Tojeiro,
  R., Jimenez, R., et al.\ 2008, \aap, 492, 631

\bibitem[Badenes et al.(2003)]{badenes03} Badenes, C., Bravo, E., 
Borkowski, K.~J., \& Dom{\'{\i}}nguez, I.\ 2003, \apj, 593, 358 

\bibitem[Badenes et al.(2008a)]{badenes08a} Badenes, C., Hughes, 
J.~P., Cassam-Chena{\"i}, G., \& Bravo, E.\ 2008a, \apj, 680, 1149 

\bibitem[Badenes et al.(2006)]{badenes06} Badenes, C., Borkowski,
  K.~J., Hughes, J.~P., Hwang, U., \& Bravo, E.\ 2006, \apj, 645, 1373

\bibitem[Badenes et al.(2008b)]{badenes08b} Badenes, C., Bravo, E., 
\& Hughes, J.~P.\ 2008b, \apj, 680, L33 

\bibitem[Bamba et al.(2003)]{Bamba03} Bamba, A., Yamazaki, R., Ueno,
  M., \& Koyama, K.\ 2003, \apj, 589, 827

\bibitem[Bamba et al.(2005)]{Bamba05a} Bamba, A., Yamazaki, R.,
  Yoshida, T., Terasawa, T., \& Koyama, K.\ 2005, \apj, 621, 793

\bibitem[Bamba et al.(2005)]{bamba05b}
Bamba, A., Yamazaki, R., \& Hiraga, J. S. 2005, ApJ, 632, 294

\bibitem[Benetti et al.(2005)]{benetti05} Benetti, S.,
Cappellaro, E., Mazzali, P.~A., et al.\ 2005, ApJ, 623, 1011

\bibitem[Berezhko et al.(2009)]{berezhko09}
Berezhko, E. G., P\"uhlhofer, G., \& V\"olk, H. J. 2009, A\&A, 505, 641

\bibitem[Blondin \& Mezzacappa(2006)]{blondin06}Blondin, J.~M., \&
  Mezzacappa, A.\ 2006, ApJ, 642, 401

\bibitem[Blondin et al.(2001)]{Blondin01} Blondin, J. M., Borkowski,
  K. J., \& Reynolds, S. P.\ 2001, ApJ, 557, 782

\bibitem[Borkowski et al.(2006)]{borkowski06} Borkowski, K.~J., 
Hendrick, S.~P., \& Reynolds, S.~P.\ 2006, \apj, 652, 1259 

\bibitem[Borkowski et al.(2013)]{borkowski13} Borkowski, K.~J., 
Reynolds, S. P., Hwang, U., Green, D. A., Petre, R. et al.
2013, \apj, 771, L9


\bibitem[Borkowski et al.(2010)]{borkowski+10} Borkowski, K. J.,
  Reynolds, S. P., Green, D. A., et al.\ 2010, \apj, 724, L161

\bibitem[Burbidge et al.(1957)]{burbidge57} Burbidge, E.~M., Burbidge,
  G.~R., Fowler, W.~A., \& Hoyle, F.\ 1957, Reviews of Modern Physics,
  29, 547

\bibitem[Burrows, Hayes, \& Fryxell(1995)]{burrows95}Burrows, A.,
  Hayes, J., \& Fryxell, B.~A.\ 1995, \apj, 450, 830

\bibitem[Burrows et al.(2007)]{burrows07}Burrows, A., Dessart, L.,
  Ott, C.~D., \& Livne, E.\ 2007, Phys.~Rep, 442, 23

\bibitem[Camilo et al.(2002)]{camilo02} Camilo, F., Manchester, R.~N.,
  Gaensler, B.~M., Lorimer, D.~R., \& Sarkissian, J.\ 2002, \apj,
  567, L71

\bibitem[Canizares \& Winkler(1981)]{canizares81}
Canizares C. R., \& Winkler, P. F.\ 1981 ApJ, 246, L33 

\bibitem[Carlton et al.(2011)]{carlton+11} Carlton, A.~K., 
Borkowski, K.~J., Reynolds, S.~P., et al.\ 2011,  \apj, 737, L22 

\bibitem[Cassam-Chena{\"i} et al.(2004)]{cc04}
 Cassam-Chena{\"i}, G., Decourchelle, A., Ballet, J., Sauvageot, J.-L., Dubner, G., \& Giacani, E.
 2004, A\&A, 427, 199

\bibitem[Cassam-Chena{\"i} et al.(2007)]{cassamchenai07}
  Cassam-Chena{\"i}, G., Hughes, J.~P., Ballet, J., \& Decourchelle,
  A.\ 2007, \apj, 665, 315

\bibitem[Cassam-Chena{\"i} et al.(2008)]{gamil08} 
Cassam-Chena{\"i}, G., Hughes, J.~P., Reynoso, E.~M., Badenes, C., 
\& Moffett, D.\ 2008, \apj, 680, 1180 

\bibitem[Chen et al.(1999)]{chen99} Chen, Y., Sun, M., Wang, Z-R, \&
  Yin, Q.~F.\ 1999, ApJ, 520, 737
  
\bibitem[Chevalier(2005)]{chevalier05} 
Chevalier, R. 2005, \apj, 619, 839


\bibitem[Chu \& Kennicutt(1988)]{chu88} Chu, Y.-H., \&
  Kennicutt, R.~C., Jr.\ 1988, AJ, 96, 1874

\bibitem[Decourchelle et al.(2001)]{decourchelle01}
Decourchelle, A, Sauvageot, J. L., Audard, M., Aschenbach, B., Semblay, S.,
Rothenflug, R., Ballet, J., Stadlbauer, T., \& West, R., 2001, A\&A, 365, L218

\bibitem[Decourchelle et al.(2000)]{decourchelle00}
Decourchelle, A, Ellison, D.~C., \& Ballet, J.\ 2000, ApJ, 543, L57

\bibitem[Dewey et al.(2012)]{dewey12} Dewey, D., Dwarkadas, 
V.~V., Haberl, F., Sturm, R., \& Canizares, C.~R.\ 2012, \apj, 752, 103 

\bibitem[Dwarkadas(2005)]{dwarkadas05} 
Dwarkadas, V. 2005, \apj, 630, 892

\bibitem[Ferrand et al.(2012)]{ferrand12}
Ferrand, G., Decourchelle, A. \& Safi-Harb, S. 2012, ApJ, 760, 34


\bibitem[Ferrand et al.(2014)]{ferrand14}
Ferrand, G., Decourchelle, A. \& Safi-Harb, S. 2014, ApJ, 789, 49

\bibitem[Foglizzo, Scheck, \& Janka(2006)]{foglizzo06}
Foglizzo, T., Scheck, L., \& Janka, H.-Th.~2006, ApJ, 652, 1436


\bibitem[Gabriel \& Phillips(1979)]{1979MNRAS.189..319G} Gabriel,
  A.~H., \& Phillips, K.~J.~H.\ 1979, \mnras, 189, 319

\bibitem[Gaensler \& Wallace(2003)]{gaensler03} Gaensler, B.~M., \&
  Wallace, B.~J.\ 2003, \apj, 594, 326

\bibitem[Gaensler et al.(2005)]{gaensler05} Gaensler, B.~M.,
McClure-Griffiths, N. M., Oey, M. S.,  Haverkorn, M., Dickey, J. M. \& Green, A. J. 2005,
\apj, 620, L95

\bibitem[Gavriil et al.(2008)]{Gavriil2008}
Gavriil, F. et al. 2008, Science, 319, 1802

\bibitem[Ghavamian et al.(2007)]{ghavamian+07}
Ghavamian, P., Blair, 
W.~P., Sankrit, R., Raymond, J.~C., \& Hughes, J.~P.\ 2007, \apj, 664, 304 

\bibitem[Gonzalez \& Safi-Harb(2003)]{gonzalez03} Gonzalez, M., \&
  Safi-Harb, S.~2003, \apj, 583, L61

\bibitem[Grefenstette et al.(2014)]{casa_nustar} Grefenstette, B.~W.,
  Harrison, F.~A., Boggs, S.~E., et al.\ 2014, \nat, 506, 339

\bibitem[Hachisu et al.(1996)]{1996ApJ...470L..97H} Hachisu, I., Kato, M.,
\& Nomoto, K.\ 1996, ApJ, 470, L97 

\bibitem[Hamilton et al.(1986)]{hamilton86}
Hamilton, A. J. S., Sarazin, C. L., \& Szymkowiak, A. E.\ 1986, ApJ, 300, 713


\bibitem[Hayato et al.(2010)]{hayato10} Hayato, A., Yamaguchi, 
H., Tamagawa, T., et al.\ 2010, \apj, 725, 894 


\bibitem[Helder \& Vink(2008)]{Helder08}
 Helder, E.~A., \& Vink, J.\ 2008, \apj, 686, 1094 

\bibitem[Helder et al.(2009)]{helder09}
 Helder, E., et al., 2009, Science, 325, 719 

\bibitem[Helder et al.(2013)]{helder13} Helder, E.~A., Broos, 
P.~S., Dewey, D., et al.\ 2013, \apj, 764, 11 

\bibitem[Heinke \& Ho(2010)]{heinke10} Heinke, C.~O., \& Ho,
  W.~C.~G.\ 2010, \apj, 719, L167

\bibitem[Herant et al.(1994)]{herant94}Herant, M., Benz, W., Hix,
  W.~R., Fryer, C.~L., \& Colgate, S.~A.~1994, \apj, 435, 339
	
\bibitem[Hiraga et al.(2005)]{hiraga05}
Hiraga, J. S., Uchiyama, Y., Takahashi, T., \& Aharonian, F. A. 2005, A\&A, 505, 157

\bibitem[Holt et al.(1994)]{holt94}
Holt, S. S., Gotthelf, E. V., Tsunemi, H., Negoro, H. 1994,
\pasj, 46, L151

\bibitem[Howell et al.(2001)]{howell01} Howell, D.~A., 
H{\"o}flich, P., Wang, L., \& Wheeler, J.~C.\ 2001, \apj, 556, 302 

\bibitem[Howell et al.(2009)]{howell09} Howell, D.~A., Sullivan,
M., Brown, E.~F., et al.\ 2009, ApJ, 691, 661 

\bibitem[Hughes et al.(2000a)]{hughes00a} Hughes, J.~P., Rakowski, 
C.~E., \& Decourchelle, A.\ 2000a, \apj, 543, L61 

\bibitem[Hughes et al.(2000b)]{hughes00b} Hughes, J.~P., Rakowski, 
C.~E., Burrows, D.~N., \& Slane, P.~O.\ 2000b, \apj, 528, L109 

\bibitem[Hughes et al.(2001)]{hughes01} Hughes, J.~P., Slane, P. O., Burrows, D. et al.
2001, \apj, 559, L153


\bibitem[Hughes et al.(2003)]{hughes03} Hughes, J.~P., 
Ghavamian, P., Rakowski, C.~E., \& Slane, P.~O.\ 2003, \apj, 582, L95


\bibitem[Hughes et al.(1995)]{hughes95} Hughes, J.~P., Hayashi, 
I., Helfand, D., et al.\ 1995, \apj, 444, L81 

\bibitem[Hughes et al.(2001)]{hughes01} Hughes, J.~P., Slane, 
P.~O., Burrows, D.~N., et al.\ 2001, \apj, 559, L153 

\bibitem[Hwang \& Gotthelf(1997)]{hwang97} Hwang, U., \& Gotthelf,
  E.~V.\ 1997, \apj, 475, 665

\bibitem[Hwang \& Laming(2003)]{hwang03} Hwang, U., \& Laming,
  J.~M.\ 2003, \apj, 597, 362

\bibitem[Hwang et al.(2002)]{hwang02}
Hwang, U., Decourchelle, A., Holt, S. S., \& Petre, R., 2002, ApJ, 581, 1101

\bibitem[Hwang et al.(2004)]{hwang04}
Hwang, U., et al., 2004, ApJ, 615, L117 

\bibitem[Hwang et al.(2000a)]{hwang00a}
Hwang, U., Holt, S.S. \& Petre, R., 2000a, ApJ, 537, L119

\bibitem[Hwang et al.(2000b)]{hwang00b} Hwang, U., Petre, R., 
\& Hughes, J.~P.\ 2000b, \apj, 532, 970 


\bibitem[Iwamoto et al.(1999)]{iwamoto99} Iwamoto, K., Brachwitz, 
F., Nomoto, K., et al.\ 1999, \apjs, 125, 439 

\bibitem[Iyudin et al.(1994)]{Iyudin94} Iyudin, A.~F., Diehl, R.,
  Bloemen, H., et al.\ 1994, \aap, 284, L1
        
\bibitem[Jiang et al.(2010)]{Jiang10} Jiang, B., Chen, Y., Wang, J.,
  Su, Y., Zhou, Xin, Safi-Harb, S., \& DeLaney, T.\ 2010, ApJ, 712,
  1147

\bibitem[Jordan et al.(2008)]{jordan08}
Jordan, G.~C., IV, Fisher, R.~T., Townsley, D.~M., Calder, A.~C.,
Graziani, C., Asida, S., Lamb, D.~Q., \& Truran, J.~W.\ 2008, \apj,
681, 1448


\bibitem[Kaastra et
al.(2009)]{2009A&A...503..373K} Kaastra, J.~S., Bykov, A.~M., \&
Werner, N.\ 2009, \aap, 503, 373

\bibitem[Kamitsukasa et al.(2014)]{kamitsukasa14}
Kamitsukasa, F., Koyama, K., Tsunemi, H. et al. 2014,
\pasj, 66, 64


\bibitem[Kalemci et al.(2006)]{2006ApJ...644..274K} Kalemci, E., Reynolds,
S.~P., Boggs, S.~E., et al.\ 2006, \apj, 644, 274

\bibitem[Kasen et al.(2009)]{kasen09} Kasen, D., R{\"o}pke, F.~K., \&
  Woosley, S.~E.\ 2009, Nature, 460, 869

\bibitem[Katsuda et al.(2013)]{katsuda13}
Katsuda, S., et al.\ 2013, ApJ, 763, 85 

\bibitem[Kifonidis et al.(2000)]{kifonidis00} Kifonidis, K., Plewa, T.,
  Janka, H.-T., M{\"u}ller, E. 2000, \apj, 531, L123

\bibitem[Khokhlov et al.(1999)]{khokhlov99} Khokhlov, A.~M.,
  H{\"o}flich, P.~A., Oran, E.~S., Wheeler, J.~C., Wang, L., \&
  Chtchelkanova, A.~Y. 1999, \apj, 524, L107

\bibitem[Koyama et al.(1995)]{koyama95}
Koyama, K., et al.\ 1995, Nature, 378, 255

\bibitem[Kuhlen et al.(2006)]{kuhlen06} Kuhlen, M., Woosley, S.~E., \&
  Glatzmaier, G.~A.\ 2006, ApJ, 640, 407

\bibitem[Kumar \& Safi-Harb(2008)]{KumarSSH2008}
Kumar, H. S. \& Safi-Harb, S. 2008, ApJ, 678, L43

\bibitem[Kumar et al.(2012)]{kumar2012}
Kumar, H. S., Safi-Harb, Gonzalez, M. E. 2012, ApJ, 754, 96

\bibitem[Kumar et al.(2014)]{kumar2014}
Kumar, H. S., Safi-Harb, S., Slane, P. O., Gotthelf, E. V. 2014, ApJ, 781, 41

\bibitem[Krause et al.(2008)]{krause08a} Krause, O., Birkmann, S.~M,
  Usuda, T., et al.\ 2008, Science, 320, 1195

\bibitem[Krause et al.(2008)]{krause08b} Krause, O., Tanaka, M., 
Usuda, T., et al.\ 2008, \nat, 456, 617 

\bibitem[Lopez et al.(2011)]{lopez11}
Lopez, L. A., Ramirez-Ruiz, E., Huppenkothen, D., Badenes, C., \& Pooley, D. A. 2011,
\apj, 732, 114

\bibitem[Laming(2014)]{laming14}
Laming, J. M. 2014, \nat, 506, 298

\bibitem[Laming \& Hwang(2003)]{laming03}
Laming, J. M., \& Hwang, U.\ 2003, ApJ, 597, 347

\bibitem[Lee et al.(2011)]{lee11} Lee, J.-J., Park, S., Hughes, J.~P.,
  Slane, P.~O., \& Burrows, D.~N.\ 2011, \apj, 731, L8

\bibitem[Lee et al.(2013)]{lee13} 
Lee, S.-H., Slane, P. O., Ellison, D. C., Nagataki, S. \& Patnaude, D. J. 2013, \apj, 767, 20

\bibitem[Lewis et al.(2003)]{lewis03} Lewis, K.~T., Burrows, 
D.~N., Hughes, J.~P., et al.\ 2003, \apj, 582, 770 

\bibitem[Long et al.(2003)]{long03}
Long, K. S., Reynolds, S. P., Raymond, J. C., Winkler, P. F., Dyer, K. K.,
\& Petre, R.\ 2003, ApJ, 586, 1162 

\bibitem[Maeda et al.(2009)]{Maeda09} Maeda, Y., Uchiyama, Y., 
Bamba, A., et al.\ 2009, \pasj, 61, 1217 

\bibitem[Maeda et al.(2010a)]{maeda10a} Maeda, K., Benetti, S.,
Stritzinger, M., et al.\ 2010a, Nature, 466, 82 

\bibitem[Maeda et al.(2010b)]{maeda10b} Maeda, K., R{\"o}pke, 
F.~K., Fink, M., et al.\ 2010b, \apj, 712, 624 

\bibitem[Maeda et al.(2012)]{maeda12} Maeda, K., Terada, Y., Kasen,
  D., et al.\ 2012, ApJ, 760, 54

\bibitem[Maggi et al.(2012)]{maggi12} Maggi, P., Haberl, F., Sturm,
  R. \& Dewey, D.\ 2012, A\&A, 548, 3

\bibitem[Mazzali et al.(2007)]{mazzali07} Mazzali, P.~A., R{\"o}pke,
  F.~K., Benetti, S., \& Hillebrandt, W.\ 2007, Science, 315, 825

\bibitem[Mereghetti(2013)]{mereghetti13}
Mereghetti, S. 2013, Brazilian Journal of Physics, 43, 356 (arXiv:1304.4825)

\bibitem[Miceli(2012)]{miceli12}
Miceli, M., Acero, F., Dubner, G., Decourchelle, A., Orlando, S., \& Bocchino, F. 2012, \apj, 782, L33

\bibitem[Morton et al.(2007)]{Morton2007}
Morton, T. D. et al. 2007, ApJ, 667, 219

\bibitem[Morlino et al.(2013)]{morlino13}
Morlino, G., Blasi, P., Bandiera, R., \& Amato, E. 2013, A\&A, 558, 25


\bibitem[Murdin \& Clark(1979)]{murdin79} Murdin, P., \& Clark,
  D.~H.\ 1979, \mnras, 189, 501

\bibitem[Nagataki et al.(1998)]{nagataki98} Nagataki, S., Shimizu, 
T.~M., \& Sato, K.\ 1998, \apj, 495, 413 

\bibitem[Nakamura et al.(1999)]{Nakamura99} Nakamura et al. 1999 ApJ,
  517, 193

\bibitem[Ng et al.(2008)]{Ng2008}
Ng, C. Y. et al. 2008, ApJ, 686, 508


\bibitem[Nomoto et al.(1984)]{nomoto84} Nomoto, K., Thielemann,
F.-K., \& Wheeler, J.~C.\ 1984, ApJ, 279, L23

\bibitem[Nomoto et al.(1997)]{nomoto97} Nomoto, K., Iwamoto, K., 
Nakasato, N., et al.\ 1997, Nuclear Physics A, 621, 467 

\bibitem[Nomoto et al.(2006)]{Nomoto06} Nomoto, K., 
Tominaga, N., Umeda, H., Kobayashi, C., \& Maeda,
K. 2006, Nucl. Phys. A, 777, 424

\bibitem[Ouyed et al.(2014)]{ouyed14}
Ouyed, R., Leahy, D., \& Koning, N. 2014, arXiv:1404.5063

\bibitem[Parizot et al.(2006)]{Parizot06} Parizot, E., Marcowith, A.,
  Ballet, J., \& Gallant, Y.~A.\ 2006, \aap, 453, 387

\bibitem[Park et al.(2007)]{park07} Park, S., Hughes, J. P., Slane, P. O. et al. 2007,
\apj, 670, L121

\bibitem[Park et al.(2013)]{parkkepler13} Park, S., Badenes, C., 
Mori, K.,  et al.\ 2013, \apj, 767, L10 

\bibitem[Park et al.(2003)]{park03} Park, S., Hughes, J.~P., Burrows,
  D.~N., et al.\ 2003, \apj, 598, L95

\bibitem[Park et al.(2007)]{park07} Park, S., Hughes, J.P., Slane,
  P.O., Burrows, D.N., Gaensler, B.M., Ghavamian, P.\ 2007, ApJ, 670,
  L121

\bibitem[Park et al.(2004)]{park04} Park, S., Hughes, J.~P., Slane,
  P.~O., et al.\ 2004, \apj, 602, L33

\bibitem[Park et al.(2002)]{park02} Park, S., Roming, P. W. A.,
  Hughes, J. P., Slane, P. O., Burrows, D. N., Garmire, G. P., \&
  Nousek, J. A.\ 2002, ApJ, 564, 39

\bibitem[Patnaude et al.(2009)]{patnaude09} Patnaude, D.~J., Ellison,
  D.~C., \& Slane, P.\ 2009, ApJ, 696, 1956

\bibitem[Patnaude et al.(2011)]{Patnaude11} Patnaude, D.~J., et al.\ 2011
  ApJL, 729, L28

\bibitem[Perlmutter et al.(1999)]{perlmutter99} Perlmutter, S.,
Aldering, G., Goldhaber, G., et al.\ 1999, ApJ, 517, 565

\bibitem[Phillips et al.(1999)]{phillips99} Phillips, M.~M., Lira,
P., Suntzeff, N.~B., et al.\ 1999, AJ, 118, 1766 

\bibitem[Rasmussen et al.(2002)]{rasmussen02} Rasmussen, A., Behar,
E., \& Vink, J.\ 2002, 34th COSPAR Scientific Assembly, 34, 

\bibitem[Rest et al.(2008a)]{rest08a} Rest, A., Matheson, T., 
Blondin, S., et al.\ 2008a, \apj, 680, 1137 

\bibitem[Rest et al.(2008b)]{rest08b} Rest, A., Welch, D.~L., 
Suntzeff, N.~B., et al.\ 2008b, \apj, 681, L81 

\bibitem[Reynolds et al.(2008)]{reynolds+08} Reynolds, S.~P.,
  Borkowski, K.~J., Green, D.~A., et al.\ 2008, \apj, 680, L41

\bibitem[Reynolds et al.(2009)]{reynolds+09} Reynolds, S.~P., 
Borkowski, K.~J., Green, D.~A., et al.\ 2009,  \apj, 695, L149 


\bibitem[Reynolds et al.(2007)]{reynolds07} Reynolds, S.~P.,
  Borkowski, K.~J., Hwang, U., et al.\ 2007, \apj, 668, L135

\bibitem[Riess et al.(1998)]{riess98} Riess, A.~G., Filippenko,
  A.~V., Challis, P., et al.\ 1998, \aj, 116, 1009

\bibitem[R{\"o}pke, \& Bruckschen(2008)]{ropke08} R{\"o}pke, F.~K., \&
  Bruckschen, R.\ 2008, New Journal of Physics, 10, 125009

\bibitem[R{\"o}pke et al.(2007)]{ropke07} R{\"o}pke F.~K.,
  Woosley S.~E., Hillebrandt W., 2007, ApJ, 660, 1344

\bibitem[Rosado et al.(1996)]{rosado96} Rosado, M.,
  Ambrocio-Cruz, P., Le Coarer, E., \& Marcelin, M.\ 1996, A\&A, 315,
  243

\bibitem[Russell \& Dopita(1992)]{1992ApJ...384..508R} Russell, S.~C.,
\& Dopita, M.~A.\ 1992, ApJ, 384, 508

\bibitem[Rutherford et al.(2013)]{casa_hetg} Rutherford, J., Dewey,
  D., Figueroa-Feliciano, E., Heine, S.~N.~T., Bastien, F.~A., Sato,
  K., \& Canizares, C.~R.\ 2013, \apj, 769, 64

\bibitem[Safi-Harb \& Kumar(2013)]{SSHKumar2013} Safi-Harb, S. \&
  Kumar, H. 2013, Proceedings of the International Astronomical Union, 
  291, 480, ed. J. van Leeuwen  (arXiv:1210.5261)

        
\bibitem[Safi-Harb et al.(2000)]{ssh00} Safi-Harb, S., Petre, R.,
  Arnaud, K. A., Keohane, J. W., Borkowski, K. J., Dyer, K. K.,
  Reynolds, S. P., \& Hughes, J. P. 2000, ApJ, 545, 922
  
\bibitem[Safi-Harb et al.(2005)]{ssh05} Safi-Harb, S., Dubner, G.,
  Petre, R., Holt, S. S., \& Durouchoux, P. 2005, ApJ, 618, 321


\bibitem[Sano et al.(2013)]{sano13}
Sano, H., Tanaka, T., Torii, K. et al. 2013, \apj, 778, 59


\bibitem[Scannapieco \& Bildsten(2005)]{scannapieco05} Scannapieco,
  E., \& Bildsten, L.\ 2005, \apj, 629, L85

\bibitem[Scheck et al.(2008)]{scheck08} Scheck, L., Janka, H.-Th,
  Foglizzo, T., \& Kifonidis, K.~2008, A\&A, 477, 931


\bibitem[Seely et al.(1987)]{1987ApJ...319..541S} Seely, J.~F., Feldman,
U., \& Doschek, G.~A.\ 1987, \apj, 319, 541

\bibitem[Seward et al.(2006)]{seward06}
Seward, F. D., Williams, R. M., Chu, Y.-H., Dickel, J. R., Smith, R. C., \& Points, S. D.
2006, \apj, 640, 327

\bibitem[Shternin et al.(2011)]{shternin11} Shternin, P.~S., Yakovlev,
  D.~G., Heinke, C.~O., Ho, W.~C.~G., \& Patnaude, D.~J.\ 2011,
  \mnras, 412, L108

\bibitem[Slane et al.(1999)]{slane99} Slane, P., Gaensler, 
B.~M., Dame, T.~M., et al.\ 1999, \apj, 525, 357 

\bibitem[Slane et al.(2014)]{slane14}
Slane, P. O.,  Lee, S.-H., Ellison, D. C.,
 Patnaude, D. J., Hughes, J. P. et al. 2014, \apj, 783, 33


\bibitem[Smith et al.(1985)]{smith85} Smith, A., Jones, L. R.,
  Peacock, A., \& Pye, J.~P., 1985, ApJ, 296, 469

\bibitem[Smith et al.(1991)]{smith91} Smith, R.~C., Kirshner, 
R.~P., Blair, W.~P., \& Winkler, P.~F.\ 1991, \apj, 375, 652 

\bibitem[Sullivan et al.(2010)]{sullivan10} Sullivan, M., Conley,
A., Howell, D.~A., et al.\ 2010, MNRAS, 406, 782

\bibitem[Takahashi et al.(2008)]{takahashi08} Takahashi, T., 
Tanaka, T., Uchiyama, Y., et al.\ 2008, \pasj, 60, 131

\bibitem[Tamagawa et al.(2009)]{tamagawa09} Tamagawa, T., et al.~2009,
  PASJ, 61, S155

\bibitem[Tananbaum(1999)]{tananbaum99} Tananbaum, H.\ 1999, IAU Circ.,
  7246, 1

\bibitem[Tanaka et al.(2008)]{tanaka08}
Tanaka, T., Uchiyama, Y., Aharonian, F. A. et al. 2008, \apj, 685, 988

\bibitem[Tatisheff et al.(1998)]{tatischeff98} Tatischeff, V., Ramaty,
  R., \& Kozlovsky, B.\ 1998, ApJ, 504, 874

\bibitem[Thielemann et al.(1996)]{thielemann96} Thielemann, F-K,
  Nomoto, K, \& Hashimoto, M~1996, \apj, 460, 408

\bibitem[Timmes et al.(1996)]{timmes96} Timmes, F.~X., Woosley, 
S.~E., Hartmann, D.~H., \& Hoffman, R.~D.\ 1996, \apj, 464, 332 

\bibitem[Timmes, Brown, \& Truran(2003)]{timmes03} Timmes, F.~X.,
  Brown, E.~F., \& Truran, J.~W.\ 2003, \apj, 590, L83

\bibitem[Tsujimoto \& Shigeyama(2012)]{2012ApJ...760L..38T} Tsujimoto,
  T., \& Shigeyama, T.\ 2012, ApJ, 760, L38

\bibitem[Tuohy et al.(1982)]{tuohy+82}Tuohy, I.~R., Dopita, M.~A.,
  Mathewson, D.~S., Long, K.~S., \& Helfand, D.~J.\ 1982, ApJ, 261,
  473

\bibitem[Uchida et al.(2013)]{uchida13} Uchida, H., Yamaguchi, H., Koyama,
  K.\ 2013, ApJ, 771, 56

\bibitem[Uchiyama et al.(2007)]{Uchiyama07} Uchiyama, Y., Aharonian,
  F.~A., Tanaka, T., Takahashi, T., \& Maeda, Y.\ 2007, \nat, 449, 576

\bibitem[Uchiyama \& Aharonian(2008)]{Uchiyama08} Uchiyama, Y., \&
  Aharonian, F.~A.\ 2008, \apj, 677, L105

\bibitem[van der Heyden et al.(2002)]{vanderheyden02} van der Heyden,
  K.~J., Behar, E., Vink, J., Rasmussen, A.~P., Kaastra, J.~S.,
  Bleeker, J.~A.~M., Kahn, S.~M., \& Mewe, R. 2002, A\&A, 392, 955

\bibitem[Vancura et al.(1995)]{vancura95} Vancura, O., Gorenstein, P.,
  \& Hughes, J.~P. 1995, ApJ, 441, 680


\bibitem[Vink(2008)]{2008A&A...486..837V} Vink, J.\ 2008, \aap, 486, 837

\bibitem[Vink et al.(2001)]{Vink01} Vink, J., Laming, J.~M., Kaastra,
  J.~S., Bleeker, J.~A.~M., Bloemen, H., \& Oberlack, U.\ 2001, \apj,
  560, L79

\bibitem[Vink et al.(2003)]{vink03} Vink, J., Laming, J.~M., Gu, M.F.,
  Rasmussen, A., \& Kaastra, J.S.\ 2003, ApJ, 587, L31

\bibitem[Vink(2012)]{Vink12} Vink, J.\ 2012,  Astronomy and
  Astrophysics Review, 20, 49

\bibitem[Wallerstein et al.(1997)]{wallerstein97} Wallerstein, G., 
Iben, I., Jr., Parker, P., et al.\ 1997, Reviews of Modern Physics, 69, 995 

\bibitem[Warren \& Blondin(2013)]{warren13}
Warren, D.~C., \& Blondin, J.~M.\ 2013, MNRAS, 429, 3099

\bibitem[Warren \& Hughes(2004)]{warrenhughes04} Warren, J.~S., \&
  Hughes, J.~P.\ 2004, \apj, 608, 261

\bibitem[Warren et al.(2005)]{warren05} Warren, J.~S., Hughes, 
J.~P., Badenes, C., et al.\ 2005, \apj, 634, 376 

\bibitem[Williams et al.(2011)]{williams11} Williams, B.~J., Blair,
  W.~P., Blondin, J.~M., et al.\ 2011, ApJ, 741, 96

\bibitem[Williams et al.(2012)]{williams12} Williams, B.~J.,
  Borkowski, K.~J., Reynolds, S.~P., et al.\ 2012, ApJ, 755, 3

\bibitem[Willingale et al.(2002)]{willingale02}
Willingale, R., Bleeker, J.~A.~M., van der Heyden, K.~J., Kaastra J.~S., \&
Vink., J.\ 2003, A\&A, 381, 1039

\bibitem[Woosley \& Heger(2007)]{woosley07} Woosley, S.~E., \& Heger,
  A.\ 2007, Phys.~Rep., 442, 269

\bibitem[Woosley \& Weaver(1995)]{WW95} Woosley, S.~E., \& Weaver,
  T.~A.\ 1995, ApJSS, 101, 181

\bibitem[Woosley et al.(2002)]{woosley02} Woosley, S.~E., Heger, 
A., \& Weaver, T.~A.\ 2002, Reviews of Modern Physics, 74, 1015 

\bibitem[Woosley et al.(2004)]{woosley04} Woosley, S.~E., Wunsch, S.,
  \& Kuhlen, M.\ 2004, ApJ, 607, 921

\bibitem[Wu et al.(1983)]{wu83} Wu, C.-C., Leventhal, M., Sarazin,
  C.L., \& Gull, T.R, 1983, ApJ, 269, L5

\bibitem[Xu, Wang, \& Miller(2011)]{2011RAA....11..537X} 
Xu, J.-L., Wang, J.-J., \& Miller, M.\ 2011, Research in Astronomy and Astrophysics, 11, 537

\bibitem[Yamaguchi et al.(2008)]{yamaguchi08} Yamaguchi, H., Koyama,
  K., Katsuda, S., et al.\ 2008, PASJ, 60, 141

\bibitem[Yamaguchi et al.(2012)]{yamaguchi12} Yamaguchi, H.,
Tanaka, M., Maeda, K., et al.\ 2012, ApJ, 749, 137

\bibitem[Yamaguchi et al.(2013)]{yamaguchi13} 
 Yamaguchi, H., Eriksen, K. A., Badenes, C., et al. 2014, ApJ, 780, 136

\bibitem[Yamaguchi et al.(2014)]{yamaguchi14} Yamaguchi, H., Badenes, C., Petre, R. et al. 2014,
\apj, 785, L27

\bibitem[Yang et al.(2009)]{Yang09} Yang, X.~J., Tsunemi, H., Lu,
  F.~J., \& Chen, L.\ 2009, \apj, 692, 894

\bibitem[Yang et al.(2013)]{yang13} Yang, X.~J., Tsunemi, H., Lu,
  F.~J., et al.\ 2013, \apj, 766, 44
  
\bibitem[Yatsu et al.(2005)]{yatsu05}
Yatsu, Y., Kawai, N., Kataoka, J., Kotani, T., Tamura, K., \& Brinkmann, W. 2005,
\apj, 631, 312

\end{thebibliography}

}
